\documentclass[10pt,final,journal,twocolumn,letterpaper]{IEEEtran}

\IEEEoverridecommandlockouts
\usepackage[dvips]{graphicx}
\usepackage{amsmath}
\usepackage{amssymb}
\usepackage[latin1]{inputenc}
\usepackage{times}
\usepackage{cite}
\usepackage{amsfonts}
\usepackage{changebar}
\usepackage{color}
\usepackage{colortbl} %ADD PACKAGE
\usepackage{multirow}
\usepackage[table]{xcolor}
\usepackage{longtable}

\makeatletter
\let\saved@longtable\longtable
\long\def\foo#1\LT@err#2#3#4!!{\def\longtable{#1#4}}
\expandafter\foo\longtable!!

\long\def\foo#1\@outputpage#2\@outputpage#3!!{%
\def\LT@output{#1\@opcol#2\@opcol#3}}
\expandafter\foo\LT@output!!
\makeatother
\newcommand{\ket}[1]{| #1 \rangle}
\newcommand{\mb}{\mathbf}

\newcommand{\eqr}[1]{Eq.~(\ref{#1})}
\newcommand{\fref}[1]{Fig.~\ref{#1}}
\newcommand{\tref}[1]{Table~\ref{#1}}
\newcommand{\etal}{\textit{et al. }}
\newcommand{\figures}{}
\newcommand{\food}{\hspace{-2.3pt}$-$ \hspace{5pt}}

\ifCLASSINFOpdf
\else
\fi

\begin{document}
\title{The Road From Classical to Quantum Codes: \\A Hashing Bound Approaching Design Procedure}
\author{Zunaira Babar, Panagiotis Botsinis, Dimitrios Alanis, Soon Xin Ng  
and Lajos Hanzo%, {\it Fellow, IEEE}%<-this % stops a space
\thanks{Z. Babar, P. Botsinis, D. Alanis, S. X. Ng, and L. Hanzo are with the School of Electronics and Computer Science, University of Southampton, 
SO17 1BJ, United Kingdom. Email: \{zb2g10,pb8g10,da4g11,sxn,lh\}@ecs.soton.ac.uk.}%
\thanks{
The financial support of EPSRC and the European Research Council under its Advanced Fellow grant is gratefully
acknowledged.  }}
%The financial support of the European Union under the auspices of the CONCERTO
%project, as well as that of the European Research Council under its Advanced Fellow Grant is gratefully
%acknowledged.  }} 

%Telephone: (800) 555--1212, Fax: (888) 555--1212}


\maketitle

\begin{abstract} 
Powerful Quantum Error Correction Codes (QECCs) are required for stabilizing and protecting fragile qubits 
against the undesirable effects of quantum decoherence.
%Similar to the classical codes, this may be achieved by exploiting concatenated code design, which invokes 
%iterative decoding.
Similar to classical codes, 
hashing bound approaching QECCs
may be designed by exploiting a concatenated code structure, which invokes 
iterative decoding.
Therefore, in this paper we provide an extensive step-by-step tutorial for designing EXtrinsic Information Transfer (EXIT) chart aided concatenated quantum codes based on the
underlying quantum-to-classical isomorphism. These design lessons are then exemplified in the context of our proposed
%Furthermore, we have conceived a 
Quantum Irregular Convolutional Code (QIRCC), which constitutes the outer 
component of a concatenated quantum code. The proposed QIRCC can be dynamically adapted to match any given inner code using EXIT
charts, hence achieving a 
%thus, facilitating 
performance close to the hashing bound. It is demonstrated that our QIRCC-based optimized design %operates 
is capable of operating within 0.4 dB of the noise limit. 
\end{abstract}

\begin{keywords}
Quantum Error Correction, Turbo Codes, EXIT Charts, Hashing Bound.
\end{keywords}

% \fi
\IEEEpeerreviewmaketitle
\addtocounter{table}{-1}
\section*{Nomenclature}
\begin{longtable}{l l}
BCH &Bose-Chaudhuri-Hocquenghem \\ 
BIBD &Balanced Incomplete Block Designs \\
BSC &Binary Symmetric Channel\\
CCC &Classical Convolutional Code \\
%CPTP &Completely Positive Trace Preserving\\
CSS &Calderbank-Shor-Steane \\
EA &Entanglement-Assisted\\
EXIT &EXtrinsic Information Transfer \\
IRCC &IRregular Convolutional Code\\
LDGM &Low Density Generator Matrix \\
LDPC &Low Density Parity Check \\
MAP &Maximum \textit{A Posteriori} \\ 
MI &Mutual Information\\ 
PCM &Parity Check Matrix \\
QBER&Qubit Error rate \\
QC&Quasi-Cyclic \\
QCC &Quantum Convolutional Code \\
QECC &Quantum Error Correction Code \\
QIRCC &Quantum IRregular Convolutional Code\\
QLDPC &Quantum Low Density Parity Check \\
QSC &Quantum Stabilizer Code \\
%QTBC &Quantum Tail-biting Block Code \\
QTC &Quantum Turbo Code \\
%QVA &Quantum Viterbi Algorithm \\
RX &Receiver\\
SISO &Soft-In Soft-Out\\
SNR &Signal-to-Noise Ratio\\
TX &Transmitter\\
WER &Word Error Rate
\end{longtable}
\section{Introduction} \label{sec:intro} 
\begin{figure*}[tb]
\begin{center}
    \includegraphics[width = \linewidth]{\figures 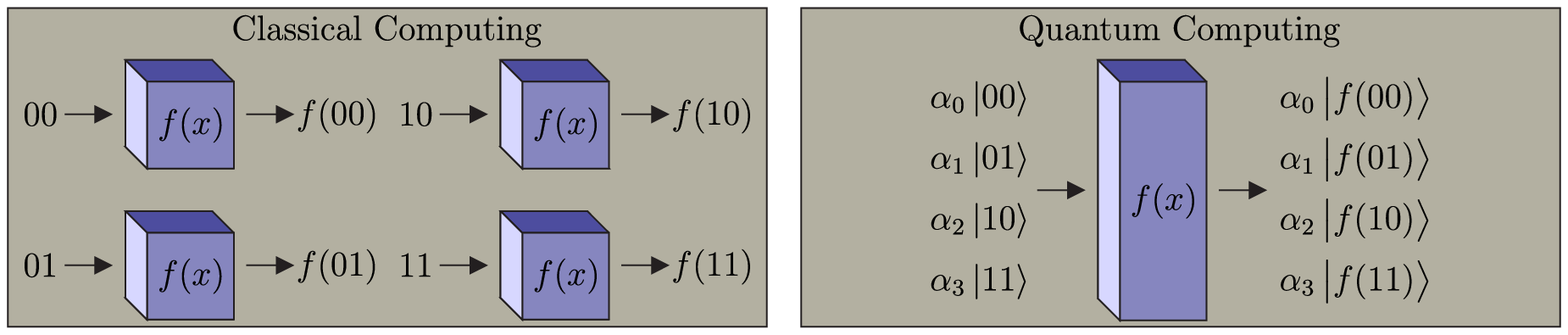}
   \caption{Quantum Parallelism: Given a function $f(x)$, which has a regular global structure such that
    $f(x): \{0,1\}^2 \rightarrow \{0,1\}^2$, a classical system requires four evaluations to compute $f(x)$ for all possible $x \in \{00,01,10,11\}$. 
    By contrast, since a $2$-qubit quantum register can be in a superposition of all the four states concurrently, i.e.
    $\ket{\psi} = \alpha_0 \ket{00} + \alpha_1 \ket{01} + \alpha_2 \ket{10} + \alpha_3 \ket{11}$, quantum computing requires only a single
    classical evaluation to yield the outcome, which is also in a superposition of all the four possibilities, i.e.
    $\alpha_0 \ket{f(00)} + \alpha_1 \ket{f(01)} + \alpha_2 \ket{f(10)} + \alpha_3 \ket{f(11)}$. However, 
    it is not possible to read all the four states because the quantum register collapses to one of the four superimposed states upon measurement.
   Nevertheless, we may manipulate the resultant superposition of the four possible states before observing the quantum register
    for the sake of determining a desired property of 
    the function, as in~\cite{Deutsch85,Deutsch92,Shor1994,Grover:1996:FQM:237814.237866}.}
  \label{fig:Quantum_fig2}
\end{center}
%\vspace{-0.5cm}
\end{figure*}

The laws of quantum mechanics provide a promising solution to our quest for miniaturization and increased processing power, 
as implicitly predicted by Moore's law formulated four decades ago~\cite{myth_hanzo_2012}. This can be attributed to the inherent parallelism associated
with the quantum bits (qubits). More explicitly, in contrast to the classical bits, which can either assume a value
of $0$ or $1$, qubits can exist in a superposition of the two states\footnote{The superimposed state of a qubit may be represented as 
$|\psi\rangle=\alpha|0\rangle+\beta|1\rangle,$
where $|$ $\rangle$ is called Dirac notation or Ket~\cite{dirac82}, which is a standard notation for states in quantum physics, while $\alpha$ and 
$\beta$ are complex numbers with $|\alpha|^2+|\beta|^2=1$. More specifically, a qubit exists in a continuum of states between $\ket{0}$
and $\ket{1}$ until it is `measured' or `observed'. Upon `measurement' it collapses to the state $\ket{0}$ with a
probability of $|\alpha|^2$ and $\ket{1}$ with a probability of $|\beta|^2$. 
%This is reminiscent of a coin spinning in the air. When the coin
%is in the air, there is $50\%$ probability of getting a `head' or a `tail' (analogous to a qubit, which is in the state $|\psi\rangle=\frac{1}{\sqrt{2}}|0\rangle+\frac{1}{\sqrt{2}}|1\rangle$). We cannot be certain 
%whether it is a `head' or a `tail' until the coin settles down, which is synonymous to `measuring' or `observing' a qubit.
}. 
Consequently, while an $N$-bit classical register
can store only a single value, an $N$-qubit quantum register can store all the $2^N$ states concurrently\footnote{A single qubit is 
essentially a vector in the $2$-dimensional Hilbert space. Consequently, %the basis states of
an $N$-qubit composite system, which consists of $N$ qubits, has a $2^N$-dimensional Hilbert space, which is the tensor product of the Hilbert space of the individual qubits.
%are given by the tensor product of the basis states of the individual qubits, which results in a $2^N$-dimensional
%Hilbert space. %For example, a $2$-qubit composite system with $\ket{\psi_1} = \alpha_1 \ket{0} + \beta_1 \ket{1}$ and 
%$\ket{\psi_2} = \alpha_2 \ket{0} + \beta_2 \ket{1}$ may be formulated as follows:
%\begin{align}
% \ket{\psi_1} \otimes \ket{\psi_2} &= (\alpha_1 \ket{0} + \beta_1 \ket{1}) \otimes (\alpha_2 \ket{0} + \beta_2 \ket{1})  \nonumber \\ 
% &= \alpha_1 \alpha_2 \ket{00} + \alpha_1 \beta_2 \ket{01} + \beta_1 \alpha_1 \ket {10} + \beta_1 \alpha_2 \ket{11}, \nonumber
%\end{align}
%where $\otimes$ denotes the tensor product. 
The resulting $N$-qubit state may be generalized as:
\begin{equation}
\alpha_0 \ket{00\dots0} + \alpha_1 \ket{00\dots1} + \dots + \alpha_{2^N-1} \ket{11\dots1}, \nonumber
\end{equation}
where $\alpha_i \in \mathbb{C}$ and $\sum\limits_{i=0}^{2^N-1} |\alpha_i|^2 = 1$.}, allowing
parallel evaluations of certain functions with regular global structure at a complexity cost that is equivalent to a single classical evaluation~\cite{Qbook3,panos2013}, as illustrated in~\fref{fig:Quantum_fig2}. 
Therefore, as exemplified by Shor's factorization algorithm~\cite{Shor1994} and Grover's search 
algorithm~\cite{Grover:1996:FQM:237814.237866}, quantum-based computation is 
capable of solving certain complex problems at a substantially lower complexity, as compared to its classical counterpart.
%Therefore, quantum-based computation is capable of solving complex problems exhibiting an exponentially escalating complexity, which were once considered intractable. 
From the perspective
of telecommunications, this quantum domain parallel processing seems to be a plausible solution for the massive parallel processing required 
for achieving joint
optimization in large-scale communication systems, e.g. quantum assisted multi-user detection~\cite{panos2013, panos_tcom_2014, panos_access_2014} and quantum-assisted
routing optimization for self-organizing networks~\cite{dimitris_access_2014}. 
Furthermore, quantum-based communication is capable of supporting secure data dissemination, where any `measurement' or 
`observation' by an eavesdropper destroys the quantum 
entanglement\footnote{Two qubits are said to be entangled if they cannot be decomposed into the tensor product of the constituent qubits. Let 
us consider the state $\ket{\psi} = \alpha \ket{00} + \beta \ket{11}$, where both $\alpha$ and $\beta$ are non-zero. 
It is not possible to decompose it into two individual qubits because we have:
\begin{equation}
\alpha \ket{00} + \beta \ket{11} \neq (\alpha_1 \ket{0} + \beta_1 \ket{1}) \otimes (\alpha_2 \ket{0} + \beta_2 \ket{1}),
\nonumber
\end{equation}
for any choice of $\alpha_i$ and $\beta_i$ subject to normalization.
Consequently, a peculiar link exists between the two qubits such that measuring one qubit also collapses the other, despite their spatial separation.
More specifically, if we measure the first qubit of $\ket{\psi}$, we may obtain a $\ket{0}$ with a probability of $|\alpha|^2$ and a $\ket{1}$
with a probability of $|\beta|^2$. If the first qubit is found to be $\ket{0}$, then the measurement of the second qubit will definitely be $\ket{0}$. Similarly,
if the first qubit is $\ket{1}$, then the second qubit will also collapse to $\ket{1}$. This mysterious correlation between the two qubits, which doesn't exist
in the classical world,
is called entanglement. It
was termed `spooky action at a distance' by Einstein~\cite{Born1971}.}, 
hence intimating the parties concerned~\cite{Qbook2,Qbook3}. 
Quantum-based communication has given rise to a new range of security paradigms, which
cannot be created using a classical communication system. In this context, quantum key distribution techniques~\cite{QKD1, QKD2}, quantum secure direct 
communication~\cite{0305-4470-35-28-103,PhysRevLett.89.187902} and
the recently proposed unconditional quantum location verification~\cite{malaney1} are of particular significance.

\begin{figure}[tb]
\begin{center}
    \includegraphics[width=0.8\linewidth]{\figures 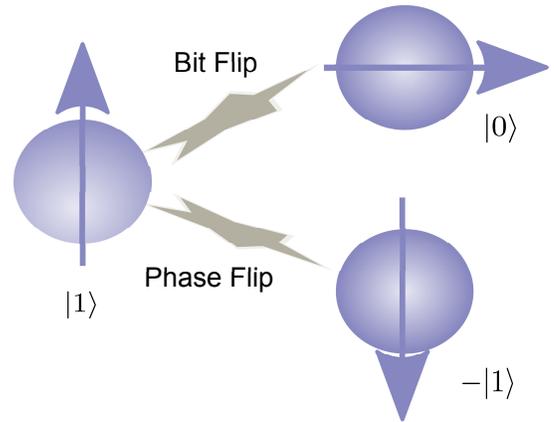}
    \caption{Quantum decoherence characterized by bit and phase flips. \textit{The vertical polarization represents the state $\ket{1}$,
    while the horizontal polarization represents the state $\ket{0}$}.}
  \label{fig:decoherence}
\end{center}
%\vspace{-0.5cm}
\end{figure}

Unfortunately, a major impediment to the practical realization of quantum 
computation as well as communication systems is quantum noise, which is conventionally termed as `decoherence' (loss of the coherent quantum state). 
More explicitly, decoherence is the undesirable interaction of the qubits with the environment~\cite{shor95,preskill1999battling}. 
%For macroscopic systems, this may be viewed as a
%continuous measurement of the qubits by the environment, thus unintentionally `collapsing' them to the classical state. This is the main reason 
%why macroscopic systems behave classically in the real world~\cite{preskill1999battling}.
%On the other hand, for microscopic systems, decoherence 
It may be viewed as the undesirable entanglement of qubits with the environment, which
perturbs the 
fragile superposition of states, %without collapsing them to the classical states, 
thus leading to the detrimental effects of
noise. The overall decoherence process may be characterized either by bit-flips or phase-flips or in fact possibly both, inflicted on the qubits~\cite{shor95}, as depicted in
\fref{fig:decoherence}\footnote{A qubit may be realized in different ways, e.g. 
    two different photon polarizations, different alignments of a nuclear spin, two electronic levels of an atom or the charge/current/energy of a Josephson
    junction.}.
The longer a qubit retains its coherent state (this period is known as the coherence time), the better. This may be achieved with the aid of 
Quantum Error Correction codes (QECCs), which also rely on the peculiar phenomenon of entanglement - hence John Preskill eloquently pointed out that
we are ``fighting entanglement with entanglement''~\cite{Preskill97}. More explicitly, analogously to the classical channel coding techniques, QECCs rectify the impact
of quantum noise (bit and phase flips) for the sake of ensuring that the qubits retain their coherent quantum state with a high 
fidelity\footnote{Fidelity is a measure of closeness of two quantum states~\cite{mark:book}.}, thus in effect beneficially increasing the coherence time of the unperturbed quantum state.
This has been experimentally demonstrated in~\cite{PhysRevLett.81.2152,citeulike:10313454,PhysRevLett.112.150801}.

Similar to the family of classical error correction codes~\cite{shuerror,tc_teq_st_2:book}, which aim for operating close to Shannon's capacity limit, 
QECCs are designed to approach the 
quantum capacity~\cite{PhysRevA.55.1613, Shor2002, 1377491}, or more specifically the hashing bound, which is a lower bound
of the achievable quantum capacity.
%Hashing bound~\cite{bennet99_5q,wilde_turbo2}, which sets the lower bound on the achievable capacity of a quantum channel. 
A significant amount of work has been carried out over the
last few decades to achieve this objective. However, the field of quantum error correction codes is still not as mature as that of their classical 
counterparts.
Recently, inspired by the family of classical near-capacity concatenated codes, which rely on iterative decoding schemes, e.g.~\cite{turbo93,brink:exit-parallel}, 
substantial efforts have been invested in~\cite{qturbo2, wilde_turbo2, babar_QTC_2014} to construct comparable quantum codes. In the light of this 
increasing interest in conceiving 
hashing bound approaching concatenated quantum code design principles, 
the contributions of this paper are: \textit{
\begin{enumerate}
%\item We survey the evolution towards constructing near-capacity concatenated quantum codes with the aid of EXtrinsic Information Transfer (EXIT) charts.
  \item We survey the evolution towards constructing 
hashing bound approaching
concatenated quantum codes with the aid of EXtrinsic Information Transfer (EXIT) charts.
  More specifically, to bridge the gap between the classical and quantum channel coding theory, we provide insights into the transformation 
  from the family of classical codes to the class of quantum codes.
  \item We propose a generically applicable structure for Quantum Irregular Convolutional Codes (QIRCCs), which
can be dynamically adapted to a specific application scenario for the sake of facilitating
hashing bound approaching performance. This is achieved with the aid of the
EXIT charts of~\cite{babar_QTC_2014}.
\item More explicitly, we provide a detailed design example by constructing a $10$-subcode QIRCC and use it as an outer code in concatenation
with the non-catastrophic and recursive inner convolutional code of~\cite{wilde_turbo,wilde_turbo2}. Our QIRCC-based optimized design outperforms both
the design of~\cite{wilde_turbo2}, as well as the exhaustive-search based optimized design of~\cite{babar_QTC_2014}.
\end{enumerate}}

This paper is organized as follows. We commence by outlining our design objectives in Section~\ref{sec:designObj}. We then provide a comprehensive historical overview 
of QECCs in Section~\ref{sec:history}.
We detail the underlying stabilizer formalism in Section~\ref{sec:stabilizer} by providing insights into constructing quantum stabilizer
codes by cross-pollinating their design with the aid of the well-known classical codes. %, which is achieved by exploiting the underlying isomorphism. 
We then proceed with the design of concatenated quantum codes in Section~\ref{sec:CQCC}, with a special emphasis on their code construction as well as 
on their decoding procedure.
In Section~\ref{sec:near-cap-design}, we will detail the EXIT-chart aided code design principles, providing insights into the application of EXIT
charts for the design of quantum codes. We will then present our proposed QIRCC design example in Section~\ref{sec:QIRCC}, followed by our simulation results
in Section~\ref{sec:results}. Finally, our conclusions and design guidelines are offered in Section~\ref{sec:conclusion}.
\section{Design Objectives} \label{sec:designObj}
Meritorious families of quantum error correction codes can be derived %by exploring the analogy between classical and quantum codes, while
from the known classical codes by exploiting the underlying quantum-to-classical isomorphism, while also
taking into account the peculiar laws of quantum mechanics. This transition from the classical to the quantum domain
must address the following challenges~\cite{Qbook2}: %applied to qubits because
\begin{itemize}
 \item \textbf{No-Cloning Theorem:} Most classical codes are based on the transmission of multiple replicas of the same bit, e.g. in a simple rate-$1/3$
repetition code each information bit is transmitted thrice. This is not possible in the quantum domain according to the 
no-cloning theorem~\cite{citeulike:507853}, which states that an 
arbitrary unknown quantum state cannot be copied/cloned\footnote{No-cloning theorem is a direct consequence of the linearity of transformations.
Let us assume that $U$ is a copying operation, which maps the arbitrary states $\ket{\psi}$ and $\ket{\phi}$ as follows:
\begin{equation}
 U \ket{\psi} = \ket{\psi} \otimes \ket{\psi}, \; \; \; \; \;
U \ket{\phi} = \ket{\phi} \otimes \ket{\phi}. \nonumber
\end{equation}
Since the transformation $U$ must be linear, we should have:
\begin{equation}
 U (\ket{\psi} + \ket{\phi}) = U \ket{\psi} + U \ket{\phi} = \ket{\psi} \otimes \ket{\psi} + \ket{\phi} \otimes \ket{\phi}. \nonumber
\end{equation}
However,
\begin{equation}
 \ket{\psi} \otimes \ket{\psi} + \ket{\phi} \otimes \ket{\phi} \neq (\ket{\psi} + \ket{\phi}) \otimes (\ket{\psi} + \ket{\phi}). \nonumber
\end{equation}

}. %as discussed in SectionXX.
\item \textbf{Continuous Nature of Quantum Errors:} In contrast to the classical errors, which are discrete with bit-flip being the only type of error, 
a qubit may experience both a bit error as well as a 
phase error or in fact both, as depicted in \fref{fig:decoherence}. These impairments have a continuous nature and the erroneous qubit may lie anywhere on the surface of the 
Bloch sphere\footnote{A qubit $\ket{\psi} = \alpha\ket{0}+\beta\ket{1}$,
whose orthogonal basis are $|0\rangle$ and $|1\rangle$, can be visualized in 
3$D$ as a unique point on the surface of a sphere (with unit radius) called Bloch sphere~\cite{Qbook2}.}.
\item \textbf{Qubits Collapse upon Measurement:} `Measurement' of the received bits is a vital step representing a hard-decision operation in the field
of classical error correction, 
but this is not feasible in the quantum domain, since qubits collapse to classical bits upon measurement.
\end{itemize}
In a nutshell, a classical $(n,k)$ binary code is designed to protect discrete-valued message sequences of length $k$ by encoding them into one of the $2^k$ discrete codewords
of length $n$. By contrast, since a quantum state of $k$ qubits is specified by $2^k$ continuous-valued complex coefficients, quantum error correction aims 
for encoding a $k$-qubit state into an $n$-qubit state, so that all the $2^k$ complex coefficients can be perfectly restored~\cite{sparse1}. For example, let 
$k = 2$, then the $2$-qubit information word $\ket{\psi}$ is given by:
\begin{equation}
 |\psi\rangle = \alpha_{0}|00\rangle + \alpha_{1}|01\rangle + \alpha_{2}|10\rangle + \alpha_{3}|11\rangle. 
 \label{eq:2qubit}
\end{equation}
Consequently, the error correction algorithm would aim for 
correctly preserving all the four coefficients, i.e. $\alpha_{0}$, $\alpha_{1}$, $\alpha_{2}$ and $\alpha_{3}$. It is interesting to note here that
although the coefficients $\alpha_{0}$, $\alpha_{1}$, $\alpha_{2}$ and $\alpha_{3}$ are continuous in nature, yet the entire continuum of 
errors can be corrected, if we can correct a discrete set of errors, 
i.e. bit (Pauli-$\mathbf{X}$)\footnote{A qubit $\ket{\psi} = \alpha\ket{0} + \beta\ket{1}$ may be represented as $\begin{pmatrix}
                                                                                                                                       \alpha \\
                                                                                                                                       \beta
                                                                                                                                      \end{pmatrix}
$ in vector notation. Consequently, $\mathbf{I}$, $\mathbf{X}$, $\mathbf{Y}$ and $\mathbf{Z}$ Pauli operators (or gates), which act on a single qubit, are 
defined as follows:
\begin{equation}
 \mathbf{I}=\begin{pmatrix}
  1& 0 \\
  0& 1
\end{pmatrix}, \ 
\mathbf{X}=\begin{pmatrix}
  0& 1 \\
  1& 0
\end{pmatrix}, \ 
\mathbf{Y}=\begin{pmatrix} 
  0& -i \\
  i& 0
\end{pmatrix}, \ 
\mathbf{Z}=\begin{pmatrix}
  1& 0 \\
  0& -1
\end{pmatrix}, \nonumber
\end{equation}
where the $\mathbf{X}$, $\mathbf{Y}$ and $\mathbf{Z}$ operators anti-commute with each other. The output of a Pauli operator 
may be computed using matrix multiplication, e.g.:
\begin{equation}
 \mathbf{X} (\alpha\ket{0} + \beta\ket{1}) = \begin{pmatrix}
  0& 1 \\
  1& 0
\end{pmatrix} \times
\begin{pmatrix}
\alpha \\
\beta
\end{pmatrix} =
\begin{pmatrix}
 \beta \\
 \alpha
\end{pmatrix}=
\beta\ket{0} + \alpha\ket{1}. \nonumber
\end{equation}
}, phase (Pauli-$\mathbf{Z}$) as well as both 
(Pauli-$\mathbf{Y}$) errors inflicted on either or both qubits~\cite{Qbook2}. 
This is because measurement results in collapsing the entire
continuum of errors to a discrete set.
More explicitly, for $\ket{\psi}$ of \eqr{eq:2qubit}, the discrete error set is as follows: 
\begin{align}
 \{&\mathbf{IX},\mathbf{IZ},\mathbf{IY},\mathbf{XI},\mathbf{XX},\mathbf{XZ},\mathbf{XY},\mathbf{ZI},\mathbf{ZX},\mathbf{ZZ}, \nonumber \\ 
 &\mathbf{ZY},\mathbf{YI},\mathbf{YX},\mathbf{YZ},\mathbf{YY}\}.
\end{align}
However, the errors $\mathbf{X}$, $\mathbf{Y}$ and $\mathbf{Z}$ may occur with varying frequencies.
In this paper, we will focus on the specific design of codes conceived for mitigating the deleterious effects of the quantum depolarizing channel, which has been extensively investigated in the
context of QECCs~\cite{sparse1,qturbo1,renes2012}. Briefly, 
a depolarizing channel, which is characterized by the probability $p$, inflicts an error $\mathcal{P} \in \mathcal{G}_n$ on $n$
qubits\footnote{A single qubit Pauli group $\mathcal{G}_1$ is a group formed by the Pauli matrices $\mathbf{I}$, $\mathbf{X}$, $\mathbf{Y}$ and 
$\mathbf{Z}$, which is closed under multiplication. Therefore, it consists of all the Pauli matrices together with the multiplicative factors 
$\pm 1$ and $\pm i$, i.e. we have:
\begin{equation}
 \mathcal{G}_1 \equiv \{\pm \mathbf{I}, \pm i \mathbf{I},\pm \mathbf{X}, \pm i \mathbf{X}, \pm \mathbf{Y}, \pm i \mathbf{Y}, \pm \mathbf{Z}, \pm i \mathbf{Z}\}. \nonumber
\end{equation}
The general Pauli group $\mathcal{G}_n$ is an $n$-fold tensor product of $\mathcal{G}_1$.}, where  
each qubit may independently experience
either a bit flip ($\mathbf{X}$), a phase flip ($\mathbf{Z}$) or both ($\mathbf{Y}$) with a probability of $p/3$.
%More precisely, a quantum depolarizing channel $\mathcal{N}_p(\rho)$ can be viewed as a Completely-Positive Trace-Preserving (CPTP) mapping, which maps a state $\rho$ onto a linear combination 
%of itself and the maximally entangled state, which is given by:
%\begin{equation}
% \mathcal{N}_p(\rho) = (1-p) \rho + \frac{p}{3}\mathbf{X} \rho \mathbf{X} + \frac{p}{3}\mathbf{Y} \rho \mathbf{Y} +\frac{p}{3}\mathbf{Z} \rho \mathbf{Z},
%\label{eq:d_ch}
%\end{equation}

An ideal code $\mathcal{C}$ designed for a depolarizing channel may be characterized in terms of the channel's depolarizing probability $p$
and its coding rate $R_Q$. Here the coding rate $R_Q$ is measured in terms of the number of qubits transmitted per channel use, i.e. we have
$R_Q = k/n$, where $k$ and $n$ are the lengths of the information word and codeword, respectively. Analogously to Shannon's classical capacity, 
the relationship between $p$ and $R_Q$ for the depolarizing channel 
is defined by the hashing bound, which sets a lower limit on the achievable quantum capacity\footnote{Quantum codes are inherently degenerate in nature because different errors may have the same impact on the quantum
state. For example, let $\ket{\psi} = \ket{00} + \ket{11}$. Both errors $\mathbf{IZ}$ and $\mathbf{ZI}$ acting on $\ket{\psi}$ yield the same corrupted 
state, i.e. $(\ket{00} - \ket{11})$, and are therefore classified as degenerate errors. Due to this degenerate nature of the channel errors,
the ultimate capacity of quantum channel can be higher than that defined by the hashing bound~\cite{q_cap_1998, PhysRevLett.98.030501}. 
However, none of the codes known to date outperform the hashing bound at practically feasible frame lengths.}.
The hashing bound is given by~\cite{bennet99_5q,wilde_turbo2}:
\begin{equation}
 C_Q(p) = 1 - H_2(p) - p \log_2(3),
 \label{eq:capacity_q}
\end{equation}
where $H_2(p)$ is the binary entropy function. More explicitly, for a given $p$, if a random code $\mathcal{C}$ of a sufficiently long codeword-length is chosen
such that its coding rate obeys $R_Q \leq C_Q(p)$, then $\mathcal{C}$ may yield an infinitesimally
low Qubit Error Rate (QBER) for a depolarizing probability of $p$. 
It must be noted here that intuitively a low QBER corresponds to a high fidelity between the transmitted and the decoded quantum state. 
More explicitly, for a given value of $p$, $C_Q(p)$ gives the hashing limit on the coding rate.
Alternatively, for a given coding rate $R_Q$, where we have $R_Q = C_Q(p^*)$, $p^*$ gives the hashing
limit on the channel's depolarizing probability. In duality to the classical domain, this may also be referred to as the noise limit. An ideal quantum code should be capable of ensuring reliable transmission close to the noise limit $p^*$.
Furthermore, for any arbitrary depolarizing probability $p$,
its discrepancy with respect to the noise limit $p^*$ may be computed in decibels (dB) as follows~\cite{wilde_turbo2}:
\begin{equation}
\text{Distance from hashing bound} \triangleq  10 \times \log_{10} \left(\frac{p^*}{p}\right).
\label{eq:distance}
\end{equation} 
Consequently, our quantum code design objective is to minimize the discrepancy with respect to the hashing bound, thereby yielding a 
hashing bound approaching code design.

\begin{figure}[tb]
\begin{center}
    \includegraphics[width=\linewidth]{\figures 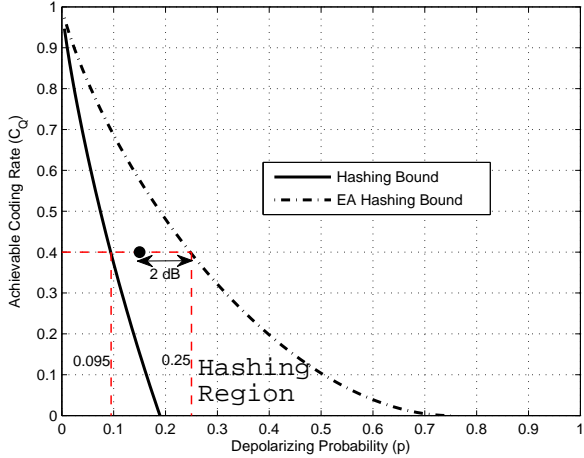}
    \caption{Unassisted and EA hashing bounds characterized by \eqr{eq:capacity_q} and \eqr{eq:capacity_q_EA:max}, respectively,
    for the quantum depolarizing channel. The region enclosed by these two bounds, which is labeled the hashing region, defines the capacity for varying number of pre-shared entangled
    qubits ($c$). At $R_Q = 0.4$, the unassisted hashing bound gives a noise limit of $p^* = 0.095$, while the maximally entangled
    hashing bound increases the limit to $p^* = 0.25$. The circle represents a maximally entangled code with $R_Q = 0.4$, which ensures reliable transmission for $p \leq 0.15$, thus operating
    at a distance of $2$~dB from the noise limit.}
  \label{fig:cap-H}
\end{center}
%\vspace{-0.5cm}
\end{figure}

\begin{figure}[tb]
\begin{center}
    \includegraphics[width=0.8\linewidth]{\figures 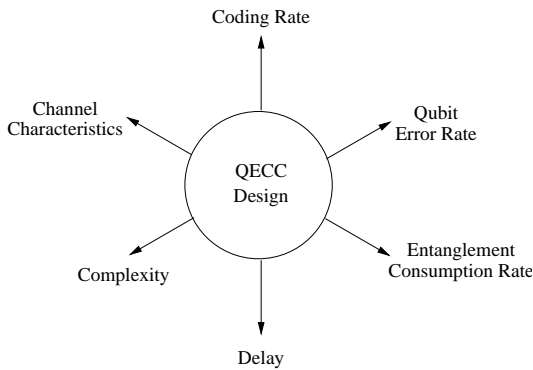}
    \caption{Stylized illustration of the conflicting design challenges, which are involved in the design of quantum codes.}
  \label{fig:Design_TO}
\end{center}
%\vspace{-0.5cm}
\end{figure}

It is pertinent to mention here the Entanglement-Assisted (EA) regime of~\cite{bowen2002,brun2006,brun2007,hsieh2007}, 
where the entanglement-assisted code $\mathcal{C}$ 
is characterized by an additional parameter $c$. Here $c$ is the number of entangled qubits pre-shared between the transmitter and the receiver, thus 
leading to the terminology of being 
entanglement-assisted\footnote{A quantum
code without pre-shared entanglement, i.e. $c = 0$, may be termed as an unassisted quantum code. EA quantum
codes will be discussed in detail in Section~\ref{sec:EA-stab}.}.
It is assumed furthermore that these pre-shared entangled qubits are transmitted over a 
noiseless quantum channel. The resultant EA hashing bound is given by~\cite{wilde_turbo2,EA-hashing-2014}:
\begin{equation}
 C_Q(p) = 1 - H_2(p) - p \log_2(3) + \textup{E},
 \label{eq:capacity_q_EA}
\end{equation}
where the so-called entanglement consumption rate is $\textup{E} = \frac{c}{n}$. Furthermore, the value of $c$ may be varied from $0$ to a maximum of 
$(n-k)$. For the family of
maximally entangled codes associated
with $c = (n - k)$, the EA hashing bound of \eqr{eq:capacity_q_EA} is reduced to~\cite{wilde_turbo2,EA-hashing-2014}:
\begin{equation}
 C_Q(p) = 1 - \frac{H_2(p) - p \log_2(3)}{2}.
 \label{eq:capacity_q_EA:max}
\end{equation}
Therefore, the resultant hashing region of the EA communication is bounded by \eqr{eq:capacity_q} and \eqr{eq:capacity_q_EA:max},
which is also illustrated in \fref{fig:cap-H}.
To elaborate a little further, let us assume that the desired coding rate is $R_Q = 0.4$. Then, as gleaned from \fref{fig:cap-H}, the noise limit for the 
`unassisted' quantum code is around
$p^* = 0.095$, which increases to around $p^* = 0.25$ with the aid of maximum entanglement, i.e. we have $\textup{E} = 1 - R_Q = 0.6$. Furthermore, 
$0 < \textup{E} < 0.6$ will result in bearing noise limits
in the range of $0.095 < p^* < 0.25$. Let us assume furthermore that we design a maximally entangled code $\mathcal{C}$ for $R_Q = 0.4$, so that it ensures reliable transmission for 
$p \leq 0.15$. Based on \eqr{eq:distance}, the performance
of this code (marked with a circle in \fref{fig:cap-H}) is around 
$[10 \times \log10 (\frac{0.25}{0.15})] = 2$~dB away from the noise limit. 
We may approach the noise limit more closely by optimizing a range of conflicting design challenges,
which are illustrated in the stylized representation of \fref{fig:Design_TO}. 
For example, we may achieve a lower QBER by increasing the code length. However, this in turn incurs longer delays. Alternatively,
we may resort to more complex code designs for reducing the QBER, which may also be reduced by
employing codes having lower coding rates or higher entanglement consumption rates, thus requiring more transmitted qubits or entangled qubits. 
Hence striking an appropriate compromise, which meets these conflicting design challenges, is
required.
%For example, we may approach closer to the
%noise limit or achieve a lower QBER by increasing the code length. However, this in turn incurs longer delays. Alternatively,
%we may resort to more complex code designs to lower the QBER. The QBER may also be reduced by
%employing codes having lower coding rates or higher entanglement consumption rates, thus imposing a higher demand on the available resources,
%i.e. transmitted qubits or entangled qubits. An optimized design, which meets these conflicting design challenges, is therefore
%required.%Hence, a range of conflicting design challenges exist, 
%as illustrated in the stylized representation of \fref{fig:Design_TO}.
\section{Historical Overview of Quantum Error Correction Codes}\label{sec:history}
A major breakthrough in the field of quantum information processing was marked by Shor's pioneering work on quantum error correction codes, which dispelled the notion
that conceiving QECCs was infeasible due to the existence of the no-cloning theorem. Inspired by the
classical $3$-bit repetition codes, Shor conceived the first quantum code in his seminal paper~\cite{shor95}, which was published in 1995.
The proposed code had a coding rate of $1/9$ and was capable of correcting only single qubit errors.
This was followed by Calderbank-Shor-Steane (CSS) codes, invented independently by Calderbank and Shor~\cite{CS} as well as by Steane~\cite{steane, steane2},
which facilitated the design of good quantum codes from the known classical binary linear codes. More explicitly, CSS codes may be defined as follows:

\textit{An $[n,k_1-k_2]$ CSS code, which is capable of correcting $t$ bit errors as well as phase errors, can be constructed from classical
linear block codes $C_1(n,k_1)$ and $C_2(n,k_2)$, if $C_2 \subset C_1$ and both $C_1$ as well as the 
dual\footnote{If $G$ and $H$ are the generator and parity check matrices for any linear block code $C$, then its dual code $C^{\bot}$ is a
unique code with $H^T$ and $G^T$ as the generator and parity check matrices respectively.} of $C_2$, i.e. $C_2^{\bot}$, 
can correct $t$ errors. Here, $C_1$ is used for correcting bit errors, while $C_2^{\bot}$ is used for phase-error correction.}

Therefore, with the aid of CSS construction, the overall problem of finding good quantum codes was reduced to finding good 
dual-containing\footnote{Code $C$ with parity check matrix $H$ is said to be dual-containing if it contains its dual code $C^{\bot}$, i.e. 
$C^{\bot} \subset C$ and $HH^T = 0$.} or self-orthogonal
classical codes. Following these principles, the classical
$[7,4,3]$ Hamming code was used to design a $7$-qubit Steane code~\cite{steane2} having a coding rate of $1/7$, which is capable of correcting single isolated errors
inflicted on the transmitted codewords. Finally, Laflamme~\etal~\cite{5qubit} and Bennett~\etal~\cite{bennet99_5q} independently proposed the optimal single error correcting code in 1996, which required only $4$ redundant 
qubits. 

\begin{figure}[tb]
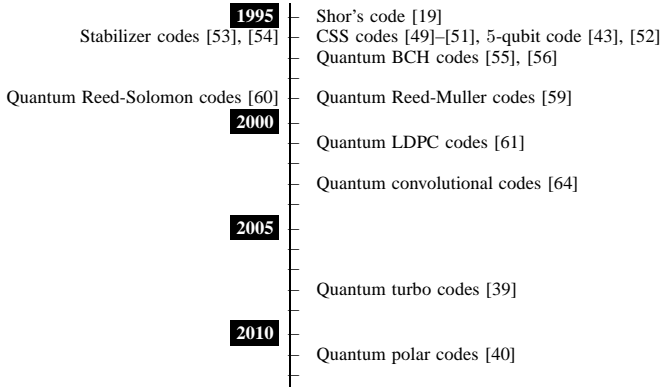

\begin{center}
\scalebox{0.7}{
\begin{tabular}{r |@{\food} l}
\colorbox{black}{\color{white} \textbf{1995}} & Shor's code~\cite{shor95}\\
Stabilizer codes~\cite{got2,G97} & CSS codes~\cite{CS,steane, steane2}, $5$-qubit code~\cite{5qubit,bennet99_5q}\\
     & Quantum BCH codes~\cite{BCH_97_1,BCH_98_1}\\
     &\\
Quantum Reed-Solomon codes~\cite{reed1} & Quantum Reed-Muller codes~\cite{reed0}\\
\colorbox{black}{\color{white} \textbf{2000}}     &\\
     &Quantum LDPC codes~\cite{qldpc2001}\\
     &\\
     &Quantum convolutional codes~\cite{ollivier2003}\\
     &\\
\colorbox{black} {\color{white} \textbf{2005}}     &\\
     &\\
     &\\
     &Quantum turbo codes~\cite{qturbo1}\\
     &\\
\colorbox{black} {\color{white} \textbf{2010}}     &\\
     &Quantum polar codes~\cite{renes2012}\\
     &\\
\end{tabular}
}
\caption{Major milestones achieved in the history of quantum error correction codes.}
\label{fig:timeline}
\end{center}
\end{figure}

Following these developments, Gottesman formalized the notion of constructing quantum codes from the classical binary and quaternary codes by establishing
the theory of Quantum Stabilizer Codes (QSCs)~\cite{got2} in his Ph.D thesis~\cite{G97}. In contrast to the CSS construction, the stabilizer formalism defines
a more general class of quantum codes, which imposes a more relaxed constraint than the CSS codes. Explicitly, the resultant quantum code structure
can either assume a CSS or a non-CSS (also called unrestricted) structure, but it has to meet the symplectic product criterion\footnote{Further details are given in Section~\ref{sec:p2b}.}. More specifically, stabilizer codes constitute a broad class of quantum codes, which subsumes CSS codes as a subclass and has undoubtedly 
provided a firm foundation for a wide variety of quantum codes developed, including for example quantum  
Bose-Chaudhuri-Hocquenghem (BCH) codes~\cite{BCH_97_1,BCH_98_1, bch1, BCH_99_1}, quantum Reed-Solomon codes~\cite{reed0, reed1}, 
Quantum
Low Density Parity Check (QLDPC) codes~\cite{qldpc2001, sparse1, camara2005, camara2007},
Quantum Convolutional Codes (QCCs)~\cite{ollivier2003,ollivier2004, forney2005,forney2007}, Quantum Turbo Codes (QTCs)~\cite{qturbo1,qturbo2} as well as quantum polar codes~\cite{renes2012,6472318,6781023}.
These major milestones achieved in the history of quantum error correction codes are chronologically arranged in \fref{fig:timeline}. 
Let us now look deeper into the development of QCCs, QLDPC codes and QTCs, which have been the prime focus of most recent research both in the classical
as well as in the quantum domain.

The inception of QCCs dates back to 1998. Inspired by the higher coding efficiencies of Classical Convolutional Codes (CCCs) as compared
to the comparable block codes and the low latency associated with the online encoding and decoding of CCCs~\cite{conv_book1}, Chau
conceived the first QCC in~\cite{chau1998}. He also generalized the classical Viterbi decoding algorithm for the class of quantum codes 
in~\cite{chau1999}, but he overlooked some crucial encoding and decoding aspects. Later, Ollivier~\etal~\cite{ollivier2003,ollivier2004} 
revisited the class of stabilizer-based convolutional codes. Similar to the classical Viterbi decoding philosophy, they also conceived
a look-up table based quantum Viterbi algorithm for the maximum likelihood decoding of QCCs, whose complexity increases linearly with the number of encoded qubits. Ollivier~\etal also 
derived the corresponding online encoding and decoding circuits having complexity which increased linearly with the number of encoded qubits. 
% having linear gate complexity. 
Unfortunately, their proposed rate-$1/5$ single-error
correcting QCC did not provide any performance or decoding complexity gain over the rate-$1/5$ single-error correcting block code of~\cite{5qubit}. 
Pursuing this line of research, Almeida~\etal~\cite{almeida2004} 
constructed a rate-$1/4$ single-error correcting Shor-type concatenated
QCC from a classical CC$(2,1,2)$ and invoked the classical syndrome-based trellis decoding for the quantum domain. Hence, the proposed QCC had a higher 
coding rate than the QCC of~\cite{ollivier2003,ollivier2004}. However, this
coding efficiency was achieved at the cost of a relatively high encoding complexity associated with the concatenated trellis structure. It must be 
pointed out here that the pair of independent trellises used for decoding the bit and phase errors impose
a lower complexity than a large joint trellis would.
%because bit and phase errors were decoded independently
%using two trellis. The associated encoding complexity was also high because of the concatenated structure. 
Finally, 
Forney~\etal~\cite{forney2005,forney2007} designed rate-$(n-2)/n$ QCCs
comparable to their classical counterparts, thus
providing higher coding efficiencies than the comparable block codes. Forney~\etal~\cite{forney2005,forney2007} achieved this by 
invoking arbitrary classical 
self-orthogonal rate-$1/n$ $\mathbb{F}_4$-linear and $\mathbb{F}_2$-linear convolutional codes for constructing unrestricted and CSS-type QCCs, respectively. 
%Similar to the classical convolutional codes, the QCCs of\cite{forney2005,forney2007} provided a higher coding efficiency
%than the comparable quantum block codes, despite their lower decoding complexity.
Forney~\etal~\cite{forney2005,forney2007} also conceived a simple decoding algorithm for single-error correcting codes.
%They also derived the corresponding Quantum Tail-biting Block Codes (QTBCs), which had the same rate, performance and decoding complexity as the parent QCCs.
Both the coding efficiency and the decoding complexity of the aforementioned QCC structures are compared in \tref{tab:qcc:comp}. 
\begin{table*}[tb] 
 \centering
 \definecolor{tcA}{rgb}{0.627451,0.627451,0.643137}
\begin{center}
\begin{tabular}{|l|l|l|}
%\begin{tabular}{|l|l|l|}
% use packages: color,colortbl
\hline
\rowcolor{tcA}
 \textbf{Author(s)}& \textbf{Coding Efficiency} & \textbf{Decoding Complexity} \\
\hline
 Ollivier and Tillich~\cite{ollivier2003,ollivier2004}&Low&Moderate\\ \hline
 Almeida and Palazzo~\cite{almeida2004}&Moderate&Moderate\\ \hline
 Forney~\etal~\cite{forney2005,forney2007}&High&Low \\ \hline   
 \end{tabular}
 \end{center}
 \caption{Comparison of the Quantum Convolutional Code (QCC) structures.} 
 \label{tab:qcc:comp}
\end{table*}
Furthermore, in the spirit of finding new constructions for QCCs, Grassl~\etal~\cite{grassal2005, grassal2007} constructed QCCs using the classical 
self-orthogonal product codes,
while Aly~\etal explored various algebraic constructions in~\cite{aly2007} and~\cite{aly2008_2}, where QCCs were derived from classical BCH codes
and Reed-Solomon and Reed-Muller codes, respectively.
Recently, Pelchat and Poulin made a major contribution to the decoding of QCCs by proposing degenerate Viterbi decoding~\cite{deg_poulin}, which runs the
Maximum \textit{A Posteriori} (MAP) algorithm~\cite{tc_teq_st_2:book} over the equivalent classes of degenerate errors, thereby improving the attainable performance.
The major contributions to the development of QCCs are summarized in \tref{tab:qcc}.
\begin{table*}[tb] 
 \centering
 \definecolor{tcA}{rgb}{0.627451,0.627451,0.643137}
\begin{center}
\begin{tabular}{|l|p{3cm}|p{12cm}|}
%\begin{tabular}{|l|l|l|}
% use packages: color,colortbl
\hline
\rowcolor{tcA}
 \textbf{Year}& \textbf{Author(s)} & \textbf{Contribution} \\
\hline
 1998& Chau~\cite{chau1998}& The first QCCs were developed. 
 Unfortunately, some important encoding/decoding aspects were ignored.\\ \hline
 1999& Chau~\cite{chau1999}& Classical Viterbi decoding algorithm was generalized to the quantum domain. However, similar to~\cite{chau1998}, some crucial
 encoding/decoding aspects were overlooked.\\ \hline
 2003& Ollivier and Tillich~\cite{ollivier2003,ollivier2004}& Stabilizer-based convolutional codes and
their maximum likelihood decoding using the Viterbi algorithm were revisited to overcome the deficiencies of~\cite{chau1998,chau1999}. Failed to provide better performance or decoding complexity than the comparable block codes.\\ \hline
 2004& Almeida and Palazzo~\cite{almeida2004}& Shor-type concatenated QCC was conceived and classical syndrome trellis was
 invoked for decoding. A high coding efficiency was achieved at the cost of a relatively high encoding complexity. \\ \hline
 2005& Forney~\etal~\cite{forney2005,forney2007}& Unrestricted and CSS-type QCCs were derived from arbitrary classical self-orthogonal $\mathbb{F}_4$ and $\mathbb{F}_2$ CCCs, 
 respectively, yielding a higher coding efficiency as well as a lower decoding complexity than the comparable block codes. 
 %QTBCs, with the same rate, performance and decoding complexity as the parent QCCs, were also conceived. 
 \\ \hline   
 2005& Grassl and Rotteler~\cite{grassal2005,grassal2007}& Conceived a new construction for QCCs from the classical self-orthogonal product codes.\\ \hline
 2007& Aly~\etal~\cite{aly2007}& Algebraic QCCs dervied from BCH codes.\\ \hline
 2008& Aly~\etal~\cite{aly2008_2}& Algebraic QCCs constructed from Reed-Solomon and Reed-Muller Codes.\\ \hline
 2013& Pelchat and Poulin~\cite{deg_poulin} & Degenerate Viterbi decoding was conceived, which runs the
MAP algorithm over the equivalent classes of degenerate errors, thereby improving the performance.\\ \hline
 %2010& Wilde and Brun& Entanglement-assisted QCC developed~\cite{mark2010}.\\ \hline
 %2012& Tan and Li& Low-complexity practical syndrome decoder for QCCs developed~\cite{tan2012}.\\ \hline
 \end{tabular}
 \end{center}
 \caption{Major contributions to the development of Quantum Convolutional Codes (QCCs).} 
 \label{tab:qcc}
\end{table*}

Although convolutional codes provide a somewhat better performance than the comparable block codes, yet they are not powerful enough to
yield a capacity approaching performance, when used on their own. Consequently,
the desire to operate close to the achievable capacity of \fref{fig:cap-H} at an affordable decoding complexity further motivated researchers to design 
beneficial quantum counterparts of the classical LDPC codes~\cite{gallager62}, which achieve information rates close to the Shannonian capacity limit with the aid
of iterative decoding schemes. Furthermore, the sparseness of the LDPC matrix is of particular
interest in the quantum domain, because it requires only a small
number of interactions per qubit during the error correction
procedure, thus facilitating fault-tolerant decoding. Moreover, this sparse nature also makes QLDPC codes highly degenerate.
In this context, Postol\cite{qldpc2001} conceived the first example of a non-dual-containing CSS-based QLDPC code from 
a finite geometry based classical LDPC in 2001. However, he did not present a generalized formalism for constructing QLDPC codes
from the corresponding classical codes.
Later, Mackay~\etal~\cite{sparse1} proposed various code structures (e.g. bicycle codes and unicycle codes) for constructing QLDPC codes from the 
family of classical 
dual-containing LDPC codes. Among the proposed constructions, the bicycle codes were found to exhibit the best performance.
It was observed that unlike good classical LDPC codes, which have at most a single overlap between the rows of the 
Parity Check Matrix (PCM), dual-containing QLDPC codes must have an even number of overlaps. This in turn results in many unavoidable length-4 cycles, which significantly
impair the attainable performance of the message passing decoding algorithm. Furthermore, the minimum distance of the proposed
codes was upper bounded by the row weight. Additionally, Mackay~\etal also proposed the class of Cayley graph-based
dual-containing codes in~\cite{qldpc_mackay2007}, which were further investigated by Couvreur~\etal in~\cite{cayley2011,couvreur2013}.
Cayley-graph based constructions yield QLDPC codes whose minimum distance has a lower bound, which is a logarithmic function 
of the code length, thus the minimum distance can be improved by extending the codeword (or block) length, albeit again, only logarithmically. However, this is achieved at the cost of an increased
decoding complexity imposed by the row weight, which also increases logarithmically with the code length. 
Aly~\etal contributed to these developments by constructing dual-containing QLDPC codes from finite geometries in~\cite{Aly2008},
while Djordjevic exploited the Balanced Incomplete Block Designs (BIBDs) in~\cite{djo2008}, albeit neither of these provided any gain
over Mackay's bicycle codes. Furthermore, Lou~\etal~\cite{Lou2005, Lou2006} 
invoked the non-dual-containing CSS structure by using both the generator and 
the PCM of classical Low Density Generator Matrix (LDGM) based codes. % for constructing non-dual-containing CSS-based QLDPCs.
Unfortunately, the proposed LDGM based constructions also suffered from length-$4$ cycles, which in turn required
a modified Tanner graph and code doping for decoding, thereby imposing a higher decoding complexity. The only exceptions to length-$4$ cycles 
were constituted by the class
of Quasi-Cyclic (QC) QLDPC codes 
conceived by Hagiwara~\etal~\cite{quasi_ldpc2007}, whereby the constituent PCMs of non-dual-containing CSS-type QLDPCs were constructed from a pair of QC-LDPC codes found using
algebraic combinatorics. The resultant codes had at minimum girth of $6$, but they did not outperform MacKay's bicycle codes conceived in~\cite{sparse1}.
Hagiwara's design of~\cite{quasi_ldpc2007} was extended to non-binary QLDPC codes in~\cite{kasai2011, hagi2012}, which operate closer to the
hashing limit than MacKay's bicycle codes. However, having an upper bounded minimum distance remains a deficiency of this construction
and the non-binary nature of the code imposes a potentially high decoding complexity. Furthermore, the performance was still not at par with that of the classical
LDPC codes. The concept of QC-QLDPC codes
was further extended to the class of spatially-coupled QC codes in~\cite{quasi_ldpc2011}, which outperformed 
the `non-coupled' design of~\cite{quasi_ldpc2007} at the cost of a small coding rate loss. The spatially-coupled QC-QLDPC was capable of
achieving a performance similar to that of the non-binary QC-LDPC code only when its block length was considerably higher. 
While all the aforementioned QLDPC
constructions were CSS-based, 
Camara~\etal~\cite{camara2007} were the first authors to conceive non-CSS QLDPC codes. They invoked group theory for deriving QLDPC codes from the 
classical self-orthogonal quaternary LDPC codes. Later, Tan~\etal~\cite{tan2010} proposed several systematic constructions
for non-CSS QLDPC codes, four of which were based on classical binary QC-LDPC codes, while one was derived from classical binary
LDPC-convolutional codes. Unfortunately, the non-CSS constructions of~\cite{camara2007,tan2010} failed to outperform Mackay's bicycle codes. 
Since most of the above-listed QLDPC constructions exhibit an upper bounded minimum distance, 
topological QLDPCs\footnote{Topological code structures are beyond the scope of this paper.} were 
derived from Kitaev's construction in~\cite{toric1,toric2,toric3}. Amidst these activities, which focused on the construction of QLDPC codes,
Poulin~\etal were the first scientists to address the decoding issues of QLDPC codes~\cite{poulin2008}. 
As mentioned above, most of the QLDPC codes consist of
unavoidable length-$4$ cycles. In fact, when QLDPC codes are viewed in the quaternary formalism, i.e. GF($4$), then they must have length-$4$
cycles, which emerge from the symplectic product criterion. These short cycles erode the performance of the classic message passing decoding 
algorithm. Furthermore, the classic message passing algorithm does not take into account the degenerate nature of quantum codes, rather
it suffers from it. This is known as the `symmetric degeneracy error'.
Hence, Poulin~\etal proposed heuristic methods 
in~\cite{poulin2008}
to alleviate the undesired affects of having short cycles and symmetric degeneracy error, which were further improved in~\cite{yun2012}.
The major contributions made in the context of QLDPC codes are summarized in \tref{tab:iter_q}, while the most promising QLDPC construction methods
are compared in \tref{tab:qldpc:comp}\footnote{The second column indicates `short cycles' in the binary formalism. Recall that all QLDPC codes
must have short cycles in the quaternary formalism.}.
\begin{table*}[tb] 
 \centering
 \definecolor{tcA}{rgb}{0.627451,0.627451,0.643137}
\begin{center}
\begin{tabular}{|c|c|l|p{2cm}|p{2cm}|p{8.5cm}|}
%\begin{tabular}{|l|l|l|}
% use packages: color,colortbl
\hline
\rowcolor{tcA}
 \multicolumn{2}{|c|}{\cellcolor{tcA}\textbf{}}&\textbf{Year}& \textbf{Author(s)} & \textbf{Code Type} &\textbf{Contribution} \\
\hline
  \multirow{30}{*}{QLDPC}&\multirow{25}{*}{Code Construction}& 2001 & Postol~\cite{qldpc2001} & \cellcolor{blue!25} Non-dual-containing CSS &
  The first example of QLDPC code constructed from a finite geometry based classical code. A generalized formalism for constructing QLDPC codes
  from the corresponding classical codes was not developed.\\ \cline{3-6}
 && 2004 & Mackay~\etal~\cite{sparse1} &\cellcolor{red!25} Dual-containing CSS &  Various code structures, e.g. bicycle codes and unicycle codes, were conceived for 
 constructing QLDPC codes from classical dual-containing LDPC codes. Performance impairment due to the presence of unavoidable length-4 cycles was
 first pointed out in this work. Minimum distance of the resulting codes was upper bounded by the row weight.\\ \cline{3-6}
 &&2005&Lou~\etal~\cite{Lou2005, Lou2006}&\cellcolor{blue!25} Non-dual-containing CSS & The generator and PCM of classical LDGM codes were exploited for
 constructing CSS codes. An increased decoding complexity was imposed and the codes had an upper bounded minimum distance.\\ \cline{3-6}
 &&\multirow{1}{*}{2007}& Mackay~\cite{qldpc_mackay2007}&\cellcolor{red!25} Dual-containing CSS& Cayley graph-based QLDPC codes were proposed,
 which had numerous length-$4$ cycles.\\ \cline{4-6}
 &&&Camara~\etal~\cite{camara2007}&\cellcolor{yellow!25} Non-CSS& QLDPC codes derived from classical self-orthogonal quaternary LDPC codes were 
 conceived, which failed to outperform MacKay's bicycle codes.\\ \cline{4-6}
 &&&Hagiwara~\etal~\cite{quasi_ldpc2007}&\cellcolor{blue!25} Non-dual-containing CSS&Quasi-cyclic QLDPC codes were constructed using a pair of quasi-cyclic LDPC codes,
 which were found using algebraic combinatorics. The resultant codes had at least a girth of $6$, but they failed to outperform MacKay's 
 constructions given in~\cite{sparse1}.\\ \cline{3-6}
 &&\multirow{1}{*}{2008}&Aly~\etal~\cite{Aly2008}&\cellcolor{red!25} Dual-containing CSS&QLDPC codes were constructed from finite geometries,
 which failed to outperform Mackay's bicycle codes.\\ \cline{4-6}
 &&&Djordjevic~\cite{djo2008}&\cellcolor{red!25} Dual-containing CSS&BIBDs were exploited to design QLDPC codes,
 which failed to outperform Mackay's bicycle codes.\\ \cline{3-6}
 &&2010& Tan~\etal~\cite{tan2010}&\cellcolor{yellow!25} Non-CSS&Several systematic constructions for non-CSS QLDPC codes were proposed, four of which were based on classical binary quasi-cyclic LDPC codes, while one 
was derived from classical binary LDPC-convolutional codes. These code designs failed to outperform Mackay's bicycle codes.\\ \cline{3-6}
  &&\multirow{1}{*}{2011}&Couvreur~\etal~\cite{cayley2011,couvreur2013}&\cellcolor{red!25} Dual-containing CSS&Cayley graph-based QLDPC codes of~\cite{qldpc_mackay2007}
 were further investigated. The lower bound on the minimum distance of the resulting QLDPC was logarithmic in the code length, but this
 was achieved at the cost of an increased decoding complexity.\\ \cline{4-6}
 &&&Kasai~\cite{kasai2011, hagi2012}&\cellcolor{blue!25} Non-dual-containing CSS& Quasi-cyclic QLDPC codes of~\cite{quasi_ldpc2007} were extended to 
 non-binary constructions, which outperformed Mackay's bicycle codes at the cost of an increased decoding complexity. Performance was still
 not at par with the classical LDPC codes and minimum distance was upper bounded.\\ \cline{3-6}
 &&&Hagiwara~\etal~\cite{quasi_ldpc2011}&\cellcolor{blue!25} Non-dual-containing CSS& Spatially-coupled QC-QLDPC codes were developed, which outperformed the `non-coupled' design of~\cite{quasi_ldpc2007}
at the cost of a small coding rate loss. Performance was similar to that of~\cite{kasai2011, hagi2012}, but larger block lengths were required.\\ \cline{2-6}
 &\multirow{3}{*}{Decoding}&2008&Poulin~\etal~\cite{poulin2008}&\cellcolor{gray!25}&Heuristic methods were developed to alleviate the 
performance degradation caused by
unavoidable length-$4$ cycles and symmetric degeneracy error.\\ \cline{3-6}
  &&2012&Wang~\etal~\cite{yun2012}&\cellcolor{gray!25}&Feedback mechanism was introduced in the context of the heuristic methods of~\cite{poulin2008} to further improve the
  performance.\\ \cline{1-6}
 \multirow{10}{*}{QTC}&\multirow{6}{*}{Code Construction}& 2008 & Poulin~\etal~\cite{qturbo1, qturbo2} & \cellcolor{yellow!25} Non-CSS &
QTCs were conceived based on the interleaved serial concatenation of QCCs. QTCs are free from the decoding issue associated with the length-$4$ cycles and they offer a 
wider range of code parameters. Degenerate iterative decoding 
algorithm was also proposed. Unfortunately, QTCs have an upper bounded minimum distance.\\ \cline{3-6}
&&2014&Babar~\etal~\cite{babar_QTC_2014}&\cellcolor{yellow!25} Non-CSS& To dispense with the time-consuming Monte Carlo simulations and
to facilitate the design of hashing bound approaching QTCs, the application of classical non-binary EXIT charts of~\cite{kliewer:exit-symbol} was extended to QTCs. \\ \cline{2-6}
&\multirow{2}{*}{Decoding}&2014&Wilde~\etal~\cite{wilde_turbo2}&\cellcolor{gray!25}& The iterative decoding algorithm of~\cite{qturbo1, qturbo2} failed to yield performance similar to the
classical turbo codes. The decoding algorithm was improved by iteratively exchanging the \textit{extrinsic} rather than the \textit{a posteriori} information. \\ \cline{1-6}
 \end{tabular}
 \end{center}
 \caption{Major contributions to the development of iterative quantum codes.} 
 \label{tab:iter_q}
\end{table*}

\begin{table*}[tb] 
 \centering
 \definecolor{tcA}{rgb}{0.627451,0.627451,0.643137}
\begin{center}
\begin{tabular}{|l|c|c|c|c|}
%\begin{tabular}{|l|l|l|}
% use packages: color,colortbl
\hline
\rowcolor{tcA}
 \textbf{Code Construction}& \textbf{Short Cycles}&\textbf{Minimum Distance}&\textbf{Delay}&\textbf{Decoding Complexity} \\ \hline
 Bicycle codes~\cite{sparse1}&Yes&Upper Bounded&Standard&Standard\\ \hline
 Cayley-graph based codes~\cite{qldpc_mackay2007,cayley2011,couvreur2013}&Yes&Increases with
 the code length&Standard&Increases with
 the code length \\ \hline
 LDGM-based codes~\cite{Lou2005, Lou2006}&Yes&Upper Bounded&Standard&High\\ \hline
 Non-binary quasi-cyclic codes~\cite{kasai2011, hagi2012}&No&Upper Bounded&Standard&High\\ \hline
 Spatially-coupled quasi-cyclic codes~\cite{quasi_ldpc2011}&No&Upper Bounded&High&High \\ \hline   
 \end{tabular}
 \end{center}
 \caption{Comparison of the Quantum Low Density Parity Check (QLDPC) code structures.} 
 \label{tab:qldpc:comp}
\end{table*}

Pursuing further the direction of iterative code structures, Poulin~\etal conceived QTCs in~\cite{qturbo1, qturbo2}, based on the interleaved 
serial concatenation of QCCs. Unlike QLDPC codes, QTCs offer a complete freedom in choosing the code parameters, 
such as the frame length, coding rate, constraint length and interleaver type. Moreover, their decoding is not impaired by the presence of length-$4$
cycles associated with the symplectic criterion.
%do not have to face the decoding issues associated with length-$4$ cycles and they
Furthermore, in contrast to QLDPC codes, the iterative decoding invoked for QTCs takes into account the inherent degeneracy associated with quantum codes.
However, it was found in~\cite{qturbo1, qturbo2, Monireh2012} that the constituent QCCs cannot be simultaneously both recursive 
and noncatastrophic. Since the recursive nature of the inner code is essential for ensuring an unbounded minimum distance\footnote{Unbounded minimum
distance of a code implies that its minimum distance increases almost linearly with the interleaver length.}, 
whereas the noncatastrophic nature is a necessary condition to be satisfied for achieving decoding convergence to a vanishingly low error rate, the QTCs designed in~\cite{qturbo1, qturbo2} had a bounded 
minimum distance. The QBER performance curves of the QTCs conceived in~\cite{qturbo1, qturbo2} also failed to match the classical turbo codes. This
issue was dealt with in~\cite{wilde_turbo2}, where the quantum turbo decoding algorithm of~\cite{qturbo2} was improved by iteratively exchanging the
\textit{extrinsic} rather than the \textit{a posteriori} information.
Furthermore, in~\cite{qturbo1, qturbo2, wilde_turbo2}, the optimal components of QTCs were found by analyzing their distance spectra, followed
by extensive Monte Carlo simulations for the sake of determining the convergence threshold of the resultant QTC. In order to circumvent
this time-consuming approach and to
facilitate the design of hashing bound approaching QTCs, the application of classical non-binary EXIT charts~\cite{kliewer:exit-symbol} was extended to QTCs in~\cite{babar_QTC_2014}.
An EXIT-chart aided exhaustive-search based optimized QTC was also presented in~\cite{babar_QTC_2014}.
The major contributions made in the domain of quantum turbo codes are summarized in \tref{tab:iter_q}.

\begin{figure}[tb]
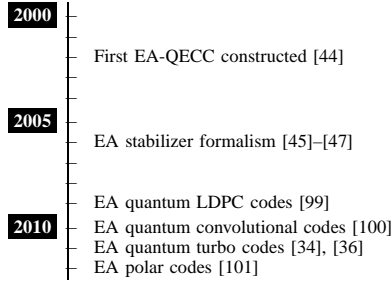

\begin{center}
\scalebox{0.7}{
\begin{tabular}{r |@{\food} l}
\colorbox{black}{\color{white} \textbf{2000}}     &\\
     &\\
     &First EA-QECC constructed~\cite{bowen2002}\\
     &\\
     &\\
\colorbox{black} {\color{white} \textbf{2005}}     &\\
     &EA stabilizer formalism~\cite{brun2006,brun2007,hsieh2007}
     \\
     &\\
     &\\
     &EA quantum LDPC codes~\cite{equasi2008}\\
\colorbox{black} {\color{white} \textbf{2010}}     &EA quantum convolutional codes~\cite{mark2010}\\
     &EA quantum turbo codes~\cite{wilde_turbo,wilde_turbo2}\\
     &EA polar codes~\cite{wilde_polar2012}\\
\end{tabular}
}
\caption{Major milestones achieved in the history of entanglement-assisted quantum error correction codes.}
\label{fig:EA-timeline}
\end{center}
\end{figure}

Some of the well-known classical codes cannot be imported into the quantum domain by invoking the aforementioned stabilizer-based
code constructions because the
stabilizer codes have to satisfy the stringent symplectic product criterion. This limitation was 
overcome in~\cite{bowen2002,brun2006,brun2007,hsieh2007} with the notion of EA quantum codes, which exploit pre-shared entanglement between the transmitter
and receiver. Later, this concept was extended to numerous other code structures, e.g. EA-QLDPC code~\cite{equasi2008},
EA-QCC~\cite{mark2010}, 
EA-QTC~\cite{wilde_turbo,wilde_turbo2} and 
EA-polar codes~\cite{wilde_polar2012}. In~\cite{wilde_turbo,wilde_turbo2}, it was also found that entanglement-assisted convolutional 
codes may be simultaneously both recursive as well as non-catastrophic. Therefore, the issue of bounded minimum distance of QTCs was resolved with the 
notion of entanglement. Furthermore, EA-QLDPC codes are free from length-$4$ cycles in the binary formalism, which in turn results in an impressive performance similar
to that of the corresponding classical LDPC codes. Hence, the concept of the entanglement-assisted regime resulted in a major breakthrough in terms
of constructing quantum codes,
whose behaviour is similar to that of the corresponding classical codes.
The major milestones achieved in the history of entanglement-assisted quantum error correction codes are chronologically arranged in \fref{fig:EA-timeline}. 

%To further enrich the family of quantum codes,
\textit{In this contribution, we design a novel QIRCC, which may be used as an outer component in
a QTC, or in fact any arbitrary concatenated quantum code structure. Explicitly, the proposed QIRCC may be invoked in conjunction with any arbitrary 
inner code (both unassisted as well as entanglement-assisted) for the sake of attaining %adapted to match any given inner code using EXIT charts; thus, facilitating 
a hashing bound approaching performance with the aid of the EXIT charts of~\cite{babar_QTC_2014}. More specifically, we construct a $10$-subcode QIRCC and use it as the
outer code in concatenation
with the non-catastrophic and recursive inner convolutional code of~\cite{wilde_turbo2}. In contrast to the concatenated code
of~\cite{wilde_turbo2}, which exhibited a performance within $0.9$~dB of the hashing bound, our QIRCC-based optimized design operates within $0.4$~dB of the noise limit. 
Furthermore, at a Word Error Rate (WER) of $10^{-3}$, our design outperforms the benchmark designed in~\cite{wilde_turbo2} by about $0.5$~dB. 
Our proposed design also yields a lower error rate than the exhaustive-search based optimized design of~\cite{babar_QTC_2014}.}
%\section{Preliminaries of Quantum Information}
%Qubits - Gates - entanglement - measurement - quantum depolarizing channel. Chap 2 of mini-thesis.
\section{Stabilizer Formalism}  \label{sec:stabilizer}%{Pauli-to-Binary Isomorphism}
Most of the quantum codes developed to date owe their existence to the theory of stabilizer codes, which allows us to import any arbitrary classical binary
as well as quaternary code to the quantum domain. Unfortunately, this is achieved at the cost of imposing restrictions on the code structure, which may 
adversely impact the performance of the code, e.g. as in QLDPC codes and QTCs, which was discussed in Section~\ref{sec:history}. In this section, we will delve deeper into
the stabilizer formalism for the sake of ensuring a smooth transition from the classical to the quantum domain. %Towards the end of this section,
%our motivated readers will be able to design both unassisted as well as entanglement-assisted quantum codes from the well-known classical codes.
\subsection{Classical Linear Block Codes} \label{sec:LBC}
The stabilizer formalism derives its existence from the theory of classical linear block codes. 
A classical linear block code $C(n,k)$ maps $k$-bit information blocks onto $n$-bit codewords. For small
values of $k$ and $n$, this can be readily achieved using a look-up table, which maps the input information blocks onto the encoded message blocks. However, 
for large values of $k$ and
$n$, the process may be simplified using an $k \times n$ generator matrix $G$ as follows:
\begin{equation}
 \overline{x}=xG,
\label{eq:clas1}
\end{equation}
where $x$ and $\overline{x}$ are row vectors for information and encoded messages, respectively. Furthermore, $G$ may be decomposed as:
\begin{equation}
 G=\left(I_k|P\right),
\label{eq:clas2}
\end{equation}
where $I_k$ is a ($k \times k$)-element identity matrix and $P$ is a $k \times (n-k)$-element matrix. This in turn implies that the first
$k$ bits of the encoded message are information bits, followed by $(n-k)$ parity bits. 

At the decoder, syndrome-based decoding is invoked, which determines the position of the channel-induced error using the observed syndromes rather than directly acting on 
the received codewords. More precisely, each generator matrix is associated with an $(n-k) \times n$-element PCM $H$ which is given by:
\begin{equation}
H=\left(P^T|I_{n-k}\right),
 \label{eq:clas3}
\end{equation}
and is defined such that $\overline{x}$ is a valid codeword only if,
\begin{equation}
\overline{x}H^T=0.
 \label{eq:clas4}
\end{equation}
For a received vector $y=\overline{x}+e$, where $e$ is the error incurred during transmission, the error syndrome of length $(n-k)$ is computed as:
\begin{equation}
 s=yH^T=(\overline{x}+e)H^T=\overline{x}H^T+eH^T=eH^T,
 \label{eq:clas5}
\end{equation}
which is then used for identifying the erroneous parity bit. 

Let us consider a simple $3$-bit repetition code, which makes three copies of the intended information bit. More precisely, $k = 1$ and $n = 3$. It is specified by the following generator matrix:
\begin{equation}
 G=\begin{pmatrix}
  1& 1& 1
\end{pmatrix},
\label{eq:g:clas6}
\end{equation}
which yields two possible codewords $[1 1 1]$ and $[0 0 0]$. At the receiver, a decision may be made on the basis of the majority voting.
For example, if $y=[0 1 1]$ is received, then we may conclude that the transmitted bit was $1$. Alternatively, we may invoke the PCM-based syndrome decoding.
According to \eqr{eq:clas3}, the corresponding PCM is given by:
\begin{equation}
H=\begin{pmatrix}
 1 & 1 & 0 \\
 1 & 0 & 1
\end{pmatrix}.
\label{eq:clas6}
\end{equation}
It can be worked out that $yH^T=0$ only for the two valid codewords $[1 1 1]$ and $[0 0 0]$. For all other received codewords, at least one of the 
two syndrome elements is set to $1$, e.g. when the first bit is corrupted, i.e. $y=[0 1 1]$ or $[1 0 0]$,  $s=[1 1]$. \tref{tab:syndrome} enlists all the
$1$-bit errors, which
may be identified using this syndrome decoding procedure.
\begin{table}[tb]
 \centering
% \definecolor{tcA}{rgb}{0.627451,0.627451,0.643137}
 \begin{center}
\begin{tabular}{|c|c|}
%\rowcolor{tcA}
\hline
\textbf{Syndrome $(s)$} & \textbf{Index of Error}\\
%\rowcolor{tcA}
%\textbf{$b_1b_2$} & \textbf{on $q_2$} & \textbf{Bell State $|\psi\rangle$}\\
\hline
 $[11]$  &$1$\\
 $[10]$  &$2$\\
 $[01]$  &$3$\\
\hline
 \end{tabular}
 \end{center}
 \caption{Single-bit errors along with the corresponding syndromes for the PCM of~\eqr{eq:clas6}.}
\label{tab:syndrome}
\end{table}

This process of error correction using generator and parity check matrices is usually preferred due to its compact nature. Generally, $C(n,k)$ code,
which encodes a $k$-bit information message into an $n$-bit codeword, would require $2^k$ $n$-bit codewords. Thus, it would required a total of $n2^k$ bits to completely specify
the code space. By contrast, the aforementioned approach only requires $kn$ bits of the generator matrix. Hence, memory resources are saved exponentially and 
encoding and decoding operations are efficiently implemented. These attractive features of classical block linear codes 
and the associated  PCM-based syndrome decoding~\cite{zbabar2013} have led to the development of quantum stabilizer codes.
%which incorporate the same underlying concept of PCM-based syndrome decoding~\cite{zbabar2013}.
\subsection{Quantum Stabilizer Codes (QSCs)} \label{sec:QSC}
Let us recall from Section~\ref{sec:designObj} that qubits collapse to classical bits upon measurement~\cite{Qbook2}. This prevents us from directly 
applying the classical error correction techniques for reliable quantum transmission. Inspired by the PCM-based syndrome decoding of classical codes,
Gottesman~\cite{got2,G97} introduced the notion of stabilizer formalism, 
which facilitates the design of quantum codes from the classical ones.
Analogous to Shor's pioneering $9$-qubit code~\cite{shor95}, stabilizer formalism overcomes the measurement issue by observing 
the error syndromes without reading the actual quantum information. More specifically, QSCs
invoke the PCM-based syndrome decoding approach of classical linear block codes for estimating the errors incurred during
transmission. 

\begin{figure*}[tb]
\begin{center}
    \includegraphics[width=0.7\linewidth]{\figures 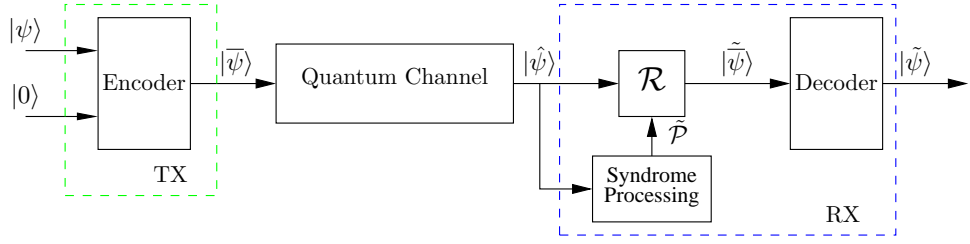}
    \caption{System Model: Quantum communication system relying on a quantum stabilizer code.}
  \label{fig:stab_sys}
\end{center}
%\vspace{-0.5cm}
\end{figure*}

\fref{fig:stab_sys} shows the general schematic of a quantum communication system relying on a quantum stabilizer code for reliable transmission. An $[n,k]$
QSC encodes the information qubits $\ket{\psi}$ into the coded sequence $\ket{\overline{\psi}}$ with the aid of $(n-k)$ auxiliary (also called ancilla) qubits,
which are initialized to the state $\ket{0}$. The noisy sequence $\ket{\hat{\psi}} = \mathcal{P} \ket{\overline{\psi}}$, where $\mathcal{P}$
is the $n$-qubit channel error, is received at the receiver (RX), which engages in a $3$-step process
for the sake of recovering the intended transmitted information. More explicitly, RX computes the syndrome of the received sequence $\ket{\hat{\psi}}$ 
and uses it to estimate the channel error $\tilde{\mathcal{P}}$. The recovery operator $\mathcal{R}$ then uses the estimated error $\tilde{\mathcal{P}}$
to restore the transmitted coded stream. Finally, the decoder, or more specifically the inverse encoder, processes the recovered coded sequence
$\ket{\tilde{\overline{\psi}}}$, yielding
the estimated transmitted information qubits $\ket{\tilde{\psi}}$. 

\textit{An $[n,k]$ quantum stabilizer code, constructed over a code space $\mathcal{C}$, which maps
the information word (logical qubits) $\ket{\psi} \in \mathbb{C}^{2^k}$ onto the codeword (physical qubits) $\ket{\overline{\psi}} \in \mathbb{C}^{2^n}$,
where $\mathbb{C}^d$ denotes the $d$-dimensional Hilbert space, is defined by a set of $(n - k)$ independent commuting $n$-tuple Pauli 
operators $g_i$, for $1 \leq i \leq (n - k)$. The corresponding stabilizer group $\mathcal{H}$ contains both
$g_i$ and all the products of $g_i$ for $1 \leq i \leq (n - k)$ and forms an abelian subgroup of $\mathcal{G}_n$. A unique feature of these operators is that 
they do not change the state of valid codewords, while yielding an eigenvalue of $-1$ for corrupted states.}

Let us now elaborate on this definition of the stabilizer code by considering a simple $3$-qubit bit-flip repetition code, which is capable of correcting
single-qubit bit-flip errors. Since the laws of quantum mechanics do not permit cloning of the information qubit, we cannot encode
$\ket{\psi}$ to $(\psi \otimes \psi \otimes \psi)$. Instead, the 
$3$-qubit bit-flip repetition code entangles two auxiliary qubits with the information qubit 
such that the basis states $\ket{0}$ and $\ket{1}$ are copied thrice in the superposition of basis states of the resulting $3$-qubit codeword,
%auxiliary qubits are in the same
%state as the information qubit, 
i.e. $\ket{0}$ and $\ket{1}$ are mapped as follows:
\begin{align}
\ket{0} &\rightarrow \ket{000}, \nonumber \\ 
\ket{1} &\rightarrow \ket{111}. 
\label{eq:3bit-rep}
\end{align}
Consequently, the information word $\ket{\psi} = \alpha \ket{0} + \beta \ket{1}$ is encoded as:
\begin{equation}
 \alpha \ket{0} + \beta \ket{1} \rightarrow \alpha \ket{000} + \beta \ket{111}.
\label{eq:trans_codeword}
\end{equation}
The resultant codeword is stabilized by the operators $g_1=\mathbf{Z}\mathbf{Z}\mathbf{I}$ and 
$g_2=\mathbf{Z}\mathbf{I}\mathbf{Z}$.
Here the term `stabilize' implies that the valid codewords are not
affected by the generators $g_1$ and $g_2$ and yield an eigenvalue of $+1$, as shown below:
\begin{align}
g_1\left[|\overline{\psi}\rangle\right]&=\alpha \ket{000} + \beta \ket{111} \equiv |\overline{\psi}\rangle, \nonumber \\
g_2\left[|\overline{\psi}\rangle\right]&=\alpha \ket{000} + \beta \ket{111} \equiv |\overline{\psi}\rangle. 
 \label{eq:stab2}
\end{align}
On the other hand, if a corrupted state $\ket{\hat{\psi}}$ is received, then the stabilizer generators yield an eigenvalue of $-1$, e.g. let 
$\ket{\hat{\psi}} = \ket{100} + \beta \ket{011}$ where $\mathcal{P} = \mathbf{X}\mathbf{I}\mathbf{I}$, then we have:
\begin{align}
g_1\left[|\hat{\psi}\rangle\right]&= - \alpha \ket{100} - \beta \ket{011} \equiv -|\hat{\psi}\rangle, \nonumber \\
g_2\left[|\hat{\psi}\rangle\right]&= - \alpha \ket{100} - \beta \ket{011} \equiv -|\hat{\psi}\rangle. 
\label{eq:stab3}
\end{align}
More specifically, the eigenvalue is $-1$ if the $n$-tuple Pauli error $\mathcal{P}$ acting on the transmitted codeword $\ket{\overline{\psi}}$ anti-commutes with the 
stabilizer $g_i$ and it is $+1$ if $\mathcal{P}$ commutes with $g_i$. Therefore, we have:
\begin{equation}
 g_i \ket{\hat{\psi}} = \left\{
\begin{array}{l l}
 \ket{\overline{\psi}}, & g_i \mathcal{P} = \mathcal{P} g_i \\
-\ket{\overline{\psi}}, & g_i \mathcal{P} = - \mathcal{P} g_i, \\
\end{array}
\right.
\label{eq:stab}
\end{equation}
where $\ket{\hat{\psi}} = \mathcal{P}\ket{\overline{\psi}}$. \tref{tab:3qubit}
enlists the eigenvalues for all possible single-qubit bit-flip errors. The resultant $\pm 1$ eigenvalue gives the corresponding error syndrome $s$, which is $0$ for an 
eigenvalue of $+1$ and $1$ for an eigenvalue of $-1$, as depicted in \tref{tab:3qubit}. 
\begin{table}[tb]
 \centering
% \definecolor{tcA}{rgb}{0.627451,0.627451,0.643137}
 \begin{center}
\begin{tabular}{|c|c|c|c|c|}
%\rowcolor{tcA}
\hline
\textbf{$\ket{\hat{\psi}} = \mathcal{P}\ket{\overline{\psi}}$} & \textbf{$g_1 \ket{\hat{\psi}}$} & \textbf{$g_2 \ket{\hat{\psi}}$} & \textbf{Syndrome ($s$)}& \textbf{Index of Error}\\
\hline
 $\alpha \ket{100} + \beta \ket{011}$  &$-1$ &$-1$ & $[11]$ &$1$\\
 $\alpha \ket{010} + \beta \ket{101}$  &$-1$ &$+1$ & $[10]$ &$2$\\
 $\alpha \ket{001} + \beta \ket{110}$  &$+1$ &$-1$ & $[01]$ &$3$\\
\hline
 \end{tabular}
 \end{center}
 \caption{Single-qubit bit-flip errors along with the corresponding eigenvalues for $3$-qubit bit-flip repetition code.}
\label{tab:3qubit}
\end{table}
%Here $X_i$ and $Z_i$ signify the application of $\mathbf{X}$ and $\mathbf{Z}$ gates to the $i^{th}$ qubit. 
%An interesting point to note is that $|\overline{\psi}\rangle$, given in \eqr{eq:stab1}, is a unique quantum state which can be 
%stabilized by the operators $g_1=\mathbf{X}\mathbf{X}$ and $g_2=\mathbf{Z}\mathbf{Z}$; thus, leading to the idea that quantum states can be completely 
%described by the stabilizers. Consequently, quantum codes can be easily described in terms of the stabilizers rather than the state vector descriptions~\cite[pg.~454]{Qbook2}. 

\begin{figure*}[tb]
\centering
\includegraphics[width=0.9\linewidth]{\figures 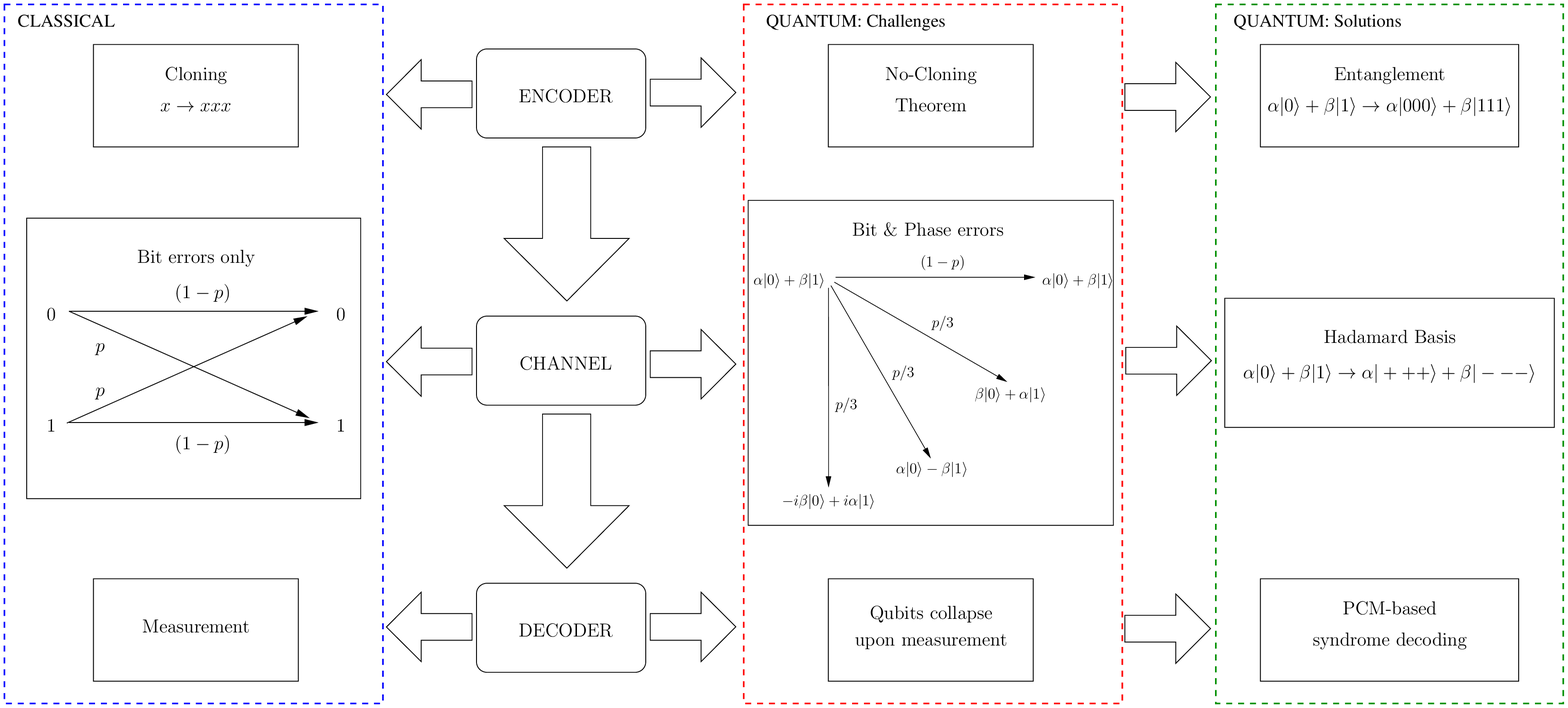}
\caption{Transition from the classical to quantum codes. \textbf{Encoder}: In classical codes, the information bit may be copied during the encoding
process, e.g. in a $3$-bit repetition code. %each information bit $x$ is copied twice. 
This is not permissible
in the quantum domain (no-cloning theorem). Alternatively, in quantum codes, the information qubit is entangled to the auxiliary qubits, e.g. in a
$3$-qubit bit-flip repetition code. \textbf{Channel}: Only bit errors occur over a classical channel, e.g. a binary symmetric channel with the channel crossover probability
$p$. By contrast, qubits may experience a bit or phase error as well as both, e.g. depolarizing channel with a probability $p$.
Since phase errors in the Hadamard basis $\{\ket{+},\ket{-}\}$ are similar to the bit errors in the
computational basis $\{\ket{0},\ket{1}\}$, phase errors may be corrected in the same way as the bit errors by exploiting the Hadamard basis. \textbf{Decoder}: In classical codes, the
received bits are measured during the decoding process, e.g. in a $3$-bit repetition code a decision may be made on the basis of majority voting.
Unfortunately, qubits collapse upon measurement. Consequently, quantum codes invoke the PCM-based syndrome decoding.}
\label{fig:comp}
\end{figure*}

A $3$-qubit phase-flip repetition code may be constructed using a similar approach. This is because phase errors in the Hadamard basis $\{\ket{+},\ket{-}\}$
are similar to the bit errors in the computational basis $\{\ket{0},\ket{1}\}$. More explicitly, the states $\ket{+}$ and $\ket{-}$
are defined as:
\begin{align}
 \ket{+} &\equiv \mathbf{H}\ket{0} = \frac{\ket{0}+\ket{1}}{\sqrt{2}}, \nonumber \\
 \ket{-} &\equiv \mathbf{H}\ket{1} = \frac{\ket{0}-\ket{1}}{\sqrt{2}}, 
 \label{eq:hadamard-basis}
\end{align}
where $\mathbf{H}$ is a single-qubit Hadamard gate, which is given by~\cite{Qbook2}:
\begin{equation}
 \mathbf{H}= \frac{1}{\sqrt{2}}\begin{pmatrix}
  1& 1 \\
  1& -1
\end{pmatrix}.
\label{eq:hadamard}
\end{equation}
Therefore, Pauli-$\mathbf{Z}$ acting on the states $\ket{+}$ and $\ket{-}$ yields:
\begin{align}
 \mathbf{Z}\ket{+} &= \ket{-}, \nonumber \\
 \mathbf{Z}\ket{-} &= \ket{+},
\end{align}
which is similar to the operation of Pauli-$\mathbf{X}$ on the states $\ket{0}$ and $\ket{1}$, i.e. we have:
\begin{align}
 \mathbf{X}\ket{0} &= \ket{1}, \nonumber \\
 \mathbf{X}\ket{1} &= \ket{0}.
 \end{align}
Consequently, analogous to \eqr{eq:3bit-rep}, a $3$-qubit phase-flip repetition code encodes $\ket{0}$ and $\ket{1}$ as follows:
\begin{align}
\ket{0} &\rightarrow \ket{+++}, \nonumber \\ 
\ket{1} &\rightarrow \ket{---}.
\label{eq:3phase-rep}
\end{align}
Based on \eqr{eq:3phase-rep}, $\ket{\psi}$ is encoded to:
\begin{equation}
 \alpha \ket{0} + \beta \ket{1} \rightarrow \alpha \ket{+++} + \beta \ket{---},
 \label{eq:psi-phase}
\end{equation}
which is stabilized by the generators $g_1=\mathbf{X}\mathbf{X}\mathbf{I}$ and 
$g_2=\mathbf{X}\mathbf{I}\mathbf{X}$. Hence, the Hadamard and Pauli-$\mathbf{X}$ operators enable a quantum code to correct phase errors.
This overall transition from the classical $3$-bit repetition code of Section~\ref{sec:LBC} to the quantum repetition code is summarized in \fref{fig:comp}.

Furthermore, the stabilizer generators $g_i$ constituting the stabilizer group $\mathcal{H}$ must exhibit the following two characteristics:
\begin{enumerate}
\item \textbf{Any two operators in the stabilizer set must commute} so that the stabilizer operators can be applied simultaneously, i.e. we have:
\begin{equation}
 g_1 g_2 |\overline{\psi}\rangle = g_2 g_1 |\overline{\psi}\rangle.
 \label{eq:stab:com}
\end{equation}
This is because the stabilizer leaves the codeword unchanged as encapsulated below:
\begin{equation}
 g_i |\overline{\psi}\rangle = |\overline{\psi}\rangle.
\end{equation}
Hence, evaluating the left-hand and right-hand sides of~\eqr{eq:stab:com} gives:
\begin{equation}
 g_1 g_2 |\overline{\psi}\rangle = g_1 |\overline{\psi}\rangle = |\overline{\psi}\rangle,
\end{equation}
and
\begin{equation}
 g_2 g_1 |\overline{\psi}\rangle = g_2 |\overline{\psi}\rangle = |\overline{\psi}\rangle,
\end{equation} 
respectively.
This further imposes the constraint that
the stabilizers should have an even number of places with different non-Identity (i.e. $\mathbf{X}$, $\mathbf{Y}$, or $\mathbf{Z}$) operations. 
This is derived from the fact that the $\mathbf{X}$, $\mathbf{Y}$ and $\mathbf{Z}$ operations anti-commute with one another as shown below:
\begin{align}
 \mathbf{X}\mathbf{Y} = i\mathbf{Z}, \; \mathbf{Y}\mathbf{X} = -i\mathbf{Z} &\rightarrow \mathbf{X}\mathbf{Y} = -\mathbf{Y}\mathbf{X} \\ \nonumber
 \mathbf{Y}\mathbf{Z} = i\mathbf{X}, \; \mathbf{Z}\mathbf{Y} = -i\mathbf{X} &\rightarrow \mathbf{Y}\mathbf{Z} = -\mathbf{Z}\mathbf{Y} \\ \nonumber
 \mathbf{Z}\mathbf{X} = i\mathbf{Y}, \; \mathbf{X}\mathbf{Z} = -i\mathbf{Y} &\rightarrow \mathbf{Z}\mathbf{X} = -\mathbf{X}\mathbf{Z} \\ \nonumber
\end{align}
Thus, for example the operators $\mathbf{ZZI}$ and $\mathbf{XYZ}$ commute, whereas $\mathbf{ZZI}$ and $\mathbf{YZI}$ anti-commute.
\item \textbf{Generators constituting the stabilizer group $\mathcal{H}$ are closed under multiplication}, i.e. multiplication of the constituent generators $g_i$ 
yields another generator, which is also part of the stabilizer group $\mathcal{H}$. For example, the full stabilizer 
group $\mathcal{H}$ of the $3$-qubit bit-flip repetition code will also include the operator $\mathbf{IZZ}$, which is the product of $g_1$ and $g_2$.
\end{enumerate}
It must be mentioned here that the Pauli errors which differ only by the stabilizer group have the same impact on all the codewords and therefore can be
corrected by the same recovery operations. This gives quantum codes the intrinsic property of degeneracy~\cite{deg_poulin}. More explicitly, the errors
$\mathcal{P}$ and $\mathcal{P}' = g_i \mathcal{P}$ have the same impact on the transmitted codeword and therefore can be corrected by the same
recovery operation. Getting back to our example of the $3$-qubit bit-flip repetition code, let $\mathcal{P} = \mathbf{IIX}$ and  $\mathcal{P}' = g_1 \mathcal{P} = \mathbf{ZZX}$.
Both $\mathcal{P}$ as well as $\mathcal{P}'$ corrupt the transmitted codeword of \eqr{eq:trans_codeword} to $\alpha \ket{001} + \beta \ket{110}$. Consequently, 
$\mathcal{P}$ and $\mathcal{P}'$ need not be differentiated and are therefore classified as degenerate errors.
\subsection{Pauli-to-Binary Isomorphism} \label{sec:p2b}
QSCs may be characterized in terms of an equivalent classical parity check matrix notation satisfying the commutativity
constraint of stabilizers~\cite{cleve97,sparse1} given in \eqr{eq:stab:com}. This is achieved by mapping the $\mathbf{I}$, $\mathbf{X}$, $\mathbf{Y}$ and $\mathbf{Z}$ Pauli operators onto 
$(\mathbb{F}_2)^2$ as follows:
\begin{align}
 \mathbf{I} &\rightarrow (0,0), \nonumber \\
 \mathbf{X} &\rightarrow (0,1), \nonumber \\
 \mathbf{Y} &\rightarrow (1,1), \nonumber \\
 \mathbf{Z} &\rightarrow (1,0).
\label{eq:mapping}
\end{align}
More explicitly, the $(n - k)$ stabilizers of an $[n,k]$ stabilizer code constitute the rows of the binary PCM $H$, which can be represented 
as a concatenation of a pair of $(n - k) \times n$ binary matrices $H_z$ and $H_x$ based on \eqr{eq:mapping}, as given below:
\begin{equation}
H = \left(H_z | H_x\right). 
\label{eq:eq-classical}
\end{equation}
Each row of $H$ corresponds to a stabilizer of $\mathcal{H}$, so that the $i$th column of $H_z$ and $H_x$
corresponds to the $i$th qubit and a binary $1$ at these locations represents a $\mathbf{Z}$ and $\mathbf{X}$ Pauli operator, respectively, in the corresponding
stabilizer. For the $3$-qubit bit-flip repetition code, which can only correct bit-flip errors, the PCM $H$ is given by:
\begin{equation}
 H = \left(\begin{array} {c c c|c c c}
  1 &1 &0 &0 &0 &0\\
  1 &0 &1 &0 &0 &0
 \end{array}\right).
\label{H:3qubit}
\end{equation}
It must be pointed out here that $H_z$ of \eqr{H:3qubit} is same as the $H$ of the classical repetition code of \eqr{eq:clas6}, yielding the same syndrome patterns
in \tref{tab:syndrome} and \tref{tab:3qubit}.

Let us further elaborate the process by considering the $[9,1]$ Shor's code, which consists of the Pauli-$\mathbf{Z}$ as well as the Pauli-$\mathbf{X}$ operators. The corresponding
stabilizer generators are given in \tref{tab:shor}. They can be mapped onto the binary matrix $H$ as follows:
\begin{equation}
 H = \left(\begin{array} {c | c}
 H_z^\prime &\mathbf{0} \\
 \mathbf{0} &H_x^\prime
\end{array}\right),
\label{eq:H_Shor}
\end{equation}
where we have $H_z = \begin{pmatrix}
                      H_z^\prime \\
                      \mathbf{0}
                     \end{pmatrix}
$, $H_x = \begin{pmatrix}
		      \mathbf{0}\\
                      H_x^\prime 
                      \end{pmatrix}
$ and :
\begin{equation}
  H_z^\prime = \begin{pmatrix}%\left[\begin{array} {c c c c c c c c c}
  1 &1 &0 &0 &0 &0 &0 &0 &0 \\
  0 &1 &1 &0 &0 &0 &0 &0 &0 \\
  0 &0 &0 &1 &1 &0 &0 &0 &0 \\
  0 &0 &0 &0 &1 &1 &0 &0 &0 \\
  0 &0 &0 &0 &0 &0 &1 &1 &0 \\
  0 &0 &0 &0 &0 &0 &0 &1 &1 \\ 
 \end{pmatrix},%\end{array}\right],
\label{eq:H_Shor_Z}
\end{equation}
\begin{equation}
  H_x^\prime = \begin{pmatrix}%\left[\begin{array} {c c c c c c c c c }
 1 &1 &1 &1 &1 &1 &0 &0 &0\\
 0 &0 &0 &1 &1 &1 &1 &1 &1
 \end{pmatrix}.%\end{array}\right].
\label{eq:H_Shor_X}
\end{equation}
\begin{table}[tb]
 \centering
 {%
%\newcommand{\mc}[3]{\multicolumn{#1}{#2}{#3}}
%\definecolor{tcB}{rgb}{0.627451,0.627451,0.643137}
%\definecolor{tcA}{rgb}{0,0,0}
\begin{center}
\begin{tabular}{c|l}
% use packages: color,colortbl
\hline
%\mc{1}{>{\columncolor{tcB}}c}{\textbf{Stabilizer}} \\
& \textbf{Stabilizer}\\
\hline
$g_1$ & \textbf{ZZIIIIIII}\\
$g_2$ & \textbf{IZZIIIIII}\\
$g_3$ & \textbf{IIIZZIIII}\\
$g_4$ & \textbf{IIIIZZIII}\\
$g_5$ & \textbf{IIIIIIZZI}\\
$g_6$ & \textbf{IIIIIIIZZ}\\
$g_7$ & \textbf{XXXXXXIII}\\
$g_8$ & \textbf{IIIXXXXXX}\\
\hline \hline
 \end{tabular}
 \end{center}
}%
 \caption{Stabilizers for $9$-qubit Shor's code.}
 \label{tab:shor}
\end{table}

With the matrix notation of \eqr{eq:eq-classical}, the commutative property of stabilizers given in~\eqr{eq:stab:com} is transformed into the orthogonality of rows with respect to the 
symplectic product (also called
twisted product). If row $m$ is $r_m = (z_m, x_m)$, where $z_m$ and $x_m$ are the binary strings for $\textbf{Z}$ and $\textbf{X}$ respectively, then the symplectic product
of rows $m$ and $m'$ is given by, 
\begin{equation}
 r_m \star r_{m'} = (z_m \cdot x_{m'} + z_{m'} \cdot x_m) \; \text{mod 2}.
\end{equation}
This symplectic product is zero if there are even number of places where the operators ($\textbf{X}$ or $\textbf{Z}$) in row $m$ and $m'$ are different; thus meeting the
commutativity requirement. In other words, if $H$ is written as $H=(H_z|H_x)$, then the symplectic product is satisfied
for all the rows only if,
\begin{equation}
 H_z H_{x}^{T} + H_x H_{z}^{T} = 0,
\label{eq:twist}
\end{equation}
which may be readily verified for the $H$ of \eqr{eq:H_Shor}.
Consequently, any classical binary codes satisfying \eqr{eq:twist} may be used to construct QSCs. A special class of these stabilizer codes are CSS codes,
which are defined as follows:

\textit{An $[n,k_1-k_2]$ CSS code, which is capable of correcting $t$ bit as well as phase errors, can be constructed from classical
linear block codes $C_1(n,k_1)$ and $C_2(n,k_2)$, if $C_2 \subset C_1$ and both $C_1$ as well as the dual of $C_2$, i.e. $C_2^{\bot}$, can correct $t$ errors.}

\begin{figure}[tb]
\centering
\includegraphics[width=\linewidth]{\figures 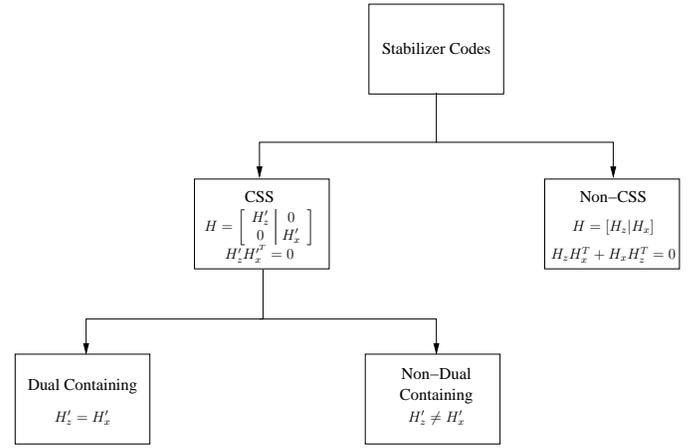}
\caption{Family of stabilizer codes.}
\label{fig:stab}
\end{figure}

In CSS construction, the PCM $H_z^\prime$ of $C_1$ is used for correcting bit errors, while the PCM $H_x^\prime$ of $C_2^{\bot}$ is used
for phase-error correction. Consequently, the PCM of the resultant CSS code takes the form of \eqr{eq:H_Shor}.
$H_z^\prime$ and $H_x^\prime$ are now the $(n-k_1) \times n$ and $k_2 \times n$ binary matrices, respectively. Furthermore, since $C_2 \subset C_1$, 
the symplectic condition of \eqr{eq:twist} is reduced to $H_z^\prime H_x^{\prime^T} = 0$. In this scenario, $(n-k_1+k_2)$ stabilizers are applied 
to $n$ qubits. Therefore, the resultant quantum code encodes 
$(k_1-k_2)$ information qubits into $n$ qubits. Furthermore, if $H_z^\prime=H_x^\prime$, the resultant structure is called dual-containing 
(or self-orthogonal) code because $H_z\prime H_z^{\prime^T} = 0$,
which is equivalent to $ C_1^{\bot} \subset C_1$.
Hence, stabilizer codes may be sub-divided into various code structures, which are summarized in \fref{fig:stab}.

Let us consider the classical $(7,4)$ Hamming code, whose PCM is given by:
\begin{equation}
 H = \begin{pmatrix}
      1 & 1 & 0 & 1 & 1 & 0 & 0\\
      1 & 0 & 1 & 1 & 0 & 1 & 0\\
      0 & 1 & 1 & 1 & 0 & 0 & 1
     \end{pmatrix}.
     \label{eq:ham}
\end{equation}
Since the $H$ of \eqr{eq:ham} yields $HH^T = 0$, it is used for constructing the dual-containing rate-$1/7$ Steane code~\cite{steane2}.

\begin{figure}[tb]
%\vspace*{-0.5cm}
  \begin{center}
  {\includegraphics[width=\linewidth]{\figures 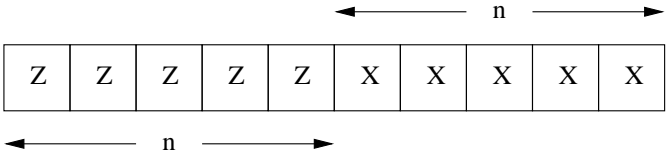}}
    \caption{Effective classical error $P$ corresponding to the error $\mathcal{P}$ imposed on an $n$-qubit frame.}
  \label{fig:eq_q_ch2}
%\vspace*{-0.5cm}
  \end{center}
\end{figure}

Based on the aforementioned Pauli-to-binary isomorphism, a quantum-based Pauli error operator $\mathcal{P}$ can be represented by the effective classical error 
pattern $P$, which is a binary vector of length $2n$. 
More specifically, $P$ is a concatenation of $n$ bits for $\mathbf{Z}$ errors, followed by another $n$ bits for $\mathbf{X}$ errors,
as depicted in \fref{fig:eq_q_ch2}.
An $\mathbf{X}$ error imposed on the $1$st qubit
will yield a $0$ and a $1$ at the $1$st and $(n+1)$th index of $P$, respectively. Similarly, a $\mathbf{Z}$ error imposed on the $1$st qubit
will give a $1$ and a $0$ at the $1$st and $(n+1)$th index of $P$, respectively, while a $Y$ error on the $1$st qubit will
result in a $1$ at both the $1$st as well as $(n+1)$th index of $P$\footnote{Since a depolarizing channel characterized by the probability $p$ incurs 
$\mathbf{X}$, $\mathbf{Y}$ and $\mathbf{Z}$ errors with an equal probability of $p/3$, the effective error-vector $P$ reduces to two Binary Symmetric Channels 
(BSCs), one channel for the $\mathbf{Z}$ errors and the other for the $\mathbf{X}$ errors. 
The crossover probability of each BSC is given by $2p/3$.}.
The resultant 
syndrome is given by the symplectic product of $\mathbf{H}$ and $P$, which is equivalent to $\mathbf{H}(P_x:P_z)^T$. Here colon ($:$) denotes the
concatenation operation.
In other words, the Pauli-$\mathbf{X}$ operator is used for correcting $\mathbf{Z}$ errors, while the Pauli-$\mathbf{Z}$ operator is used for correcting $\mathbf{X}$ errors~\cite{Qbook2}.
Thus, the quantum-domain syndrome is equivalent to the classical-domain binary syndrome 
and a basic quantum-domain decoding procedure is similar to syndrome based decoding of the equivalent classical code~\cite{sparse1}. 
However, due to the degenerate nature of quantum codes, quantum decoding aims for finding the most likely error coset, while the classical syndrome decoding~\cite{zbabar2013} finds the most likely error. 

Hence, an $[n,k]$ quantum stabilizer code associated with $(n-k)$ stabilizers can be effectively modeled using an $(n-k) \times 2n$-element classical PCM 
satisfying \eqr{eq:twist}. The coding rate of the equivalent classical code $R_c$ can be determined as follows:
\begin{align}
 R_c &= \frac{2n-(n-k)}{2n} \nonumber \\
     &= \frac{n+k}{2n} \nonumber \\
     &= \frac{1}{2} \left(1 + \frac{k}{n}\right) \nonumber \\
     &= \frac{1}{2} \left(1 + R_Q\right),
\label{eq:rc}
\end{align}
where $R_Q$ is its quantum coding rate. Using \eqr{eq:rc}, the coding rate of the classical equivalent of Shor's rate-$1/9$ quantum code is $5/9$.
\subsection{Stabilizer Formalism of Quantum Convolutional Codes} \label{sec:stab-QCC}
Quantum convolutional codes are derived from the corresponding classical convolutional codes using stabilizer formalism. This is based on the equivalence between
the classical convolutional codes and the classical linear block codes with semi-infinite length, which is derived below~\cite{shuerror}.

\begin{figure}[tb]
\centering
\includegraphics[scale=0.75]{\figures 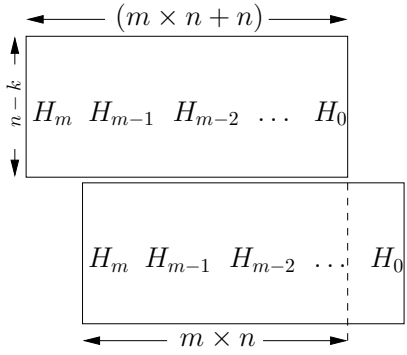}
\caption{Block-band structure of the semi-infinite classical PCM $H$.}
\label{fig:H_bb}
\end{figure}

Consider a $(2,1,m)$ classical convolutional code with generators,
\begin{align}
 g^{(0)} &= (g^{(0)}_0, g^{(0)}_1, \dots , g^{(0)}_m), \nonumber \\
 g^{(1)} &= (g^{(1)}_0, g^{(1)}_1, \dots , g^{(1)}_m).
\end{align}
For an input sequence $[u = (u_0, u_1, u_2, \dots)]$, the output sequences $[v^{(0)} = (v^{(0)}_0, v^{(0)}_1, v^{(0)}_2, \dots)]$ and
$[v^{(1)} = (v^{(1)}_0, v^{(1)}_1, v^{(1)}_2, \dots)]$are given as follows:
\begin{align}
 v^{(0)} &= u \circledast g^{(0)}, \nonumber \\
 v^{(1)} &= u \circledast g^{(1)}, 
\end{align}
where $\circledast$ denotes discrete convolution (modulo 2), which implies that for all $l \geq 0$ we have:
\begin{equation}
 v_l^{(j)} = \sum_{i = 0}^{m} u_{l - i} g_i^{(j)} = u_l g_0^{(j)} + u_{l-1} g_1^{(j)} + \dots + u_{l-m} g_m^{(j)}, 
\label{eq:enc}
\end{equation}
where $j = 0,1$ and $u_{l-i} \triangleq  0$ for all $l < i$. The two encoded sequences are multiplexed into a single codeword sequence $v$ given by:
\begin{equation}
 v = (v_0^{(0)}, v_0^{(1)}, v_1^{(0)}, v_1^{(1)}, v_2^{(0)}, v_2^{(1)}, \dots)
\end{equation}
This encoding process can also be represented in matrix notation by interlacing the generators $g^{(0)}$ and $g^{(1)}$ and arranging them in matrix form as follows\footnote{Blank spaces in the matrix indicate zeros.},
\begin{equation}
 G = \begin{pmatrix} %\left[\begin{array} {c c c c c c}
  g^{(0)(1)}_0&g^{(0)(1)}_1&\dots &g^{(0)(1)}_m\\
   &g^{(0)(1)}_0&g^{(0)(1)}_1&\dots &g^{(0)(1)}_m\\
   & &g^{(0)(1)}_0&g^{(0)(1)}_1&\dots &g^{(0)(1)}_m\\
    &  &\ddots             &                    &  \dots   &\ddots 
    \end{pmatrix},%\end{array}\right],
\end{equation}
where $ g^{(0)(1)}_i \triangleq \left(g^{(0)}_i  g^{(1)}_i\right)$.
The encoding operation of \eqr{eq:enc} is therefore equivalent to,
\begin{equation}
 v = uG.
\end{equation}
Since the information sequence $u$ is of arbitrary length, $G$ is semi-infinite. Furthermore, each row of $G$ is identical to the previous row, but is shifted
to the right by two places (since $n = 2$). In practice, $u$ has a finite length $N$. Therefore, $G$ has $N$ rows and $2(m+N)$ columns for CC$(2,1,m)$. For
CC$(n,k,m)$, $G$ can be generalized as follows:
\begin{equation}
 G = \begin{pmatrix} %\left[\begin{array} {c c c c c c}
  G_0 &G_1 &\dots &G_m \\
   &G_0 &G_1 &\dots &G_m\\
   & &G_0 &G_1 &\dots &G_m \\
    &  &\ddots         &                    &  \dots    &\ddots 
    \end{pmatrix}, %\end{array}\right],
\end{equation}
where $G_l$ is a ($k$ x $n$) submatrix with entries,
\begin{equation}
 G_l = \begin{pmatrix} %\left[\begin{array} {c c c c}
  g_{1,l}^{(0)} &g_{1,l}^{(1)} &\dots &g_{1,l}^{(n-1)}\\
  g_{2,l}^{(0)} &g_{2,l}^{(1)} &\dots &g_{2,l}^{(n-1)}\\
 \vdots         &\vdots        &      &\vdots\\
  g_{k,l}^{(0)} &g_{k,l}^{(1)} &\dots &g_{k,l}^{(n-1)} 
  \end{pmatrix}.  %\end{array}\right].
\end{equation}
The corresponding PCM $H$ can be represented as a semi-infinite matrix consisting of submatrices $H_l$ with dimensions of $(n-k) \times n$. For a
convolutional code
with constraint length\footnote{Constraint length is the number of memory units (shift registers) plus 1.} $(m+1)$, $H$ is given by:
\begin{equation}
  H = \begin{pmatrix} %\left[\begin{array} {c c c c c c c c}
  H_0 \\
  H_1 &H_0\\
  H_2 &H_1 &H_0\\
  \vdots &\vdots &\vdots\\
  H_m &H_{m-1} & H_{m-2} &\dots &H_0\\
  &H_m &H_{m-1} & H_{m-2} &\dots &H_0\\
  %&&H_m &H_{m-1} & H_{m-2} &\dots &H_0\\
  %&&&H_m &H_{m-1} & H_{m-2} &\dots &H_0\\
  &&\vdots  &\vdots   & &\vdots 
  \end{pmatrix}.  %\end{array}\right].
\end{equation}
Therefore, a CCC can be represented as a linear block code with semi-infinite block length. Furthermore, if each row of the submatrices $H_l$ is considered as a single block 
and $h_{j,i}$ is the $i$th row of the $j$th block, then $H$ has a block-band structure after the first $m$ blocks, whereby the successive blocks
are time-shifted versions of the first block $(j = 0)$ and the adjacent blocks have an overlap of $m$ submatrices. This has been depicted in \fref{fig:H_bb}
and can be mathematically represented as follows:
\begin{equation}
 h_{j,i} = [\mathbf{0}^{j \times n},h_{0,i}], \; 1 \leq i \leq(n-k), \; 0 \leq j,
\end{equation}
where $\mathbf{0}^{j \times n}$ is a row-vector with $(j \times n)$ zeros.

As discussed in Section~\ref{sec:p2b}, the rows of a classical PCM correspond to the stabilizers of a quantum code. Hence, the quantum stabilizer
group $\mathcal{H}$ of an $[n,k,m]$ stabilizer convolutional code is given by~\cite{ollivier2004}:
\begin{equation}
 \mathcal{H} = sp\{g_{j,i} = I^{\otimes jn} \otimes g_{0,i}\}, \; 1 \leq i \leq(n-k), \; 0 \leq j,
\label{stab}
\end{equation}
where $g_{j,i}$ is the $i$th stabilizer of the $j$th block of the stabilizer group $\mathcal{H}$. Furthermore, $sp$ represents a symplectic group, thus implying
that all the stabilizers $g_{j,i}$ must be independent and must commute with each other. 

As proposed by Forney in~\cite{forney2005,forney2007}, CSS-type QCCs can be derived from the classical self-orthogonal binary convolution codes. 
Let us consider the rate $1/3$ QCC of~\cite{forney2005,forney2007}, which is constructed from a binary rate-$1/3$ CCC with generators:
\begin{equation}
 G = \left(\begin{array} {c c c | c c c | c c c| c c c|c}
  1 & 1 & 1 & 1 & 0 & 0 & 1 & 1 & 0 & 0 & 0 & 0 &\dots\\
0 & 0 & 0 & 1 & 1 & 1 & 1 & 0 & 0 & 1 & 1 & 0 &\dots\\
  &   &   &   &  \hdots  &   &   &   &         &&&&
  
    \end{array}\right).
\label{qcc2}
\end{equation}
In $D$-transform notation, these generators are represented as $(1 + D + D^2, 1 + D^2,1)$. Each generator is orthogonal to all other generators under
the binary inner product, making it a self-orthogonal code. Moreover, the dual $C^{\bot}$ has the capability of correcting $1$ bit. 
Therefore, based on the CSS construction, the basic stabilizers of the corresponding single-error correcting $[3,1]$ QCC are as follows:
\begin{align}
 g_{0,1} &= [\mathbf{XXX},\mathbf{XII},\mathbf{XXI}], \\
 g_{0,2} &= [\mathbf{ZZZ},\mathbf{ZII},\mathbf{ZZI}].
\end{align}
Other stabilizers of $\mathcal{H}$ are the time-shifted versions of these basic stabilizers as depicted in \eqr{stab}.

Let us further consider a non-CSS QCC construction given by Forney in~\cite{forney2005,forney2007}. It is derived from the
classical self-orthogonal rate-$1/3$ quaternary $(\mathbb{F}_4)$ convolutional code $C$ having generators $(1 + D, 1 + wD, 1 + \bar{w}D)$,
where $\mathbb{F}_4 = \{0,1,w,\overline{w}\}$. These generators can also be represented as follows:
\begin{equation}
 G = \left(\begin{array} {c c c | c c c | c c c|c}
  1 & 1 & 1 & 1 & w & \bar{w} & 0 & 0 & 0 & \dots\\
0 & 0 & 0 & 1 & 1 & 1 & 1 & w & \bar{w} & \dots\\
  &   &   &   &  \hdots  &   &   &   &         &

\end{array}\right).
\label{qcc1}
\end{equation}
Since all these generators are orthogonal under the Hermitian inner product, $C$ is self-orthogonal. Therefore, a $[3,1]$ QCC can be derived from this classical code. The basic generators $g_{0,i}$, for $1 \leq i \leq 2$,
of the corresponding stabilizer group, $\mathcal{H}$, are generated by multiplying the generators of \eqr{qcc1} with $w$ and $\bar{w}$, and mapping $0$, $w$, $1$, $\bar{w}$
onto $\mathbf{I}$, $\mathbf{X}$, $\mathbf{Y}$ and $\mathbf{Z}$ respectively. The resultant basic stabilizers are as follows:
\begin{align}
 g_{0,1} &= \left(\mathbf{XXX},\mathbf{XZY}\right), \\
 g_{0,2} &= \left(\mathbf{ZZZ},\mathbf{ZYX}\right),
\end{align}
and all other constituent stabilizers of $\mathcal{H}$ can be derived using \eqr{stab}.
\subsection{Entanglement-Assisted Stabilizer Formalism} \label{sec:EA-stab}
Let us recall that the classical binary and quaternary codes may be 
used for constructing stabilizer codes only if they satisfy the symplectic criterion of \eqr{eq:twist}. 
Consequently, some of the well-known classical codes cannot be explored in the quantum domain. This limitation can be readily overcome by using the
entanglement-assisted stabilizer formalism, which exploits pre-shared entanglement between the transmitter and receiver
to embed a set of non-commuting stabilizer generators into a larger set of commuting generators. 

\begin{figure*}[tb]
\begin{center}
    \includegraphics[width=0.7\linewidth]{\figures 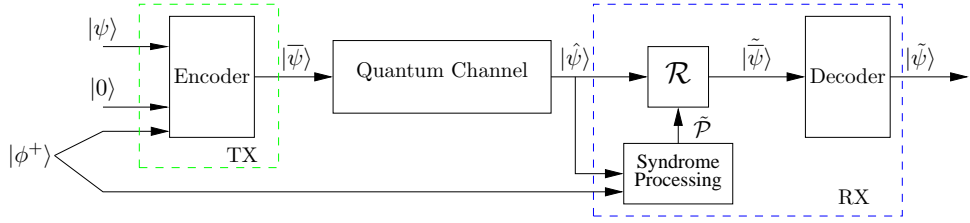}
    \caption{System Model: Quantum communication system relying on an entanglement-assisted quantum stabilizer code}.
  \label{fig:EA_stab_sys}
\end{center}
%\vspace{-0.5cm}
\end{figure*}

\fref{fig:EA_stab_sys} shows the general schematic
of a quantum communication system, which incorporates an Entanglement-Assisted Quantum Stabilizer Code (EA-QSC).
An $[n,k,c]$ EA-QSC encodes the information qubits $\ket{\psi}$ into the coded sequence $\ket{\overline{\psi}}$ with the aid of $(n-k-c)$ 
auxiliary qubits, which are initialized to the state $\ket{0}$. Furthermore, the transmitter and receiver share $c$ entangled qubits (ebits) before actual
transmission takes place.                 
This may be carried out during the off-peak hours, when the channel is under-utilized, thus efficiently distributing 
the transmission requirements in time.
More specifically, the state $\ket{\phi^+}$ of an ebit is given by the following Bell state:
\begin{equation}
 \ket{\phi^+} = \frac{\ket{00}^{T_XR_X} + \ket{11}^{T_XR_X}}{\sqrt{2}}, 
 \label{eq:ebit}
\end{equation}
where $T_X$ and $R_X$ denotes the transmitter's and receiver's half of the ebit, respectively. Similar to the superdense coding protocol of~\cite{SD92},
it is assumed that the receiver's half of the $c$ ebits are transmitted over a noiseless quantum channel, while the transmitter's half
of the $c$ ebits together with the $(n-k-c)$ auxiliary qubits are used to encode the intended $k$ information qubits into $n$ coded qubits. The resultant
$n$-qubit codewords $\ket{\overline{\psi}}$ are transmitted over a noisy quantum channel. The receiver then combines his half of the $c$
noiseless ebits with the received $n$-qubit noisy codewords $\ket{\hat{\psi}}$ to compute the syndrome, which is used for estimating the error $\tilde{\mathcal{P}}$ 
incurred on the $n$-qubit codewords. The rest of the processing at the receiver is the same as that in \fref{fig:stab_sys}.

The entangled state of \eqr{eq:ebit} has unique commutativity properties, which aid in transforming a set of non-abelian generators into
an abelian set. The state $\ket{\phi^+}$ is stabilized by the operators $\mathbf{X}^{T_X}\mathbf{X}^{R_X}$ and $\mathbf{Z}^{T_X}\mathbf{Z}^{R_X}$, which commute
with each other. Therefore, we have\footnote{$[a,b]$ represents the commutative relation between $a$ and $b$, 
while $\{a,b\}$ denotes the anti-commutative relation.}:
\begin{equation}
 [\mathbf{X}^{T_X}\mathbf{X}^{R_X},\mathbf{Z}^{T_X}\mathbf{Z}^{R_X}] = 0.
\end{equation}
However, local operators acting on either of the qubits anti-commute, i.e. we have:
\begin{equation}
  \{\mathbf{X}^{T_X},\mathbf{Z}^{T_X}\}=\{\mathbf{X}^{R_X},\mathbf{Z}^{R_X}\} = 0.
\end{equation}
Therefore, if we have two single qubit operators $\mathbf{X}^{T_X}$ and $\mathbf{Z}^{T_X}$, which anti-commute with each other, then we can resolve
the anti-commutativity by entangling another qubit and choosing the local operators on this additional qubit such that the resultant two-qubit 
generators ($\mathbf{X}^{T_X}\mathbf{X}^{R_X}$ and $\mathbf{Z}^{T_X}\mathbf{Z}^{R_X}$ for this case) commute. 
This additional qubit constitutes the receiver half of the ebit.
In other words, we entangle an additional
qubit for the sake of ensuring that the resultant two-qubit operators have an even number of places with different non-identity operators, which
in turn ensures commutativity.

Let us consider a pair of classical binary codes associated with the following PCMs:
\begin{equation}
 H_z = \begin{pmatrix}
        0 & 1 & 0 & 0 \\
        0 & 0 & 0 & 0 \\
        1 & 1 & 1 & 0 \\
        0 & 1 & 1 & 1 
       \end{pmatrix},
\end{equation}
and
\begin{equation}
 H_x = \begin{pmatrix}
        1 & 0 & 1 & 0 \\
        1 & 1 & 0 & 1 \\
        1 & 0 & 0 & 1 \\
        1 & 1 & 1 & 0
       \end{pmatrix},
\end{equation}
which are used to construct a non-CSS quantum code having $H = (H_z|H_x)$. The PCM $H$ does not satisfy the symplectic criterion.
The resultant non-abelian set of Pauli generators are as follows:
\begin{equation}
 H_Q = \left(\begin{array} {c c c c}
  \mathbf{X}& \mathbf{Z} & \mathbf{X} & \mathbf{I} \\
  \mathbf{X}& \mathbf{X} & \mathbf{I} & \mathbf{X} \\
  \mathbf{Y}& \mathbf{Z} & \mathbf{Z} & \mathbf{X} \\
  \mathbf{X}& \mathbf{Y} & \mathbf{Y} & \mathbf{Z}
 \end{array}\right).
  \label{eq:Ea-eg}
\end{equation}
In \eqr{eq:Ea-eg}, the first two generators (i.e. the first and second row) anti-commute, while all other generators commute with each other. This is because the local
operators acting on the second qubit in the first two generators anti-commute, while the local operators acting on all other qubits in these two generators commute.
In other words, there is a single index (i.e. $2$) with different non-Identity operators. To transform this non-abelian set  into an abelian set,
we may extend the generators of \eqr{eq:Ea-eg} with a single additional qubit, whose local operators also anti-commute for the sake of
ensuring that the resultant extended generators commute. Therefore, we get:
\begin{equation}
 H_Q = \left(\begin{array} {c c c c | c}
  \mathbf{X}& \mathbf{Z} & \mathbf{X} & \mathbf{I} & \mathbf{Z} \\
  \mathbf{X}& \mathbf{X} & \mathbf{I} & \mathbf{X} & \mathbf{X} \\
  \mathbf{Y}& \mathbf{Z} & \mathbf{Z} & \mathbf{X} & \mathbf{I} \\
  \mathbf{X}& \mathbf{Y} & \mathbf{Y} & \mathbf{Z} & \mathbf{I} 
 \end{array}\right),
  \label{eq:Ea-eg2}
\end{equation}
where the operators to the left of the vertical bar $(|)$ act on the transmitted
$n$-qubit codewords, while those on the right of the vertical bar act on the receiver's half of the ebits.
\section{Concatenated Quantum Codes} \label{sec:CQCC}
In this section, we will lay out the structure of a concatenated quantum code, with a special emphasis on the encoder structure and the decoding
algorithm. We commence with the circuit-based representation of quantum stabilizer codes, followed by the system model and then the decoding algorithm.
\subsection{Circuit-Based Representation of Stabilizer Codes} \label{sec:cct-QCC}
Circuit-based representation of quantum codes facilitates the design of concatenated code structures. More specifically, for decoding
concatenated quantum codes it is more convenient to exploit the circuit-based representation of the constituent codes, rather than the conventional 
PCM-based syndrome decoding. Therefore, in this section, we will review the circuit-based representation of 
quantum codes. This discussion is based on~\cite{qturbo2}.

\begin{figure}[tb]
%\vspace*{-0.5cm}
  \begin{center}
  {\includegraphics[width=0.6\linewidth]{\figures 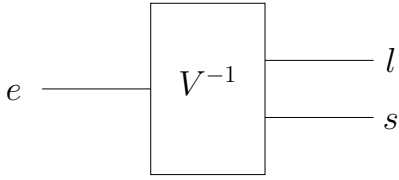}}
    \caption{Circuit representation of the inverse encoder $eV^{-1} = (l : s)$.}
  \label{fig:cct_V2}
%\vspace*{-0.5cm}
  \end{center}
\end{figure}

Let us recall from Section~\ref{sec:LBC} that an $(n,k)$ classical linear block code constructed over the code space $C$ maps the information word $x \in \mathbb{F}^k_2$ 
onto the corresponding codeword $\overline{x} \in \mathbb{F}^n_2$. In the circuit-based representation, this encoding procedure can be encapsulated as follows:
\begin{equation}
 C = \{\overline{x} = \left(x:0_{n-k}\right)V\},
\label{eq:V-classical}
\end{equation}
where $V$ is an $(n \times n)$-element invertible encoding matrix over $\mathbb{F}_2$ and $0_{n-k}$ is an $(n-k)$-bit vector initialized to $0$. Furthermore, given the generator matrix $G$ and the PCM $H$,
the encoding matrix $V$ may be specified as:
\begin{equation}
 V = \begin{pmatrix}
      G \\
      \left(H^{-1}\right)^T
     \end{pmatrix},
\label{eq:V1}
\end{equation}
and its inverse is given by:
\begin{equation}
 V^{-1} = \begin{pmatrix}
      G^{-1} &H^T
     \end{pmatrix}.
\label{eq:V-inv}
\end{equation}
The encoding matrix $V$ specifies both the code space as well as the encoding operation, while its inverse $V^{-1}$ specifies the error syndrome.
More specifically, let $y = \overline{x} + e$ be the received codeword, where $e$ is the $n$-bit error incurred during transmission. Then, passing the received codeword
$y$ through the inverse encoder $V^{-1}$ yields:
\begin{equation}
 y V^{-1} = \left(\tilde{x}:s \right),
\label{eq:V-inv-opr}
\end{equation}
where $\tilde{x} = x + l$ for the logical error $l \in \mathbb{F}^k_2$ inflicted on the information word $x$ and $s \in \mathbb{F}^{n-k}_2$ is the syndrome, which is equivalent to $yH^T$.
\eqr{eq:V-inv-opr} may be further decomposed to:
\begin{align}
 \left(\overline{x}+e\right) V^{-1} &= \left(x+l:s\right), \nonumber \\
  \overline{x}V^{-1} + eV^{-1} &= \left(x:0_{n-k} \right) + \left(l:s \right),
\label{eq:V-inv-opr2}
\end{align}
which is a linear superposition of the inverse of \eqr{eq:V-classical} and $eV^{-1} = \left(l:s\right)$.
Hence, the inverse encoder $V^{-1}$ decomposes the channel error $e$ into the logical error $l$ and error syndrome $s$,
which is also depicted in \fref{fig:cct_V2}.

Analogously to \eqr{eq:V-classical}, the unitary encoding operation $\mathcal{V}$ of an $[n,k]$ QSC, constructed over a code space $\mathcal{C}$, 
which maps the information word (logical qubits) $\ket{\psi} \in \mathbb{C}^{2^k}$ onto the codeword (physical qubits) $\ket{\overline{\psi}} \in \mathbb{C}^{2^n}$,
may be mathematically encapsulated as follows:
\begin{equation}
 \mathcal{C} = \{\ket{\overline{\psi}} = \mathcal{V}(\ket{\psi} \otimes \ket{0_{n-k}}) \},
\label{eq:unitary}
\end{equation}
where $\ket{0_{n-k}}$ are $(n-k)$ auxiliary qubits initialized to the state $\ket{0}$.
The unitary encoder $\mathcal{V}$ of \eqr{eq:unitary}
carries out an $n$-qubit Clifford transformation, which maps an $n$-qubit Pauli group $\mathcal{G}_n$ onto itself under conjugation~\cite{clif2003},
i.e. we have:
\begin{equation}
 \mathcal{V} \mathcal{G}_n \mathcal{V}^\dag = \mathcal{G}_n. 
\label{eq:bin_V}
\end{equation}
In other word, a Clifford operation preserves the elements of the Pauli group under conjugation such that for $\mathcal{P} \in \mathcal{G}_n$, 
$\mathcal{V} \mathcal{P} \mathcal{V}^\dag \in \mathcal{G}_n$. Furthermore,
any Clifford unitary matrix is completely specified by a combination of Hadamard ($\mathbf{H}$) gates, phase ($\mathbf{S}$) gates
and controlled-NOT (C-NOT) gates, which are defined as follows~\cite{Qbook2}:
\begin{align}
&\mathbf{H}= \frac{1}{\sqrt{2}}\begin{pmatrix}
  1& 1 \\
  1& -1
\end{pmatrix}, \
\mathbf{S}=\begin{pmatrix} 
  1& 0 \\
  0& i
\end{pmatrix}, \nonumber \\
&\text{C-NOT}=\begin{pmatrix}
  1& 0 & 0 & 0 \\
  0& 1 & 0 & 0 \\
  0& 0 & 0 & 1 \\
  0& 0 & 1 & 0
\end{pmatrix}.
\label{eq:Cliff_gates}
\end{align}
Hadamard gate preserves the elements of a single-qubit Pauli group $\mathcal{G}_1$ as follows:
\begin{align}
 \mathbf{X} &\rightarrow \mathbf{H}\mathbf{X}\mathbf{H}^\dag = \mathbf{Z}, \nonumber \\
 \mathbf{Z} &\rightarrow \mathbf{H}\mathbf{Z}\mathbf{H}^\dag = \mathbf{X}, \nonumber \\
 \mathbf{Y} &\rightarrow \mathbf{H}\mathbf{Y}\mathbf{H}^\dag = -\mathbf{Y},
\label{eq:had:trans}
\end{align}
while phase gate preserves them as:
\begin{align}
 \mathbf{X} &\rightarrow \mathbf{S}\mathbf{X}\mathbf{S}^\dag = \mathbf{Y}, \nonumber \\
 \mathbf{Z} &\rightarrow \mathbf{S}\mathbf{Z}\mathbf{S}^\dag = \mathbf{Z},  \nonumber \\
 \mathbf{Y} &\rightarrow \mathbf{S}\mathbf{Y}\mathbf{S}^\dag = -\mathbf{X},
\label{eq:phase:trans}
\end{align}
Since C-NOT is a $2$-qubit gate, it acts on the elements of $\mathcal{G}_2$, transforming the standard basis of $\mathcal{G}_2$ as given below:
\begin{align}
 \mathbf{X} \otimes \mathbf{I} &\rightarrow \mathbf{X} \otimes \mathbf{X},  \nonumber \\
 \mathbf{I} \otimes \mathbf{X} &\rightarrow \mathbf{I} \otimes \mathbf{X},  \nonumber \\
 \mathbf{Z} \otimes \mathbf{I} &\rightarrow \mathbf{Z} \otimes \mathbf{I},  \nonumber \\ 
 \mathbf{I} \otimes \mathbf{Z} &\rightarrow \mathbf{Z} \otimes \mathbf{Z}. 
\label{eq:cnot:trans}
\end{align}
Let us further emphasize on the significance of Clifford encoding operation. \textit{Since $\mathcal{V}$ belongs to the Clifford group, it preserves the elements of the
stabilizer group $\mathcal{H}$ under conjugation}. If $g'_i$ is the $i$th stabilizer of the unencoded state $\ket{\psi}$, then this may
be proved as follows:
\begin{equation}
 \ket{\psi} \otimes \ket{0_{n-k}} = g'_i \left(\ket{\psi} \otimes \ket{0_{n-k}}\right).
\end{equation}
Encoding $\ket{\psi}$ with $\mathcal{V}$ yields:
\begin{equation}
  \mathcal{V}\left(\ket{\psi} \otimes \ket{0_{n-k}}\right) = \mathcal{V} \left(g'_i \left(\ket{\psi} \otimes \ket{0_{n-k}}\right)\right),
\end{equation}
which is equivalent to:
\begin{equation}
  \mathcal{V}\left(\ket{\psi} \otimes \ket{0_{n-k}}\right) = \mathcal{V} \left(g'_i \mathcal{V}^\dag \mathcal{V}\left(\ket{\psi} \otimes \ket{0_{n-k}}\right)\right),
\label{eq:proof_clif1}
\end{equation}
since $\mathcal{V}^\dag \mathcal{V} = \mathbb{I}_n$. Substituting \eqr{eq:unitary} into \eqr{eq:proof_clif1} gives:
\begin{equation}
  \ket{\overline{\psi}} = \left( \mathcal{V} g'_i \mathcal{V}^\dag \right) \ket{\overline{\psi}}.
\label{eq:proof_clif2}
\end{equation}
Hence, the encoded state $\ket{\overline{\psi}}$ is stabilized by $g_i = \mathcal{V} g'_i \mathcal{V}^\dag$. From this it appears as if any
arbitrary $\mathcal{V}$ (not necessarily Clifford) can be used to preserve the stabilizer subspace, which is not true. Since we assume that the stabilizer group
$\mathcal{H}$ is a subgroup of the Pauli group, we impose the additional constraint that $\mathcal{V}$ must yield the elements of Pauli group under conjugation as 
in \eqr{eq:bin_V}, which is only true for Clifford operations. 

Furthermore, \textit{the Clifford encoding operation also preserves the commutativity relation of stabilizers}.
Let $g'_i$ and $g'_j$ be a pair of unencoded stabilizers. Then the above statement can be proved as follows:
\begin{equation}
g_i g_j = \left(\mathcal{V} g'_i \mathcal{V}^\dag \right) \left(\mathcal{V} g'_j \mathcal{V}^\dag \right) = \mathcal{V} g'_i g'_j \mathcal{V}^\dag.
\end{equation}
Since $g'_i$ and $g'_j$ commute, we have:
\begin{equation}
 \mathcal{V} g'_i g'_j \mathcal{V}^\dag = \mathcal{V} g'_j g'_i \mathcal{V}^\dag.
\end{equation}
Using $\mathcal{V}^\dag \mathcal{V} = \mathbb{I}_n$, gives:
\begin{equation}
 \mathcal{V} g'_j \mathcal{V}^\dag \mathcal{V} g'_i \mathcal{V}^\dag. = g_j g_i. 
\end{equation}

Since the $n$-qubit Pauli group forms a basis for the ($2^n\times 2^n$)-element matrices of \eqr{eq:Cliff_gates}, the Clifford encoder $\mathcal{V}$, which acts on the $2^n$-dimensional Hilbert space, 
can be completely defined by specifying its action under conjugation on the Pauli-$\mathbf{X}$ and $\mathbf{Z}$ operators acting on each of the $n$ qubits,
as seen in \eqr{eq:had:trans} to~\eqref{eq:cnot:trans}. However,
$\mathcal{V}$ and $\mathcal{V}'$, which differ only through a global phase such that $\mathcal{V}' = e^{j\theta}\mathcal{V}$, have the same impact under
conjugation. Therefore, global phase has no physical significance in the context of \eqr{eq:bin_V} and the $n$-qubit encoder $\mathcal{V}$ can be completely
specified by its action on the binary equivalent of the Pauli operators. More specifically, for an $n$-qubit Clifford transformation, there is an equivalent $2n \times 2n$
binary symplectic matrix $V$, which is given by:
\begin{equation}
 [ \mathcal{V} \mathcal{P} \mathcal{V}^\dag ] = [\mathcal{P}] V = PV,
\end{equation}
where $[.]$ denotes the effective Pauli group $G_n$ such that $P = [\mathcal{P}]$ differs from $\mathcal{P}$ by a multiplicative constant, i.e.
we have $P = \mathcal{P}/\{\pm 1 \pm i\}$, and the elements of $G_n$ are represented by $2n$-tuple binary vectors based on the mapping given
in \eqr{eq:mapping}.
%\footnote{In Chapter~\ref{C4} and~\ref{C5}, $P$ was defined as $P = (P_x : P_z)$. However, in this chapter we have used the standard notation of
%\eqr{eq:mapping}, i.e. $P = (P_z : P_x)$, where colon ($:$) denotes the concatenation operation. Consequently, syndrome $s$, which was previously defined as $HP$, is now equivalent to the symplectic product of
%$P$ and $H$, i.e. $s = \left(P \star H_i\right)_{1\leq i \leq n-k}$, where $\star$ represents the symplectic inner product.}. %, as discussed in Section~\ref{c4:relation_qc}.
As a consequence of this equivalence, any Clifford unitary can be efficiently simulated on a classical system as stated in the Gottesman-Knill theorem~\cite{gottesman-knill}.

We next define $\mathcal{V}$ by specifying its action on the elements of the Pauli group $\mathcal{G}_n$. More precisely, we consider 
$2n$ $n$-qubit unencoded operators $Z_i, X_i, \dots, Z_n, X_n$, where $Z_i$ and $X_i$ represents the Pauli $\mathbf{Z}$ and $\mathbf{X}$ operator, respectively, 
acting on the $i$th qubit and the identity $\mathbf{I}$ on all other qubits. The unecoded operators $Z_{k+1}, \dots, Z_n$ stabilizes the unencoded state of \eqr{eq:unitary}, 
i.e. $(\ket{\psi} \otimes \ket{0_{n-k}})$, and are therefore called the unencoded stabilizer generators. On the other hand, $X_{k+1}, \dots, X_n$ are the unencoded
pure errors since $X_i$ anti-commutes with the corresponding unencoded stabilizer generator $Z_i$, yielding an error syndrome of $1$.
Furthermore, the unencoded logical operators acting on the information qubits are $Z_i, X_i, \dots, Z_k, X_k$, which commute with the unencoded stabilizers 
$Z_{k+1}, \dots, Z_n$. The encoder $\mathcal{V}$ maps the unencoded operators $Z_i, X_i, \dots, Z_n, X_n$ onto the encoded operators 
$\overline{Z}_i, \overline{X}_i, \dots, \overline{Z}_n, \overline{X}_n$, which may be represented as follows:
\begin{equation}
 \overline{X}_i = \left[\mathcal{V} X_i \mathcal{V}^\dag\right] = \left[X_i\right] V, \; \; \; \; \; \; \overline{Z}_i = \left[\mathcal{V} Z_i \mathcal{V}^\dag\right] = \left[Z_i\right] V.
\end{equation}
Since Clifford transformations do not perturb the commutativity relation of the operators, the resultant encoded stabilizers 
$\overline{Z}_{k+1}, \dots, \overline{Z}_n$ are equivalent to the stabilizers $g_i$ of \eqr{eq:stab}, while $\overline{X}_{k+1}, \dots, \overline{X}_n$ are the pure
errors $t_i$ of the resultant stabilizer code, which trigger a non-trivial syndrome. Moreover, $\overline{Z}_i, \overline{X}_i, \dots, \overline{Z}_k, \overline{X}_k$
are the encoded logical operators, which commute with the stabilizers $g_i$. Logical operators merely map one codeword onto the other, without affecting
the codespace $\mathcal{C}$ of the stabilizer code. It also has to be mentioned here that the stabilizer generators $g_i$ together with the encoded logical
operations constitute the normalizer of the stabilizer code. 
The ($2n \times 2n$)-element binary symplectic encoding matrix $V$ is therefore given by:
\begin{equation}
 V =  \begin{pmatrix}
      \overline{Z}_1 \\
      \vdots \\
      \overline{Z}_k \\
      \overline{Z}_{k+1} \\
      \vdots \\
      \overline{Z}_n\\
      \overline{X}_1 \\
      \vdots \\
      \overline{X}_k \\
      \overline{X}_{k+1} \\
      \vdots \\
      \overline{X}_n 
     \end{pmatrix} =
 \begin{pmatrix}
      \overline{Z}_1 \\
      \vdots \\
      \overline{Z}_k \\
      g_1 \\
      \vdots \\
      g_{n-k} \\
      \overline{X}_1 \\
      \vdots \\
      \overline{X}_k \\
      t_1 \\
      \vdots \\
      t_{n-k} 
     \end{pmatrix},
\label{eq:V}
\end{equation}
where the Pauli $\mathbf{Z}$ and $\mathbf{X}$ operators are mapped onto the classical bits using the Pauli-to-binary isomorphism of Section~\ref{sec:p2b}.

Analogously to the classical inverse encoder of \eqr{eq:V-inv-opr}, the inverse encoder of a quantum code is the Hermitian conjugate $\mathcal{V}^\dag$.
Let $\ket{\hat{\psi}} = \mathcal{P} \ket{\overline{\psi}}$ be the received codeword such that $\mathcal{P}$ is the $n$-qubit channel error. Then,
passing the received codeword $\ket{\hat{\psi}}$ through the inverse encoder $\mathcal{V}^{\dag}$ yields:
\begin{align}
 \mathcal{V}^{\dag}\mathcal{P}|\overline{\psi}\rangle &= \mathcal{V}^{\dag}\mathcal{P}\mathcal{V}(|\psi\rangle \otimes |0_{(n-k)}\rangle) \nonumber \\
& = (\mathcal{L}|\psi\rangle) \otimes (\mathcal{S}|0_{(n-k)}\rangle),
\label{eq:inverse-cct}
\end{align}
where $\mathcal{V}^{\dag}\mathcal{P}\mathcal{V} \equiv (\mathcal{L} \otimes \mathcal{S})$ and $\mathcal{L} \in \mathcal{G}_k$ denotes the error imposed on the information word, while
$\mathcal{S} \in \mathcal{G}_{n-k}$ represents the error inflicted on the remaining $(n - k)$ auxiliary qubits. In the equivalent binary representation, \eqr{eq:inverse-cct}
may be modeled as follows:
\begin{equation}
 PV^{-1} = \left(L:S\right),
\end{equation}
where we have $P = [\mathcal{P}]$, $L = [\mathcal{L}]$ and $S = [\mathcal{S}]$.

\begin{figure}[tb]
%\vspace*{-0.5cm}
  \begin{center}
  {\includegraphics[width=0.6\linewidth]{\figures 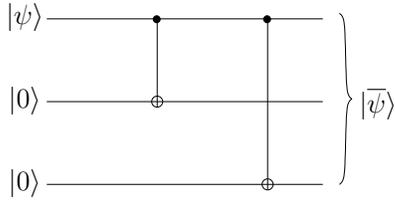}}
    \caption{Encoding Circuit for $3$-qubit bit-flip repetition code.}
  \label{fig:cct-3qubit}
\vspace*{-0.5cm}
  \end{center}
\end{figure}

Let us now derive the encoding matrix $V$ for the $3$-qubit bit-flip repetition code, which has a binary PCM $H$ given by:
\begin{equation}
 H = \left(\begin{array} {c c c|c c c}
  1 &1 &0 &0 &0 &0\\
  1 &0 &1 &0 &0 &0
 \end{array}\right).
\label{H:3qubit-2}
\end{equation}
The corresponding encoding circuit is depicted in \fref{fig:cct-3qubit}. Its unencoded operators are as follows:
\begin{equation}
\begin{pmatrix}
      Z_1 \\
      Z_2 \\
      Z_3 \\
      X_1 \\
      X_2 \\
      X_3
     \end{pmatrix} =
\begin{pmatrix}
      \mathbf{ZII} \\
      \mathbf{IZI} \\
      \mathbf{IIZ} \\
      \mathbf{XII}\\
      \mathbf{IXI}\\
      \mathbf{IIX}
     \end{pmatrix} \equiv
\left(\begin{array}{c c c|c c c}
  1 &0 &0 &0 &0 &0 \\
  0 &1 &0 &0 &0 &0 \\
  0 &0 &1 &0 &0 &0 \\ \hline
  0 &0 &0 &1 &0 &0 \\
  0 &0 &0 &0 &1 &0 \\
  0 &0 &0 &0 &0 &1 
 \end{array} \right).
\end{equation}
A C-NOT gate is then applied to the second qubit, which is controlled by the first. As seen in \eqr{eq:cnot:trans}, the C-NOT gate copies Pauli $\mathbf{X}$ operator
forward from the control qubit to the target qubit, while $\mathbf{Z}$ is copied in the opposite direction. Therefore, we get:
\begin{equation}
\begin{pmatrix}
      \mathbf{ZII} \\
      \mathbf{IZI} \\
      \mathbf{IIZ} \\
      \mathbf{XII}\\
      \mathbf{IXI}\\
      \mathbf{IIX}
     \end{pmatrix} \underrightarrow{\text{C-NOT}(1,2)}
\begin{pmatrix}
      \mathbf{ZII} \\
      \mathbf{ZZI} \\
      \mathbf{IIZ} \\
      \mathbf{XXI}\\
      \mathbf{IXI}\\
      \mathbf{IIX}
     \end{pmatrix} \equiv
\left(\begin{array}{c c c|c c c}
  1 &0 &0 &0 &0 &0 \\
  1 &1 &0 &0 &0 &0 \\
  0 &0 &1 &0 &0 &0 \\ \hline
  0 &0 &0 &1 &1 &0 \\
  0 &0 &0 &0 &1 &0 \\
  0 &0 &0 &0 &0 &1 
 \end{array} \right).
\end{equation}
Another C-NOT gate is then applied to the third qubit, which is also controlled by the first, yielding:
\begin{align}
\begin{pmatrix}
      \mathbf{ZII} \\
      \mathbf{ZZI} \\
      \mathbf{IIZ} \\
      \mathbf{XXI}\\
      \mathbf{IXI}\\
      \mathbf{IIX}
     \end{pmatrix} \underrightarrow{\text{C-NOT}(1,3)}
\begin{pmatrix} 
      \mathbf{ZII} \\
      \mathbf{ZZI} \\
      \mathbf{ZIZ} \\
      \mathbf{XXX}\\
      \mathbf{IXI}\\
      \mathbf{IIX}
     \end{pmatrix} &\equiv
\left(\begin{array}{c c c|c c c}
  1 &0 &0 &0 &0 &0 \\
  1 &1 &0 &0 &0 &0 \\
  1 &0 &1 &0 &0 &0 \\ \hline
  0 &0 &0 &1 &1 &1 \\
  0 &0 &0 &0 &1 &0 \\
  0 &0 &0 &0 &0 &1 
 \end{array} \right) \nonumber \\ &= V.
\label{eq:3qubit2}
\end{align}
As gleaned from \eqr{eq:3qubit2}, the stabilizer generators of the $3$-qubit bit-flip repetition code are $g_1 = \mathbf{ZZI}$ and $g_2 = \mathbf{ZIZ}$.
More explicitly, rows $2$ and $3$ of $V$ constitute the PCM $H$ of \eqr{H:3qubit-2}. The encoded logical operators are
$\overline{Z}_1 = \mathbf{ZII}$ and $\overline{X}_1 = \mathbf{XXX}$, which commute with the stabilizers $g_1$ and $g_2$. Finally, the pure errors are
$t_1 = \mathbf{IXI}$ and $t_2 = \mathbf{IIX}$, which anti-commute with $g_1$ and $g_2$, respectively, yielding a non-trivial syndrome. 

\begin{figure}[tb]
%\vspace*{-0.5cm}
  \begin{center}
  {\includegraphics[width=\linewidth]{\figures 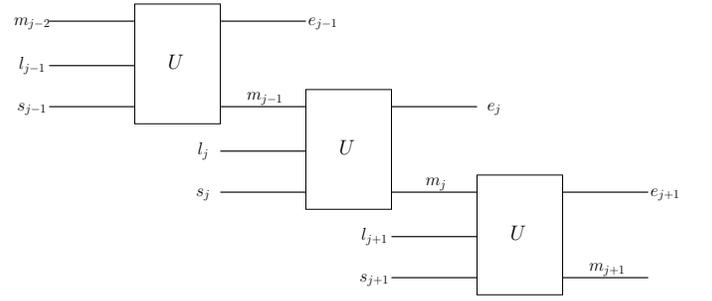}}
    \caption{Circuit representation of the encoder $V$ of a convolutional code~\cite{qturbo2}.}
  \label{fig:cct_U}
%\vspace*{-0.5cm}
  \end{center}
\end{figure}

Based on the above discussion, we now proceed to lay out the circuit-based model for a convolutional code, which is given in~\cite{qturbo2}. 
As discussed in Section~\ref{sec:stab-QCC}, convolutional codes are equivalent to linear
block codes associated with semi-infinite block lengths. More specifically, as illustrated in \fref{fig:H_bb}, the PCM $H$ of an $(n,k,m)$ convolutional code has a block-band
structure, where the adjacent blocks have an overlap of $m$ submatrices. Similarly, the encoder $V$ of a classical convolutional code can be built from repeated
applications of a linear invertible seed transformation $U$, which is an $(n+m) \times (n+m)$-element encoding matrix, as shown in \fref{fig:cct_U}. The inverse encoder
$V^{-1}$ can be easily obtained by moving backwards in time, i.e. by reading \fref{fig:cct_U} from right to left. Let us further elaborate by stating that at time instant $j$, 
the seed transformation matrix $U$ takes as its input the memory bits $m_{j-1} \in \mathbb{F}_2^m$, the logical bits $l_j \in \mathbb{F}_2^k$ and the syndrome bits 
$s_j \in \mathbb{F}_2^{n-k}$ to generate the output bits $e_j \in \mathbb{F}_2^n$ and the memory state $m_j$. More explicitly,
we have:
\begin{equation}
 \left(m_{j}:e_j\right) = \left(m_{j-1}:l_j:s_j\right) U,
\label{eq:cct:cc}
\end{equation}
and the overall encoder is formulated as~\cite{qturbo2}:
\begin{align}
 V &= U_{[1,\dots,n+m]} U_{[n+1,\dots,2n+m]} \dots U_{[(N-1)n+1,\dots,Nn+m]}, \nonumber \\
   &= \prod_{j=1}^{N} U_{[(j-1)n+1,\dots,jn+m]},
\end{align}
where $N$ denotes the length of the convolutional code and $U_{[(j-1)n+1,\dots,jn+m]}$ acts on $(n+m)$ bits, i.e. $\left(m_{j-2}:l_{j-1}:s_{j-1}\right)$.
For an $[n,k,m]$ quantum convolutional code, the seed transformation $U$ is a $2(n+m) \times 2(n+m)$-element symplectic matrix and 
\eqr{eq:cct:cc} may be re-written as:
\begin{equation}
 \left(M_{j}:P_j\right) = \left(M_{j-1}:L_j:S_j\right) U,
\label{eq:cct:qcc}
\end{equation}
where $M$ represents the memory state with an $m$-qubit Pauli operator.

The aforementioned methodology conceived for constructing the circuit-based model of unassisted quantum codes may be readily extended to the class of
entanglement-assisted codes~\cite{wilde_turbo2}. The unitary encoding operation $\mathcal{V}$ of an $[n,k,c]$ EA-QSC, 
which acts only on the $n$ transmitter qubits, may be mathematically modeled as follows:
\begin{equation}
 \mathcal{C} = \{\ket{\overline{\psi}} = \mathcal{V}(\ket{\psi}^{T_X} \otimes \ket{0_{a}}^{T_X} \otimes \ket{\phi^+_{c}}^{T_XR_X}) \},
\label{eq:unitary-EA}
\end{equation}
where the superscripts $T_X$ and $R_X$ denote the transmitter's and receiver's qubits, respectively. Furthermore, $\ket{0_{a}}^{T_X}$ are $a$
auxiliary qubits initialized to the state $\ket{0}$, where $a = ( n - k -c)$, and $\ket{\phi^+_{c}}^{T_XR_X}$ are the $c$
entangled qubits. Analogously to \eqr{eq:inverse-cct}, the inverse encoder of an entanglement-assisted quantum code $\mathcal{V}^{^\dag}$ gives:
\begin{align}
 \mathcal{V}^{\dag}\mathcal{P}|\overline{\psi}\rangle &= \mathcal{V}^{\dag}\mathcal{P}\mathcal{V}(\ket{\psi}^{T_X} \otimes \ket{0_{a}}^{T_X} \otimes \ket{\phi^+_{c}}^{T_XR_X}) \nonumber \\
& = (\mathcal{L}^{T_X}\ket{\psi}^{T_X}) \otimes (\mathcal{S}^{T_X}\ket{0_{a}}^{T_X} \otimes (\mathcal{E}^{T_X}\ket{\phi^+_{c}}^{T_XR_X}),
\label{eq:inverse-cct-EA}
\end{align}
where $\mathcal{L}^{T_X} \in \mathcal{G}_k$ denotes the error imposed on the information word, while
$\mathcal{S}^{T_X} \in \mathcal{G}_{a}$ represents the error inflicted on the transmitter's $a$ auxiliary qubits and 
$\mathcal{E}^{T_X} \in \mathcal{G}_{c}$ is the error corrupting the transmitter's half of $c$ ebits. The equivalent binary representation
of \eqr{eq:inverse-cct-EA} is given by:
\begin{equation}
 PV^{-1} = \left(L:S:E\right),
\end{equation}
where we have $P = [\mathcal{P}^{T_X}]$, $L = [\mathcal{L}^{T_X}]$, $S = [\mathcal{S}^{T_X}]$ and $E = [\mathcal{E}^{T_X}]$.
Similarly, \eqr{eq:cct:qcc} can be re-modeled as follows:
\begin{equation}
 \left(M_{j}:P_j\right) = \left(M_{j-1}:L_j:S_j:E_j\right) U.
\label{eq:cct:qcc-EA}
\end{equation}

\subsection{System Model: Concatenated Quantum Codes} \label{sec:system-model}
\begin{figure*}[tb]
\begin{center}
    \includegraphics[width=0.6\linewidth]{\figures 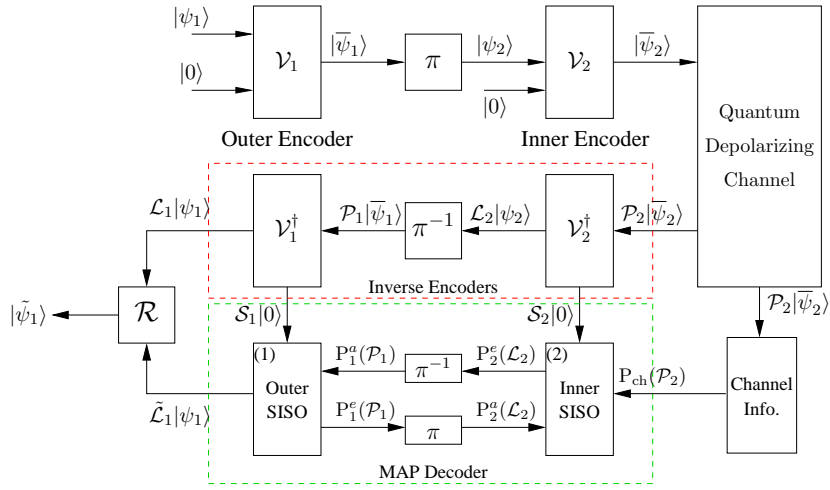}
    \caption{System Model: Quantum communication system relying on concatenated quantum stabilizer codes.
\textit{$\textup{P}^a_i(.)$, $\textup{P}^e_i(.)$ and $\textup{P}^o_i(.)$ denote
the \textit{a-priori}, \textit{extrinsic} and \textit{a-posteriori} probabilities related to the $i$th decoder.}}
  \label{fig:system-model}
\end{center}
%\vspace{-0.5cm}
\end{figure*}
\fref{fig:system-model} shows the general schematic of a quantum communication system relying on a pair of concatenated quantum stabilizer codes. 
In this contribution, both the inner as well as the outer codes are assumed to be convolutional codes. Furthermore, analogously to the classical concatenated codes,
the inner code must be recursive, while both the inner as well the outer code must be non-catastrophic. Having a recursive nature of the inner code is essential
for the sake of ensuring that the resultant families of codes have an unbounded minimum distance.
%, i.e. their minimum distance increases almost linearly with the interleaver length.
On the other hand, the non-catastrophic nature of both the inner and the outer codes guarantees that a decoding convergence to an infinitesimally low error rate is achieved.
It was found in~\cite{qturbo1,Monireh2012} that QCCs cannot be simultaneously recursive and non-catastrophic. In order to overcome this problem,
Wilde \textit{et al.}~\cite{wilde_turbo,wilde_turbo2} proposed to employ entanglement-assisted inner codes, which are
recursive as well as non-catastrophic. Therefore, the inner code should be an entanglement-assisted recursive and non-catastrophic code,
while the outer code can be either an unassisted or an entanglement-assisted non-catastrophic code.

At the transmitter, the intended quantum information $\ket{\psi_1}$ is encoded by an $[n_1,k_1]$ outer encoder $\mathcal{V}_1$ using $(n_1-k_1)$ auxiliary qubits, which are initialized 
to the state $\ket{0}$, as depicted in \eqr{eq:unitary}. The encoded qubits $\ket{\overline{\psi}_1}$ are passed through a quantum interleaver ($\pi$). The resultant permuted qubits $\ket{\psi_2}$ are fed to an $[n_2,k_2]$ inner encoder $\mathcal{V}_2$, which encodes them into the codewords
$\ket{\overline{\psi}_2}$ using $(n_2-k_2)$ auxiliary qubits initialized to the state $\ket{0}$\footnote{Please note that this is a general schematic.
The inner code can be either an un-assisted or an entanglement-assisted code. However, it is advisable to use an entanglement-assisted inner code for
the sake of ensuring an unbounded minimum distance of the resultant concatenated code.}.
The $n$-qubit codewords $\ket{\overline{\psi}_2}$, where we have $n = n_1n_2$, 
are then serially transmitted over a quantum depolarizing channel, which imposes an $n$-tuple error $\mathcal{P}_2 \in \mathcal{G}_n$ on the transmitted codewords.

At the receiver, the received codeword $\ket{\hat{\psi}_2} = \mathcal{P}_2 \ket{\overline{\psi}_2}$ is passed through the inverse encoder $\mathcal{V}_2^{\dag}$,
which yields the corrupted information word of the inner encoder $\mathcal{L}_2\ket{\psi_2}$ and the associated $(n_2-k_2)$-qubit syndrome 
$\mathcal{S}_2\ket{0_{(n_2-k_2)}}$ as depicted previously in \eqr{eq:inverse-cct},
%\begin{align}
% \mathcal{V}_2^{\dag}\mathcal{P}_2|\overline{\psi}_2\rangle &= \mathcal{V}_2^{\dag}\mathcal{P}_2\mathcal{V}_2(|\psi_2\rangle \otimes |0_{(n_2-k_2)}\rangle) \nonumber \\
%& = (\mathcal{L}_2|\psi_2\rangle) \otimes (\mathcal{S}_2|0_{(n_2-k_2)}\rangle),
%\label{eq:inverse}
%\end{align}
where $\mathcal{L}_2$ denotes the error imposed on the logical qubits of the inner encoder, while
$\mathcal{S}_2$ represents the error inflicted on the remaining $(n_2 - k_2)$ qubits. The corrupted logical qubits of the inner encoder are de-interleaved,
resulting in $\mathcal{P}_1\ket{\overline{\psi}_1}$, which is then passed through the inverse outer encoder $\mathcal{V}_1^\dag$. % similar to \eqr{eq:inverse}.
This gives the corrupted information word of the outer encoder $\mathcal{L}_1\ket{\psi_1}$ and the associated $(n_1-k_1)$-qubit syndrome 
$\mathcal{S}_1\ket{0_{(n_1-k_1)}}$. 

The next step is to estimate the error $\mathcal{L}_1$ for the sake of ensuring that the original logical qubit $\ket{\psi_1}$ can
be restored by applying the recovery operation $\mathcal{R}$. For estimating $\mathcal{L}_1$, both the syndromes $\mathcal{S}_2\ket{0_{(n_2-k_2)}}$
and $\mathcal{S}_1\ket{0_{(n_1-k_1)}}$ are fed to the inner and outer Soft-In Soft-Out (SISO) decoders~\cite{tc_teq_st_2:book}, respectively, which engage in iterative
decoding~\cite{qturbo2,wilde_turbo2} in order to yield the estimated error $\tilde{\mathcal{L}}_1$. 
The corresponding block is marked as `MAP Decoder' in \fref{fig:system-model}.
Here, $\textup{P}^a_i(.)$, $\textup{P}^e_i(.)$ and $\textup{P}^o_i(.)$ denote
the \textit{a-priori}, \textit{extrinsic} and \textit{a-posteriori} probabilities~\cite{tc_teq_st_2:book} related to the $i$th decoder. 
Based on this notation, the turbo decoding process can be summarized as follows: 
\begin{itemize}
\item The inner SISO decoder of \fref{fig:system-model} uses the channel information $\textup{P}_\text{ch}(\mathcal{P}_2)$, the \textit{a-priori} information gleaned from the 
outer decoder $\textup{P}_2^{a}(\mathcal{L}_2)$ (initialized to be equiprobable for the first iteration) and the syndrome $\mathcal{S}_{2}$ 
to compute the \textit{extrinsic} information $\textup{P}_2^{e}(\mathcal{L}_2)$.
For a coded sequence of length $N$, we have $\mathcal{P}_2 = [\mathcal{P}_{2,1}, \mathcal{P}_{2,2}, \dots, \mathcal{P}_{2,t}, \dots, \mathcal{P}_{2,N}]$, 
where $\mathcal{P}_{2,t} = [\mathcal{P}_{2,t}^1, \mathcal{P}_{2,t}^2, \dots, \mathcal{P}_{2,t}^n ]$. The channel information 
$\mathbf{P}_{\text{ch}}\left(P_{2,t}\right)$ is computed assuming that each qubit
is independently transmitted over a quantum depolarizing channel having a depolarizing probability of $p$, whose channel transition probabilities
are given by~\cite{qturbo2}:
\begin{equation}
\mathbf{P}_{\text{ch}}\left(P_{2,t}^i\right) = \left\{
\begin{array}{l l}
 1-p, & \text{if } \mathcal{P}_{2,t}^i = \mathbf{I} \\
 p/3, & \text{if } \mathcal{P}_{2,t}^i \in \{\mathbf{X}, \mathbf{Z}, \mathbf{Y}\}. \\
\end{array}
\right.
\label{eq:dp_ch}
\end{equation} 
\item $\textup{P}_2^{e}(\mathcal{L}_2)$ is passed through the quantum de-interleaver $(\pi^{-1})$ of \fref{fig:system-model} to generate the \textit{a-priori} information for the outer decoder 
$\textup{P}_1^{a}(\mathcal{P}_1)$.
\item Based on both the \textit{a-priori} information $\textup{P}_1^{a}(\mathcal{P}_1)$ and on the syndrome $\mathcal{S}_{1}$, the outer SISO decoder
of \fref{fig:system-model} computes both
the \textit{a-posteriori} information $\textup{P}_1^{o}(\mathcal{L}_1)$ and the \textit{extrinsic} information $\textup{P}_1^{e}(\mathcal{P}_1)$.
\item $\textup{P}_1^{e}(\mathcal{P}_1)$ is then interleaved to obtain $\textup{P}_2^{a}(\mathcal{L}_2)$, which is fed back to the inner SISO 
decoder of \fref{fig:system-model}. This iterative procedure continues,
until either convergence is achieved or the maximum affordable number of iterations is reached.
\item Finally, a qubit-based MAP decision is made for determining the most likely error coset $\mathcal{L}_1$.
It must be mentioned here that both the inner and outer SISO decoders employ the degenerate decoding approach of~\cite{qturbo2}, which
aims for finding the `most likely error coset' rather than the `most likely error' acting on the logical qubits $\mathcal{L}_i$, as we will discuss in the next section.
\end{itemize}
%It must
%be mentioned here that the syndrome sequence $\ket{0_{(n_i-k_i)}}$ is invariant to the $\mathbf{Z}$-component of the error on the syndrome $\mathcal{S}_i$. Therefore,
%the inner and outer SISO decoders employ the degenerate decoding approach of~\cite{qturbo2}, which
%aims for finding the most likely error coset acting on the logical qubits $\mathcal{L}_i$ satisfying only the $\mathbf{X}$-component of $\mathcal{S}_i$.
\subsection{Degenerate Iterative Decoding} \label{sec:deg-dec}
As discussed in Section~\ref{sec:QSC}, quantum codes exhibit the intrinsic property of degeneracy, which is also obvious from \eqr{eq:inverse-cct}. More explicitly,
we have:
\begin{equation}
  \mathcal{S}\ket{0_{n-k}} = \mathcal{S}_1\ket{0} \otimes \dots \otimes \mathcal{S}_{n-k}\ket{0}.
\label{eq:syn1}
\end{equation}
Since, we have $\mathcal{S}_i \in \{\mathbf{I},\mathbf{X},\mathbf{Y},\mathbf{Z}\}$, we can re-write \eqr{eq:syn1} as follows~\cite{qturbo2}:
\begin{equation}
 \mathcal{S}\ket{0_{n-k}} \equiv \epsilon \ket{s_1} \otimes \dots \otimes \ket{s_{n-k}},
\label{eq:syn2}
\end{equation}
where $\epsilon \in \{\pm 1, \pm i\}$, and:
\begin{align}
 s_i &= 0 \;\;\;\; \text{if } \mathcal{S}_i = \mathbf{I} \; \text{or} \; \mathcal{S}_i = \mathbf{Z}, \nonumber \\%\mathcal{S}_i \in \{\mathbf{I},\mathbf{Z}\}, \nonumber \\
 s_i &= 1 \;\;\;\; \text{otherwise}.
\end{align}
For example, if $\mathcal{S}_1 = \mathbf{Y}$ and $\mathcal{S}_i = \mathbf{I}$ for $i \neq 1$, since $\mathbf{Y} = i \mathbf{X}\mathbf{Z}$, we get 
$\mathcal{S}\ket{0_{n-k}} = i \ket{1} \otimes \ket{0} \otimes \dots \otimes \ket{0}$.

Observing the $(n-k)$ syndrome qubits of \eqr{eq:syn2} collapses them to the classical syndrome $s = \{s_1, \dots, s_{n-k}\}$, which is equivalent to the symplectic product of
$P$ and $H$, i.e. $s = \left(P \star H_i\right)_{1\leq i \leq n-k}$. More precisely, the syndrome sequence $\ket{0_{n-k}}$ is invariant to the $\mathbf{Z}$-component
of $\mathcal{S}$ since $\mathbf{Z}\ket{0} = \ket{0}$. Let $S$ be the effective $2(n-k)$-bit error on the syndrome, which may
be decomposed as $S = S^x + S^z$, where $S^x$ and $S^z$ are the $\mathbf{X}$ and $\mathbf{Z}$ components of $S$, respectively. Then $s$ only reveals $S^x$. Hence, two distinct error sequences $P = (L:S^x+S^z)V$ and $P' = (L:S^x+S'^z)V$,
which only differ in the $\mathbf{Z}$-component of $\mathcal{S}$, yield the same syndrome 
$s$. Furthermore, it must be noted that both $P$ and $P'$ have the same logical error $L$. Therefore, $P$ and $P'$ differ only by the stabilizer group
and are known as degenerate errors, which do not have to be distinguished, since they can
be corrected by the same recovery operation $L^{-1}$.

Recall that a classical syndrome-based MAP decoder aims for finding the most likely error for a given syndrome, which may be modeled as:
\begin{equation}
 L(S) = \text{argmax}_{L} \textup{P}(L|S),
\end{equation}
where $\textup{P}(L|S)$ denotes the probability of experiencing the logical error $L$ imposed on the transmitted qubits, given that the syndrome of the received qubits is $S$.
By contrast, quantum codes employ degenerate decoding, which aims for finding the most likely error coset $C(L,S^x)$ associated with the observed syndrome $S^x$.
The coset $C(L,S^x)$ is defined as~\cite{qturbo2}:
\begin{equation}
 C(L,S^x) = \{ P = (L:S^x+S^z)V\} \; \; \; \forall S^z \in \{\mathbf{I},\mathbf{Z}\}^{n-k}.
\end{equation}
Therefore, a degenerate MAP decoder yields:
\begin{equation}
 L(S^x) = \text{argmax}_{L} \textup{P}(L|S^x),
\end{equation}
where we have:
\begin{equation}
 \textup{P}(L|S^x) \equiv \sum_{S^z \in \{\mathbf{I},\mathbf{Z}\}^{n-k}} \textup{P}(L|(S^x+S^z)).
\end{equation}

\begin{figure}[tp]
\begin{center}
    \includegraphics[width=\linewidth]{\figures 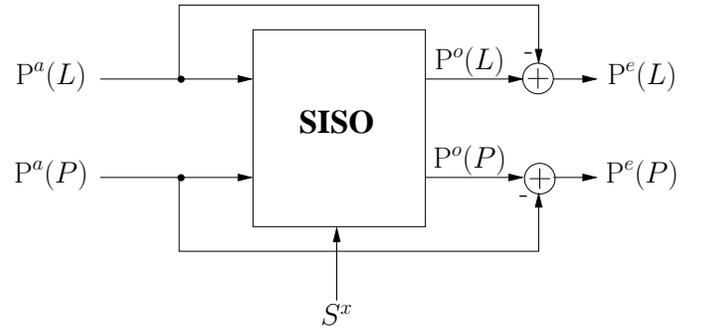}
    \caption{General schematic of a SISO decoder.
\textit{$\textup{P}^a(.)$, $\textup{P}^e(.)$ and $\textup{P}^o(.)$ denote
the \textit{a-priori}, \textit{extrinsic} and \textit{a-posteriori} probabilities.}}
  \label{fig:SISO1}
\end{center}
%\vspace{-0.5cm}
\end{figure}
The MAP decoder of \fref{fig:system-model} consists of two serially concatenated SISO decoders, which employ the aforementioned degenerate decoding approach. 
\fref{fig:SISO1} shows the general schematic of a SISO decoder, where the Pauli operators $\mathcal{P}$, $\mathcal{L}$ and $\mathcal{S}$ are replaced by the effective
operators $P$, $L$ and $S^x$, respectively. The SISO decoder of \fref{fig:SISO1} yields the \textit{a-posteriori} 
information pertaining to $L$ and $P$ based on the classic forward-backward recursive coefficients $\alpha$ and $\beta$, as follows~\cite{qturbo2}:
\begin{itemize}
\item For a coded sequence of duration $N$, let $P = [P_{1}, P_{2}, \dots, P_{t}, \dots, P_{N}]$ 
and $L = [L_{1}, L_{2}, \dots, L_{t}, \dots, L_{N}]$, where $P_{t} \in G_n$ and $L_{t} \in G_k$. More explicitly, 
$P_{t} = [P_{t}^1, P_{t}^2, \dots, P_{t}^n ]$ and 
$L_{t} = [L_{t}^1, L_{t}^2, \dots, L_{t}^k ]$.  
 \item Let us decompose the seed transformation as $U = (U_M:U_P)$, where $U_M$ is the binary matrix formed by the first $2m$ columns of $U$, while
$U_P$ is the binary matrix formed by the last $2n$ columns of $U$. Therefore, we have:
\begin{equation}
 M_t = \left(M_{t-1}:L_{t}:S_t\right)U_M,
\label{eq:Um}
\end{equation}
\begin{equation}
 P_t = \left(M_{t-1}:L_{t}:S_t\right)U_P.
\label{eq:Up}
\end{equation}
\item Let $\alpha_t\left(M_t\right)$ be the forward recursive coefficient, which is defined as follows:
\begin{align}
\alpha_t\left(M_t\right) &\triangleq \textup{P}\left(M_t|S^x_{\leq t}\right), \nonumber \\
& \propto \sum_{\substack{\mu, \lambda, \sigma}} \textup{P}^a\left(L_{t} = \lambda\right) \textup{P}^{a}\left(P_{t}\right) \alpha_{t-1}\left(\mu\right),
\label{eq:alpha}
\end{align}
where $S^x_{\leq t} \triangleq \left(S^x_j\right)_{0 \leq j \leq t}$, $\mu \in G_m$, $\lambda \in G_{k}$ and $\sigma \in G_{n-k}$, while $\sigma = \sigma_x + \sigma_z$, having $\sigma_x = S^x_t$. 
Furthermore, we have $P_{t} = (\mu:\lambda:\sigma)U_P$ and $M_t = \left(\mu:\lambda:\sigma \right)U_M$.
\item Let $\beta_{t}\left(M_{t}\right)$ be the backward recursive coefficient, which is defined as:
\begin{align}
\beta_t\left(M_t\right) &\triangleq \textup{P}\left(M_t|S^x_{> t}\right), \nonumber \\
& \propto \sum_{\lambda, \sigma} \textup{P}^a\left(L_{t} = \lambda\right) \textup{P}^{a}\left(P_{t+1}\right) \beta_{t+1}\left(M_{t+1}\right),
\label{eq:beta}
\end{align}
where $S^x_{> t} \triangleq \left(S^x_j\right)_{t < j \leq N}$, $P_{t+1} = (M_t:\lambda:\sigma)U_P$ and
$M_{t+1} = \left(M_t:\lambda:\sigma \right)U_M$.
\item Finally, we have the \textit{a-posteriori} probabilities $\textup{P}^o(L_{t})$ and $\textup{P}^o(P_{t})$, which are given by:
\begin{align}
 \textup{P}^o(L_{t}) &\triangleq \textup{P}(L_{t}|S^x), \nonumber \\
& \propto \sum_{\mu, \sigma} \textup{P}^a(L_t)\textup{P}^{a}(P_t) \alpha_{t-1}\left(\mu\right) \beta_{t}\left(M_t\right), \\
\textup{P}^o(P_{t}) &\triangleq \textup{P}(P_{t}|S^x), \nonumber \\
& \propto \sum_{\substack{\mu, \lambda, \sigma}} \textup{P}^a(P_t)\textup{P}^a(L_t=\lambda) \alpha_{t-1}\left(\mu\right) \beta_{t}\left(M_t\right),
\label{eq:app}
\end{align}
where $S^x \triangleq \left(S^x_t\right)_{0 \leq t \leq N}$, $P_{t} = (\mu:L_{t}:\sigma)U_P$ and
$M_t = \left(\mu:L_{t}:\sigma \right)U_M$.
\item The marginalized probabilities $\textup{P}^o(L^{j}_{t})$, for $j \in \{0,k-1\}$, and $\textup{P}^o(P^{j}_{t})$, for $j \in \{0,n-1\}$, are then computed from 
$\textup{P}^o(L^{j}_{t})$ and $\textup{P}^o(P^{j}_{t})$, respectively. The
\textit{a-priori} information is then removed in order to yield the \textit{extrinsic} probabilities~\cite{wilde_turbo2}, i.e we have:
\begin{align}
 \ln [\textup{P}^e (L^{j}_{t})] &= \ln [\textup{P}^o (L^{j}_{t})] - \ln [\textup{P}^a (L^{j}_{t})], \\
 \ln [\textup{P}^e (P^{j}_{t})] &= \ln [\textup{P}^o (P^{j}_{t})] - \ln [\textup{P}^a (P^{j}_{t})].
\label{eq:ext}
\end{align}
\end{itemize}
It has to be mentioned here that the property of degeneracy is only an attribute of auxiliary qubits and the ebits of an entanglement-assisted code do not contribute
to it. This is because both $\mathbf{X}$ as well as $\mathbf{Z}$ errors acting on the transmitter's half of ebits give distinct results
when measured in the Bell basis, i.e. $\mathcal{E}^{T_X}\ket{\phi_c^+}^{T_XR_X}$ gives four distinct Bell states for 
$\mathcal{E}_j^{T_X} \in \{\mathbf{I},\mathbf{X},\mathbf{Z},\mathbf{Y}\}$. Consequently, the degeneracy is a function of $a$ and reduces to zero
for $a=0$.
\section{EXIT-Chart Aided Code Design} \label{sec:near-cap-design}
%\subsection{EXIT Chart for Concatenated Quantum Codes} \label{sec:EXIT}
EXIT charts~\cite{brink:exit-parallel,tc_teq_st_2:book,Hajjar2013} are capable of visualizing the convergence behaviour of iterative decoding
schemes by exploiting the input/output relations of the constituent decoders in terms of their average Mutual Information (MI) characteristics. 
The EXIT chart analysis not only allows us to dispense with the time-consuming Monte-Carlo simulations, but also facilitates the
design of capacity approaching codes without resorting to the tedious analysis of their distance spectra. Therefore,
they have been extensively employed for designing near-capacity
classical codes~\cite{near_cap_brink,near_cap_2010,near_cap_2007,zbabar2013_2}. 
Let us recall that the EXIT chart of a serially concatenated scheme visualizes the exchange of 
four MI terms, i.e. average \textit{a-priori} MI of the outer decoder $I_A^1$, average \textit{a-priori} MI of the inner decoder $I_A^2$,
average \textit{extrinsic} MI of the outer decoder $I_E^1$, and average \textit{extrinsic} MI of the inner decoder $I_E^2$.
More specifically, $I_A^1$ and $I_E^1$ constitute the EXIT curve of the outer decoder, while $I_A^2$ and $I_E^2$ yield the EXIT curve of the inner
decoder. The MI transfer characteristics of both the decoders are plotted in the same graph, with the $x$ and $y$ axes
of the outer decoder swapped. The resultant EXIT chart quantifies the improvement in the mutual information as the iterations proceed,
which can be viewed as a stair-case-shaped decoding trajectory. An open tunnel between the two EXIT curves ensures that the decoding trajectory
reaches the $(1,y)$ point of perfect convergence.

In our prior work~\cite{babar_QTC_2014}, we extended the application of EXIT charts to the quantum domain by appropriately
adapting the conventional non-binary EXIT chart generation technique for the quantum syndrome decoding approach.
Recall from Section~\ref{sec:p2b} that a quantum code is equivalent to a classical code. More specifically, the decoding of a quantum code is essentially 
carried out with the aid of the equivalent classical code by exploiting the additional property
of degeneracy, as discussed in Section~\ref{sec:deg-dec}. Quantum codes employ syndrome decoding, which yields information about the error-sequence rather than about the information-sequence 
or coded qubits, hence avoiding the observation of the latter sequences, which would collapse them back to the classical domain.
Since a quantum code has an equivalent classical representation and the depolarizing channel is analogous to
a Binary Symmetric Channel (BSC), we employ the EXIT chart technique to 
design hashing bound approaching concatenated quantum codes. %the conventional EXIT chart approach is equally applicable to QTCs. 
The major difference between the EXIT charts conceived for the classical and quantum domains is that while the former models the \textit{a-priori}
information concerning the input bits of the inner encoder (and similarly the output bits of the outer encoder), the latter models the
\textit{a-priori} information concerning the corresponding error-sequence, i.e. the error-sequence related to the input qubits of the inner encoder $L_2$
(and similarly the error-sequence related to the output qubits of the outer encoder $P_1$).

Similar to the classical EXIT charts, it is assumed that
the interleaver length is sufficiently high to ensure that~\cite{brink:exit-parallel,tc_teq_st_2:book}:
\begin{itemize}
 \item the \textit{a-priori} values are fairly uncorrelated; and
 \item the \textit{a-priori} information has a Gaussian distribution.
\end{itemize}
 \begin{figure}[tb]
%\vspace*{-0.5cm}
  \begin{center}
  {\includegraphics[width=\linewidth]{\figures 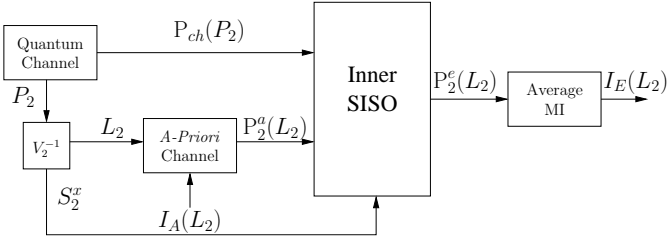}}
  \caption{System model for generating the EXIT chart of the inner decoder~\cite{babar_QTC_2014}.}
  \label{fig:exit_inner}
%\vspace*{-0.5cm}
  \end{center}
\end{figure}
\fref{fig:exit_inner} shows the system model used for generating the EXIT chart of the inner decoder. Here, a quantum
depolarizing channel having a depolarizing probability of $p$ generates the error sequence $P_2$, which is passed through the inverse inner encoder $V_2^{-1}$. This
yields both the error imposed on the logical qubits $L_2$ and the syndrome $S^x_2$. % according to \eqr{eq:cct}. 
The \textit{a-priori} channel block then models the 
\textit{a-priori} information $\textup{P}^{a}_1(L_2)$ such that the average MI between the actual error $L_2$ and the \textit{a-priori} 
probabilities $\textup{P}^{a}_2(L_2)$ is given by $I_A(L_2)$~\cite{brink:exit-parallel,tc_teq_st_2:book,Hajjar2013}. 
More explicitly, we have $I_A(L_2) = I[L_2,\textup{P}^{a}_2(L_2)]$, where $I$ denotes the average MI function.
Moreover, the $i$th and $(N+i)$th bits of the effective error vector $L_2$ can be visualized as $4$-ary symbols. 
Consequently, similar to classical non-binary EXIT charts~\cite{grant2001,kliewer:exit-symbol}, the \textit{a-priori} information is modeled using an independent Gaussian distribution 
with a mean of zero and variance of $\sigma_A^2$, assuming that the $\mathbf{X}$ and $\mathbf{Z}$ errors constituting the $4$-ary 
symbols are 
independent\footnote{Under the idealized asymptotic conditions of having an infinite-length interleaver, $I_A(L_2)$
may be accurately modeled by the Gaussian distribution. As and when shorter interleavers are used, the Gaussian assumption becomes
less accurate, hence in practice a histogram-based approximation may be relied upon.}.
Based on the channel information $\textup{P}_{\text{ch}}(P_2)$, on the syndrome $S^x_2$ and on the \textit{a-priori} information, the inner SISO decoder generates the \textit{extrinsic} 
information $\textup{P}^{e}_2(L_2)$ by using the degenerate decoding approach of Section~\ref{sec:deg-dec}.
Finally, the
\textit{extrinsic} average MI $I_E(L_2) = I[L_2,\textup{P}^{e}_2(L_2)]$ between $L_2$ and $\textup{P}^{e}_2(L_2)$ is computed.
Since the equivalent classical capacity of a quantum channel is given by the capacity achievable over each half of the $4$-ary symmetric channel,
%as depicted in \eqr{eq:c4},
$I_E(L_2)$ is the normalized MI of the $4$-ary symbols, which can be computed based on~\cite{kliewer:exit-symbol,mike3} as:
\begin{equation}
 I_E(L_2) = \frac{1}{2}\left(2 + \mbox{E} \left[ \sum_{\text{m} = 0}^{3} \textup{P}^{e}_2(L_2^{j(\text{m})}) \log_2\textup{P}^{e}_2(L_2^{j(\text{m})}) \right]\right),
\label{eq:MI}
\end{equation}
where \mbox{E} is the expectation (or time average) operator and $L_2^{j(\text{m})}$ is the $\text{m}^{\text{th}}$ hypothetical error imposed on the 
logical qubits. More explicitly, since the error on each qubit is represented by an equivalent pair of classical bits, $L_2^{j(\text{m})}$ is a 
$4$-ary classical symbol associated
with $\text{m} \in \{0,3\}$. The process is repeated for a range of $I_A(L_2) \in [0,1]$ values for the sake of obtaining the \textit{extrinsic} information transfer characteristics 
at the depolarizing probability $p$. The resultant inner EXIT transfer function $T_{2}$ of the specific inner decoder may be defined as follows: 
\begin{equation}
 I_E(L_2) = T_{2}[I_A(L_2),p],
\label{eq:t1}
\end{equation}
which is a function of the channel's depolarizing probability $p$. 

\begin{figure}[tb]
%\vspace*{-0.5cm}
  \begin{center}
  {\includegraphics[width=\linewidth]{\figures 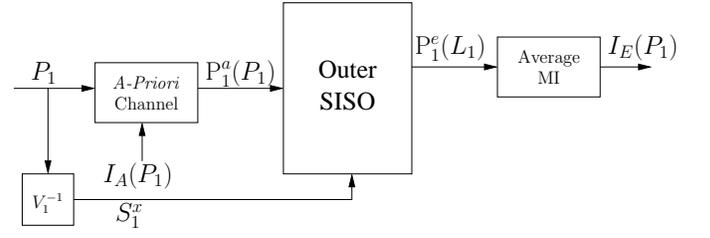}}
    \caption{System model for generating the EXIT chart of the outer decoder~\cite{babar_QTC_2014}.}
  \label{fig:exit_outer}
%\vspace*{-0.5cm}
  \end{center}
\end{figure}

The system model used for generating the EXIT chart of the outer decoder is depicted in \fref{fig:exit_outer}. As inferred from \fref{fig:exit_outer}, the EXIT curve
of the outer decoder is independent of the channel's output information. The \textit{a-priori} information is generated by the \textit{a-priori}
channel based on $P_1$ (error on the physical qubits of the second decoder) and $I_A(P_1)$, which is the average MI
between $P_1$ and $\textup{P}^a_1(P_1)$. Furthermore, as for the inner decoder, $P_1$ is passed through the inverse outer encoder $V_1^{-1}$ to
compute $S^x_1$, which is fed to the outer SISO decoder to yield the \textit{extrinsic} information $\textup{P}^e_1(P_1)$. 
The average MI between $P_1$ and 
$\textup{P}^e_1(P_1)$ is then calculated using \eqr{eq:MI}. The resultant EXIT chart is characterized by the following MI transfer function:
\begin{equation}
 I_E(P_1) = T_{1}[I_A(P_1)],
\label{eq:t2}
\end{equation}
where $T_{1}$ is the outer EXIT transfer function, which is dependent on the specific outer decoder, but it is independent of the depolarizing probability $p$.

Finally, the MI transfer characteristics of both decoders characterized by \eqr{eq:t1} and \eqr{eq:t2} are plotted in the same graph,
with the $x$ and $y$ axes of the outer decoder swapped. 
%The area under the EXIT curve of the inner decoder is approximately equal to the
%attainable channel capacity~\cite{ashikhmin04:exit}, provided that the channel's input symbols are equiprobable, while the area under the EXIT curve of the 
%outer decoder is approximately equivalent to $(1 - R)$~\cite{ashikhmin04:exit}, where $R$ is the coding rate of the equivalent classical code given by~\eqr{eq:rc}.
%Hence, 
For the sake of approaching the achievable capacity of \fref{fig:cap-H}, \textit{our EXIT-chart aided design aims for creating a narrow, but marginally open tunnel between the EXIT curves of the inner and outer decoders at the 
highest possible depolarizing probability (analogous to the lowest possible SNR for a classical channel)}. 
For a given noise limit $p^*$ and the desired code parameters, this may be achieved in two steps. We first find that specific inner code, which yields
the largest area under its EXIT-curve at the noise limit $p^*$. Once the optimal inner code is selected, we find the optimal outer code, whose EXIT-curve gives the best match
with the chosen inner code. 
The narrower the tunnel-area between the inner and outer decoder's EXIT curve, the lower is the deviation from the
achievable capacity, which may be quantified using \eqr{eq:distance}.
%The resultant EXIT chart is capable of visualizing the exchange of extrinsic MI as a stair-case-shaped decoding trajectory as the iterations proceed.
\section{A Key to Hashing Bound: Quantum Irregular Convolutional Codes} \label{sec:QIRCC}
In this section, we exploit the EXIT-chart aided design criterion of Section~\ref{sec:near-cap-design} to design concatenated codes, which
operate arbitrarily close to the hashing bound. Here, we assume that we already have the optimal inner code. \textit{More explicitly, our design objective is to find the optimal outer code $\mathcal{C}$
having a coding rate $R_Q$, which gives the best match with the given inner code, i.e. whose EXIT curve yields a marginally open tunnel with the given inner decoder's EXIT curve at a 
depolarizing probability close to the hashing bound.}
For the sake of achieving this objective, a feasible design option could be to create the outer EXIT curves of all the possible convolutional codes to find the
optimal code $\mathcal{C}$, which gives the best match, as we did in our prior work~\cite{babar_QTC_2014}.
To circumvent this exhaustive code search, in this contribution we propose to invoke Quantum Irregular Convolutional Codes (QIRCCs) for achieving 
EXIT-curve matching. 

\begin{figure}[tb]
\begin{center}
    \includegraphics[width=\linewidth]{\figures 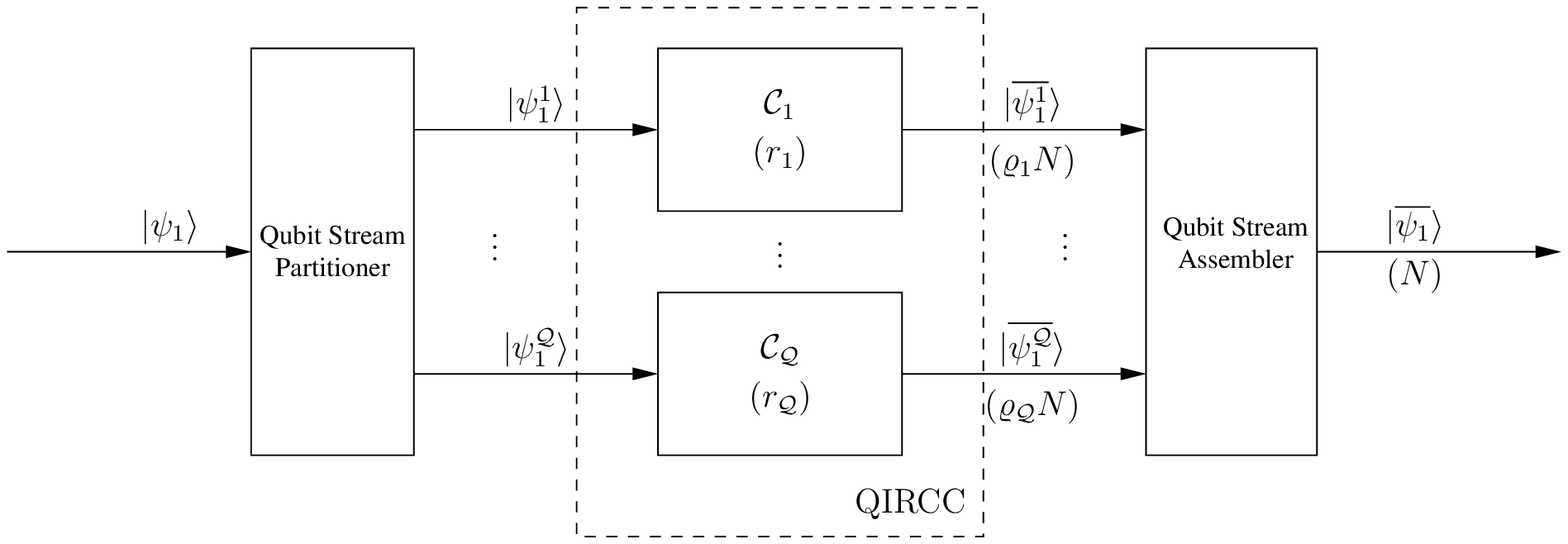}
    \caption{Structure of a $\mathcal{Q}$-subcode QIRCC encoder.}
  \label{fig:qircc}
\end{center}
\vspace{-0.5cm}
\end{figure}

Similar to the classical
Irregular Convolutional Code (IRCC) of~\cite{tuchler04:design-scc}, our proposed QIRCC employs a family of $\mathcal{Q}$ subcodes 
$\mathcal{C}_q$, $q \in \{1, 2, \dots, \mathcal{Q}\}$, for constructing the target code $\mathcal{C}$. Due to its inherent flexibility, the 
resultant QIRCC provides a better
EXIT-curve match than any single code, when used as the outer component in the concatenated structure of \fref{fig:system-model}. 
The $q^{th}$ subcode has a coding rate of $r_q$ and it encodes a specifically designed
fraction of the original information qubits to $\varrho_q N$ encoded
qubits. Here, $N$ is the total length of the coded frame. More specifically, for a $\mathcal{Q}$-subcode IRCC, $\varrho_q$ is the $q^{th}$ IRCC weighting coefficient
satisfying the following constraints~\cite{ircc:tuchler_hagenauer,tuchler04:design-scc}:
\begin{eqnarray}
\sum_{q=1}^{\mathcal{Q}}\varrho_q=1\ , \ R_Q = \sum_{q=1}^{\mathcal{Q}} \varrho_q r_q \ , 
\ \varrho_q \in [0,1], \forall q \ ,
\label{eq:constraint2}
\end{eqnarray}
which can be conveniently represented in the following matrix form:
\begin{eqnarray}
\begin{pmatrix}
1   & 1   & \ldots & 1 \\
r_1 & r_2 & \ldots & r_\mathcal{Q} 
\end{pmatrix} \ 
\begin{pmatrix}
\varrho_1 & \varrho_2\ldots & \varrho_\mathcal{Q}
\end{pmatrix}^T
&=& 
\begin{pmatrix}
1   \\
R_Q 
\end{pmatrix} \nonumber \\
\mb{r} \ \boldsymbol\varrho &=& \mb{R} \ .
\label{eq:constraint_vec2}
\end{eqnarray}
Hence, as shown in \fref{fig:qircc}, the input stream is partitioned into $\mathcal{Q}$ sub-frames\footnote{This is only true if all subcodes are active.
If $\varrho_q = 0$ for the $q^{th}$ subcode, then $\mathcal{C}_q$ is not activated. Therefore, the input stream is only divided among the active subcodes.}, 
which are assembled back into a single 
$N$-qubit stream after encoding.

\begin{figure}[tb]
\begin{center}
\includegraphics[scale=0.4]{\figures 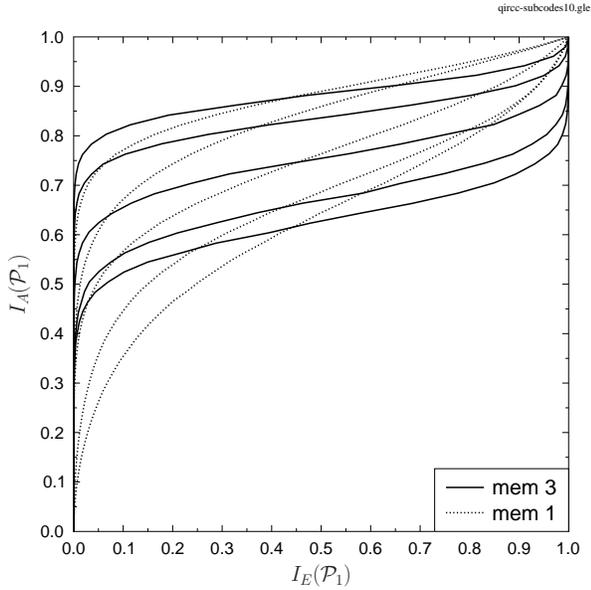}
%\vspace*{-0.3cm}
\caption{Outer EXIT curves (inverted) of our QIRCC subcodes having code rates $\{1/4, 1/3, 1/2, 2/3, 3/4\}$ for both memory-$3$ as well as memory-$1$.}
\label{exit:qircc}
\end{center}
%\vspace*{-0.5cm}
\end{figure}

In the context of classical IRCCs, the subcodes $\mathcal{C}_q$ are constructed from a mother code~\cite{ircc:tuchler_hagenauer,tuchler04:design-scc}. More specifically, high-rate subcodes are obtained
by puncturing the mother code, while the lower rates are obtained by adding more generators. 
However, unlike classical codes, puncturing is not easy to implement for quantum codes, since the resultant punctured code must satisfy the symplectic criterion, as in~\cite{non_qc_rains}. 
In this context, in order to design the constituent subcodes of our proposed QIRCC, we selected $5$ strong randomly-constructed memory-$3$ quantum convolutional
codes with quantum code rates $\{1/4, 1/3, 1/2, 2/3, 3/4\}$, which met the non-catastrophic criterion of~\cite{qturbo2}. More explicitly, for the sake of achieving
a random construction for the Clifford encoder specifying the quantum convolutional code, we used the classical random walk algorithm over the $(n+m)$-qubit
Clifford group as in~\cite{Divincenzo2002}.
The seed transformations of the resultant subcodes having rates $\{1/4, 1/3, 1/2, 2/3, 3/4\}$ are given below:
\begin{align}
U_1 &= \{9600, 691, 11713, 4863, 1013, 6907, 1125, 828, 10372,  \nonumber \\
    &6337, 5590, 11024, 12339, 3439\}, \nonumber \\
U_2 &= \{3968, 1463, 2596, 3451, 1134, 3474, 657, 686, 3113,  \nonumber \\
    &1866, 2608, 2570\}, \nonumber \\
U_3 &= \{848, 1000, 930, 278, 611, 263, 744, 260, 356, 880\}, \nonumber \\
U_4 &= \{529, 807, 253, 1950, 3979, 2794, 956, 1892, 3359, 2127, \nonumber \\
&3812, 1580\}, \nonumber \\
U_5 &= \{62, 6173, 4409, 12688, 7654, 10804, 1763, 15590, 6304, \nonumber \\
&3120, 2349, 1470, 9063, 4020\}.
\label{eq:subcode1-5}
\end{align}
The EXIT curves of these QIRCC subcodes are shown in \fref{exit:qircc}, whereby the memory-$3$ subcodes of \eqr{eq:subcode1-5} are indicated by solid lines.
Furthermore, in order to facilitate accurate EXIT curve matching with a sufficiently versatile and diverse set of inner EXIT functions, we also selected $5$ weak randomly-constructed  memory-$1$ subcodes for the same range of
coding rates, i.e. $\{1/4, 1/3, 1/2, 2/3, 3/4\}$. The corresponding seed transformations are as follows:
\begin{align}
U_6 &= \{475, 194, 526, 422, 417, 988, 426, 611, 831, 84\}, \nonumber \\
U_7 &= \{26, 147, 149, 99, 112, 184, 64, 139\}, \nonumber \\
U_8 &= \{37, 55, 58, 35, 57, 54\}, \nonumber \\
U_9 &= \{57, 248, 99, 226, 37, 93, 244,54\}, \nonumber \\
U_{10} &= \{469, 634, 146, 70, 186, 969, 387, 398, 807, 452\},
\label{eq:subcode6-10}
\end{align}
and their EXIT curves are plotted in \fref{exit:qircc} with the aid of dotted lines. It must be mentioned here that the range of coding rates chosen for the QIRCC subcodes
can be expanded such that the EXIT curves cover a larger portion of the EXIT plot, which further improves curve matching. However, this increases the  
encoding and decoding complexity. 

Based on our proposed QIRCC, relying on the $10$ subcodes specified by \eqr{eq:subcode1-5} and \eqref{eq:subcode6-10},
the input bit stream is divided into $10$ fractions corresponding to the $10$ different-rate subcodes. The specific 
optimum fractions to be encoded by these codes are found by dynamic programming. More specifically, since the QCCs belong to the class of linear codes,
the EXIT curves of the $10$ subcodes, given in \fref{exit:qircc}, 
are superimposed onto each other
after weighting by the appropriate fraction-based weighting coefficients, which are determined by minimizing the area of the open EXIT-tunnel. To elaborate a
little further, the transfer function of the QIRCC is given by the weighted sum of each subcode's transfer function as shown below:
\begin{eqnarray}
I_{E}(P_1) =  T_{1}[I_{A}(P_1)] = \sum_{q=1}^{\mathcal{Q}} \varrho_q \ T_{1}^q\left[I_{A}(P_1)\right], 
\label{eq:IE_ircc2}
\end{eqnarray}
where $T_{1}^q\left[I_{A}(P_1)\right]$ is the transfer function of the
$q^{th}$ subcode. 
For a given inner EXIT curve and outer code rate $R_Q$, we employ the curve matching algorithm of~\cite{ircc:tuchler_hagenauer,tuchler04:design-scc} for optimizing the
weighting coefficients $\boldsymbol\varrho$ 
of our proposed QIRCC such that the square of the error between the inner and inverted outer EXIT curves is minimized subject to \eqr{eq:constraint2}. More explicitly, the error function
may be modeled as:
\begin{equation}
 e(i) = T_2[i,p] - T_1^{-1}[i],
\end{equation}
where $p = (p^* - \epsilon)$ given that $p^*$ is the noise limit defined by the hashing bound and $\epsilon$ is an arbitrarily small
number. The corresponding matrix-based notation may be formulated as~\cite{ircc:tuchler_hagenauer,tuchler04:design-scc}:
\begin{equation}
 \mathbf{e} = \mathbf{b} - \mathbf{A} \boldsymbol \varrho,
\end{equation}
where we have:
\begin{align}
\mathbf{b} &= 
\begin{pmatrix}
T_2[i_1,p]\\
T_2[i_2,p] \\
\vdots \\
T_2[i_{\text{N}},p]
\end{pmatrix}, \text{and} \nonumber \\
\mathbf{A} &=
\begin{pmatrix}
T^{1^{-1}}_1[i_1] &T^{2^{-1}}_1[i_1] &\hdots &T^{\mathcal{Q}^{-1}}_1[i_1] \\
T^{1^{-1}}_1[i_2] &T^{2^{-1}}_1[i_2] &\hdots &T^{\mathcal{Q}^{-1}}_1[i_2] \\
\vdots &\vdots &\vdots &\vdots \\
T^{1^{-1}}_1[i_\text{N}] &T^{2^{-1}}_1[i_\text{N}] &\hdots &T^{\mathcal{Q}^{-1}}_1[i_\text{N}]
\end{pmatrix}.
\end{align}
Here, N denotes the number of sample points such that $i \in \{i_1, i_2, \dots, i_\text{N} \}$ and it is assumed that $\text{N} > \mathcal{Q}$.
Furthermore, the error should be greater than zero for the sake of ensuring an open tunnel, i.e. we have:
\begin{equation}
e(i) > 0, \ \forall i \in [0,1].
\label{eq:err}
\end{equation}
The resultant cost function, i.e. sum of the square of the errors, is given by\cite{ircc:tuchler_hagenauer}:
\begin{equation}
 \mathcal{J}(\varrho_1, \dots,\varrho_{\mathcal{Q}}) = \int_0^1 e(i)^2 di,
\label{eq:er}
\end{equation}
which may also be written as:
\begin{equation}
 \mathcal{J}(\boldsymbol \varrho) = \mathbf{e}^T \mathbf{e}.
\end{equation}
The overall process may be encapsulated as follows:
\begin{equation}
 \boldsymbol \varrho_{opt} = \arg \min_{\boldsymbol\varrho} \mathcal{J}(\boldsymbol \varrho),
\label{eq:opt}
\end{equation} 
subject to \eqr{eq:constraint2} and \eqref{eq:err}, which is a convex optimization problem. The unconstrained optimal solution for
\eqr{eq:opt} is found iteratively using steepest descent approach with a gradient of 
$\partial \mathcal{J}(\boldsymbol \varrho)/\partial \boldsymbol \varrho = 2 \mathbf{e}$, which is then projected
onto the constraints defined in \eqr{eq:constraint2} and \eqref{eq:err}. Further details of this optimization algorithm can be found in~\cite{ircc:tuchler_hagenauer,tuchler04:design-scc}.
\section{Results and Discussions} \label{sec:results}
\begin{figure}[tb]
\begin{center}
\includegraphics[scale=0.4]{\figures 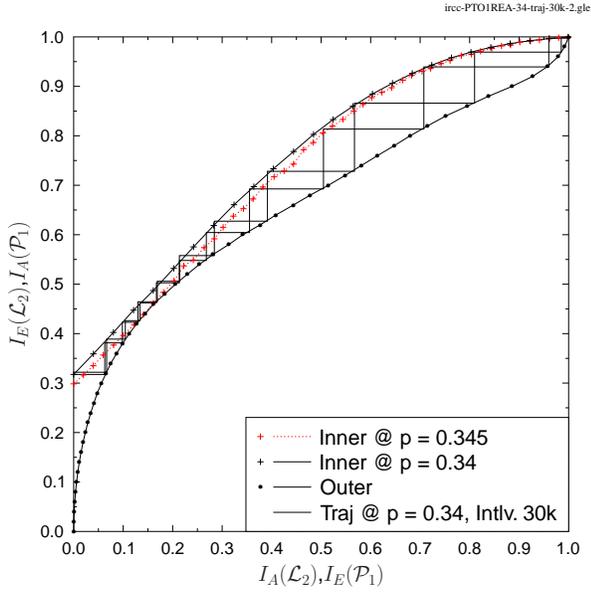}
%\vspace*{-0.3cm}
\caption{EXIT curves of the concatenated rate-$1/9$ system, with PTO1REA as the inner code and QIRCC as the outer, at $p = 0.345$ and $p = 0.34$.}
\label{exit:PTO1REA-match}
\end{center}
%\vspace*{-0.5cm}
\end{figure}

For the sake of demonstrating the curve matching capability of our proposed QIRCC, we designed a rate-$1/9$ concatenated code relying on the rate-$1/3$
entanglement-assisted inner code of~\cite{wilde_turbo,wilde_turbo2}, namely ``PTO1REA'', with our proposed QIRCC as the outer code. 
Since the entanglement consumption rate of  ``PTO1REA'' is $2/3$, the resultant code has an entanglement consumption rate of $6/9$, for which the corresponding noise limit
is $p^* = 0.3779$ according to \eqr{eq:capacity_q_EA}~\cite{wilde_turbo2}. Furthermore, since we intend to design a rate-$1/9$ system with a rate-$1/3$
inner code, we have $R_Q = 1/3$. 
Hence, for
a target coding rate of $1/3$, we used the optimization algorithm discussed in Section~\ref{sec:QIRCC} for the sake of finding the
optimum weighting coefficients of \eqr{eq:opt} at the highest possible depolarizing probability $p = p^* - \epsilon$. It was found that
we only need to invoke two subcodes out of the $10$ possible subcodes, based on
$\boldsymbol \varrho$ = [0 0 0 0 0.168 0.832 0 0 0 0]$^T$, for attaining a marginally open tunnel, which occurs at $p = 0.345$, as shown in \fref{exit:PTO1REA-match}. 
Hence, the resultant code has a convergence threshold of $p = 0.345$, which is only $\left[10 \times \log_{10} (\frac{0.345}{0.3779})\right] = 0.4$ dB
from the 
noise limit of $0.3779$. 
\fref{exit:PTO1REA-match} also shows two decoding trajectories at $p = 0.34$ for a $30,000$ qubit long interleaver. As gleaned from
the figure, the decoding trajectories closely follow the EXIT curves reaching the $(1,1)$ point of perfect convergence.

\begin{figure}[tb]
   \begin{center}
  {\includegraphics[width=\linewidth]{\figures 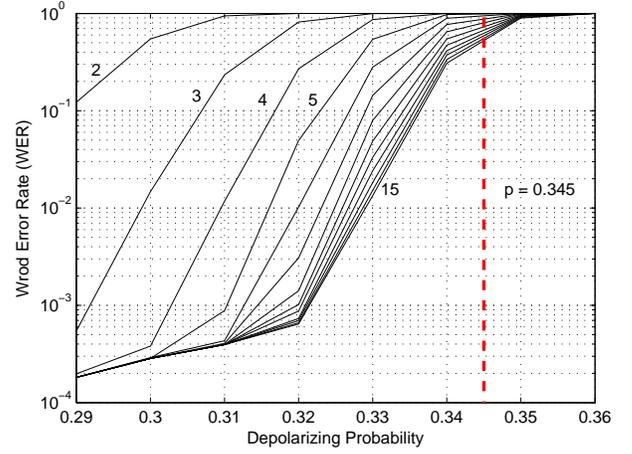}}
    \caption{WER performance curves with increasing iteration number for an interleaver length of $3,000$ qubits. \textit{Rate-1/9 concatenated code,
relying on PTO1REA as the inner code and the proposed QIRCC as the outer code, was used.}}
\label{fig:qircc_wer_ber_1}
%\vspace*{-0.5cm}
  \end{center}
\end{figure}

The corresponding Word Error Rate (WER) performance curves recorded for our QIRCC-based optimized design using a $3,000$ qubit long interleaver are 
seen in \fref{fig:qircc_wer_ber_1}, where the WER is reduced upon increasing the number of iterations. 
More explicitly, our code converges to a low WER for $p \leq 0.345$.
Thus, this convergence threshold matches the one predicted using EXIT charts in \fref{exit:PTO1REA-match}.
More explicitly, since the EXIT chart tunnel closes for $p > 0.345$, the system fails to
converge, if the depolarizing probability is increased beyond $0.345$.
Hence, the performance does not improve upon increasing the
number of iterations if the depolarizing probability exceeds
the threshold. By contrast, when the depolarizing probability
is below the threshold, the WER improves at each successive
iteration. It should also be
noted that the performance improves with diminishing returns
at a higher number of iterations. 

\begin{figure}[tb]
  \begin{center}
  {\includegraphics[width=\linewidth]{\figures 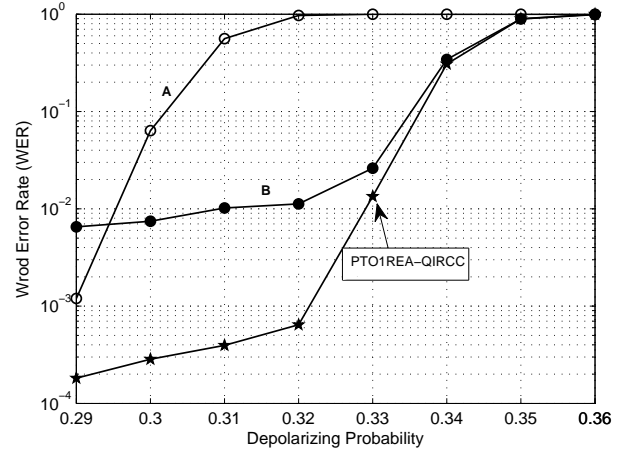}}
    \caption{Comparison of WER performance of our QIRCC-based optimized rate-$1/9$ QTC with the PTO1REA-PTO1R configuration of~\cite{wilde_turbo2} 
    (labeled ``A'') and the exhaustive-search based optimized QTC of~\cite{babar_QTC_2014} (labeled ``B'') for an interleaver length of $3,000$ qubits and a maximum of $15$ iterations.}
  \label{fig:qircc_wer_comp}
%\vspace*{-0.5cm}
  \end{center}
\end{figure}

\fref{fig:qircc_wer_comp} compares our QIRCC-based optimized design with the rate-$1/9$ ``PTO1REA-PTO1R'' configuration of~\cite{wilde_turbo2}, 
which is labeled ``A'' in the figure. An interleaver length of $3000$ qubits was used. For the ``PTO1REA-PTO1R'' configuration, the turbo cliff region 
emerges around $0.31$, which is within $0.9$ dB of the noise limit. Therefore, our QIRCC-based design outperforms the ``PTO1REA-PTO1R'' configuration 
of~\cite{wilde_turbo2}. %by around $\left[0.9 - 0.3 \right] = 0.6$ dB. 
More specifically, the ``PTO1REA-PTO1R'' configuration yields a WER of $10^{-3}$ at $p = 0.29$, while our design
gives a WER of $10^{-3}$ at $p = 0.322$. Hence, our optimized design outperforms the `PTO1REA-PTO1R'' configuration by about 
$\left[10 \times \log_{10} (\frac{0.29}{0.322})\right] = 0.5$ dB at a WER of $10^{-3}$. 
It must be mentioned here that the ``PTO1REA-PTO1R'' configuration may have a lower error floor than our design, yet our design exhibits a better performance in the turbo
cliff region.
We further compare our QIRCC-based optimized design with
the exhaustive-search based optimized turbo code of~\cite{babar_QTC_2014}, which is labeled ``B'' in \fref{fig:qircc_wer_comp}. Both code designs have similar convergence
threshold. However, our QIRCC-based design has a much lower error rate, resulting in a lower error floor as gleaned from \fref{fig:qircc_wer_comp}.
\section{Conclusions and Design Guidelines} \label{sec:conclusion}
Powerful QECCs are required for stabilizing and protecting the fragile constituent qubits of quantum computation as well as 
communication systems against the undesirable decoherence. In line with the developments in the field of classical channel coding theory,
this may be achieved by exploiting concatenated codes designs, which invoke iterative decoding. Therefore, in this paper we have laid out a slow-paced
tutorial for designing hashing bound approaching concatenated quantum codes using EXIT charts. To bridge the gap between the quantum and classical 
channel coding theory, we have provided insights into the transition from the classical to the quantum code design.
%by giving a tutorial insight into constructing 
%quantum stabilizer codes from the classical linear codes. 
More specifically, with the help of toy examples, we have illustrated that quantum block codes as 
well as convolutional codes may be constructed from arbitrary classical linear codes. %if the resulting PCM satisfies the symplectic criterion. A special
%class of such codes is called dual-containing CSS codes, whereby the symplectic criterion is simplified at the expense of imposing
%additional constraints on the code structure.
We then move onto the construction of concatenated quantum codes, focusing specifically on the
circuit-based structure of the constituent encoders and their equivalent classical representation as well as the degenerate iterative decoding. Finally,
we have detailed the procedure for generating EXIT charts for quantum codes and the principles of EXIT-chart aided design.
Our design guidelines may be summarized as follows:
\begin{itemize}
 \item As discussed in the context of our design objectives in Section~\ref{sec:designObj}, we commence our design by determining the noise limit
 $p^*$ for the desired code parameters, i.e the coding rate and the entanglement consumption rate of the resultant
 concatenated quantum code, which was introduced in Section~\ref{sec:designObj}.
 \item We then proceed with the selection of the inner stabilizer code of \fref{fig:system-model}, which has to be both recursive as well as 
 non-catastrophic, as argued in Section~\ref{sec:system-model}. Since the unassisted
 quantum codes cannot be simultaneously both recursive as well as non-catastrophic, we employ an entanglement-assisted code. Furthermore,
 the EA inner code of \fref{fig:system-model} may be either derived from the family of known classical codes, 
 as discussed in Section~\ref{sec:stabilizer} or it may be constructed using random Clifford operations,
 which were discussed in Section~\ref{sec:cct-QCC}. At this point, the EXIT curves of Section~\ref{sec:near-cap-design} may be invoked 
 for the sake of finding that specific inner code, which yields
 the largest area under its EXIT-curve at the noise limit $p^*$.
 \item Finally, we find the optimal non-catastrophic outer code of \fref{fig:system-model}, which gives the best EXIT-curve match with 
 that of the chosen inner code. 
 In this context, our EXIT-chart aided design of 
 Section~\ref{sec:near-cap-design} aims for creating a narrow, but marginally open tunnel between the EXIT curves of the inner and outer decoders at the 
highest possible depolarizing probability. The narrower the tunnel-area, the lower is the deviation from the
hashing bound, which may be quantified using \eqr{eq:distance}.
\end{itemize}
Recall that the desired code structure may also be optimized on the basis of a range of conflicting design challenges, which were illustrated in \fref{fig:Design_TO}.

Furthermore, for the sake of facilitating the hashing bound approaching code design, we have proposed the structure of QIRCC, which
constitutes the outer component of a concatenated quantum code. The proposed QIRCC allows us to dispense with the exhaustive code-search methods,
since it can be dynamically adapted to match any given inner code using 
EXIT charts. We have constructed a $10$-subcode QIRCC and used it as an outer code in concatenation
with a non-catastrophic and recursive inner convolutional code of~\cite{wilde_turbo,wilde_turbo2}. In contrast to the concatenated codes
of~\cite{wilde_turbo2}, whose performance is within $0.9$ dB of the hashing bound, our QIRCC-based optimized design operates within $0.4$ dB of the noise limit. 
Furthermore, at a WER of $10^{-3}$, our design outperforms the design of~\cite{wilde_turbo2} by around $0.5$ dB. Our proposed design also yields lower error rate
as compared to the exhaustive-search based optimized design of~\cite{babar_QTC_2014}.

\section*{Acknowledgement}
The expert advice of Dr. Mark Wilde is gratefully acknowledged.
\bibliographystyle{IEEEtran}

\bibliography{reference}

\end{document}